\documentclass[10pt]{article}
\usepackage{graphicx}
\usepackage[english]{babel}
\usepackage[autostyle]{csquotes}
\usepackage{amsmath,amsfonts}
\usepackage{amsthm}
\usepackage{geometry}
\usepackage{multirow}
\usepackage{subcaption}
\usepackage[toc,page]{appendix}
\usepackage{hyperref}
\hypersetup{
	colorlinks=true,
	linkcolor=blue,
	filecolor=blue,      
	urlcolor=blue,
	citecolor = blue,
	anchorcolor = blue 
}
\newtheorem{theorem}{Theorem}

\newtheorem{remark}{Remark}
\newtheorem{assumption}{Assumption}

\providecommand{\keywords}[1]{\textbf{\textbf{Key Words:}} #1}

\begin{document}

\title{\bf A Componentwise Estimation Procedure for Multivariate Location and Scatter: Robustness, Efficiency and Scalability}
\author{Soumya Chakraborty$^{1}\footnote{Corresponding author. Email address: \url{chak.soumya95@gmail.com}}$ , Ayanendranath Basu$^{2}$ and Abhik Ghosh$^{2}$
\\
$^{1}$Bethune College, $^{2}$Indian Statistical Institute, Kolkata, India.   
}

\maketitle
\begin{abstract}
\noindent
	Covariance matrix estimation is an important problem in multivariate data analysis, both from theoretical as well as applied points of view. Many simple and popular covariance matrix estimators are known to be severely affected by model misspecification and the presence of outliers in the data; on the other hand robust estimators with reasonably high efficiency are often computationally challenging for modern large and complex datasets. In this work, we propose a new, simple, robust and highly efficient method for estimation of the location vector and the scatter matrix for elliptically symmetric distributions. The proposed estimation procedure is designed in the spirit of the minimum density power divergence (DPD) estimation approach with appropriate modifications which makes our proposal (componentwise minimum DPD estimation) computationally very economical and scalable to large as well as higher dimensional datasets. Consistency and asymptotic normality of the proposed componentwise estimators of the multivariate location and scatter are established along with asymptotic positive definiteness of the estimated scatter matrix. Robustness of our estimators are studied by means of influence functions. All theoretical results are illustrated further under multivariate normality. A large-scale simulation study is presented to assess finite sample performances and scalability of our method in comparison to the usual maximum likelihood estimator (MLE), the ordinary minimum DPD estimator (MDPDE) and other popular non-parametric methods. The applicability of our method is further illustrated with a real dataset on credit card transactions.
%\url{mathscinet.ams.org/mathscinet/freeTools.html?version=2}.
\\
\\

\end{abstract}

\section{Introduction\label{sec:1}}
\noindent
Efficient estimation of covariance and correlation matrices is of prime interest in the analysis of multivariate data arising across all scientific disciplines. Many celebrated statistical tools, such as, discriminant analysis (linear as well as quadratic), principal component analysis, factor analysis and cluster analysis are based on the covariance structures of different attributes; see, e.g., Mardia et al.~$(1979) $ \cite{mardia}. That is why numerous estimation procedures for multivariate location and scatter have been proposed in the literature. 
The most eminent classical covariance estimator based on a random sample of multivariate observations,  $\{\boldsymbol{X}_{1},\ldots,\boldsymbol{X}_n\}$, is the sample covariance matrix $\widehat{\boldsymbol{\Sigma}}=\frac{1}{n}\sum_{i=1}^{n} (\boldsymbol{X}_{i}-\bar{\boldsymbol{X}})(\boldsymbol{X}_{i}-\bar{\boldsymbol{X}})^\top,$ where $\bar{\boldsymbol{X}}$ is the sample mean. It is indeed the method of moments estimator which is consistent, location invariant and scale equivariant. Under normality of the data, it is also the maximum likelihood estimator (MLE) and thus asymptotically the most efficient one under standard regularity conditions. But, this estimator is not resistant to outliers as it depends on the absolute values of the deviations of sample observations from the non-robust sample mean. The asymptotic breakdown point of this estimator is indeed zero indicating its extreme non-robust nature. However, the subjective complexity of real-life practical problems as well as the volume of data are increasing by the day and hence the analyses of such data are often expected to face the issues of model misspecification and sensitivity to outliers. Although non-parametric methods based on ranks, signs and depth measures provide possible solutions for non-robustness against model misspecification, they often lead to a major compromise in asymptotic efficiency along with the curse of dimensionality. Alternatively, when the majority of the data can be well-fitted by a parametric model, the effects of the outlying minority can be handled robustly, without a major loss in asymptotic efficiency, by using appropriate parametric robust procedures. 

Bickel $(1964)$ \cite{bickel_old}, and Sen and Puri $(1971)$ \cite{Sen and Puri} proposed coordinatewise median and R estimators for multivariate location estimation, respectively. Later Maronna (1976) \cite{maronna76} proposed the pioneering idea of simultaneous M-estimation of multivariate location and scatter for the elliptically symmetric distributions. The breakdown bounds and influence functions are calculated to show strong robustness properties of these multivariate M-estimators, although the upper bounds of their breakdown points ($1/(p+1)$ in dimension $p$, Donoho $(1982)$ \cite {donoho}$\Big)$ decrease to zero as the dimension grows to infinity. Tyler (\cite{tyler1,tyler3,tyler2}) and Kent and Tyler (\cite{tyler4,tyler5}) made significant contributions to the area of M-estimation of multivariate location and scatter, which include the constrained M-estimators (Kent and Tyler $(1996)$ \cite{tyler5}), distribution free M-estimators (Tyler $(1987)$ \cite{tyler3}) and redescending M-estimators (Kent and Tyler $(1991)$ \cite{tyler4}). Theoretical properties, indicating their robustness and efficiencies, have also been discussed in some of the aforesaid works. In particular, Tyler $(2014)$ \cite{tyler2} has shown that the asymptotic breakdown point of the multivariate scatter M-estimator is bounded above by $1/(p+1)$ only under certain \enquote{coplanar} contamination of the data. Under the absence of such \enquote{coplanar} contamination, the breakdown point of the scatter M-estimator is close to $1/2$ which fixes the problem of poor breakdown values of the scatter M-estimators in higher dimensions (a drawback of Maronna's method \cite{maronna76}). Another popular approach to construct robust estimates of location and scatter is to perform trimming on sample observations, leading to the popular minimum covariance determinant (MCD) and minimum volume ellipsoid (MVE) estimators. A detailed exposition of MCD and MVE estimators can be found in Rousseeuw and Driessen $(1999)$ \cite{mcd} and Rousseeuw $(1985)$ \cite{mcd1}, respectively. A major drawback of the MVE estimators is its slower convergence rate of $o(n^{-\frac{1}{3}})$. However, the MCD estimators are  $n^{\frac{1}{2}}$-consistent but compromise significantly in asymptotic efficiency compared to maximum likelihood estimators to achieve robustness (higher breakdown value) under normality assumption. Stahel (1981) \cite{Stahel} and Donoho $(1982)$ \cite{donoho} have independently proposed estimators of multivariate location and scatter as weighted means and weighted covariance matrices of the data where the weights are based on reasonable measures of \enquote{outlyingness}. Some other recent works in the field of robust estimation of multivariate location and scatter matrix include Danilov et al.~$(2012)$ \cite{missingjasa}, who have dealt with data contamination and missing data at the same time, Maronna and Yohai $(2014)$ \cite{recent_marona} and Agostinelli et al.~$(2015)$ \cite{recent}, who have proposed estimates of multivariate location and scatter that are resistant to both cellwise and casewise contamination, and Agostinelli and Greco $(2019)$ \cite{recent1}, who have proposed a weighted likelihood based approach for estimating multivariate location and scatter. Minimum distance methodology have also been utilized to produce robust and efficient estimators of multivariate location and scatter matrices; see Basu et al.~$(2011)$ \cite{basubook}, Ghosh et al.~$(2017)$ \cite{bernoulliayb} and Martin $(2020)$ \cite{nirian}.

One of the crucial challenges behind constructing robust estimators in multivariate set-ups is the resulting lower asymptotic efficiency along with significantly higher computational costs as dimension increases. Most existing robust estimators that enjoy high asymptotic efficiency (e.g., methods based on sophisticated iterative algorithms) are computationally expensive and, hence, it becomes really problematic to compute them for large and high-dimensional multivariate  datasets. A satisfactory solution to this problem is extremely important for analyzing modern large datasets that we are frequently encountering in the twenty-first century. One possible approach to settle this issue could be the incorporation of a two-stage procedure where a highly robust but inefficient estimator is first taken as the initial choice and subsequently an efficient estimator is produced in the second stage by solving suitable estimating equations. Gervini $(2003)$ \cite{jmv2003} has presented such a two-stage procedure that uses the reweighted estimators of multivariate location and scatter with data-driven weights, starting from a highly robust and fast (but not necessarily efficient) estimator at the first step. Componentwise estimation of location and scatter is another possible solution, which is computationally much simpler and often can maintain desirable robustness and asymptotic efficiency, simultaneously. Mehrotra $(1995)$ \cite{biometrics}, Ma and Genton $(2001)$ \cite{jmv2001} are two such examples where elementwise estimation of scatter matrix is done, respectively, using the modified A-estimator (Lax $(1985)$ \cite{lax}) of scale and using a highly robust estimator of scale by bypassing the estimation of location vector. Other notable (and recent) works in the field of robust estimation of location vectors and scale matrices in multivariate and high-dimensional set-ups include Raymaekers and Rousseeuw $(2019)$ \cite{cellmcd} and $(2024)$ \cite{frc}, Rousseeuw and Bossche $(2018)$ \cite{ref_cell_bossche}, Leung et al.~$(2017)$ \cite{leung_3sgs}, D\"umbgen $(1998)$ \cite{tyler_hd}, D\"umbgen and Nordhausen $(2024)$ \cite{ustat} and Miettinen et al.~$(2016)$ \cite{adv_rob_stat}.

In this article, we propose a new robust estimation procedure for the  multivariate location vector and the scatter matrix for the class of elliptically symmetric distributions, which is simultaneously fast and easy to compute via a suitably parallelised computational algorithm and also has significantly high asymptotic efficiency at the model distribution. In particular, the component means and variances are first estimated separately for each variable (in parallel), which are subsequently used to estimate the correlation coefficients between each pair of variables (in parallel). The covariances are then computed from the estimated variances and correlation values. In each step of estimation, we utilize nice properties, e.g., strong robustness and high efficiency, of the popular minimum density power divergence (DPD) estimator (MDPDE, Basu et al.~$(1998)$ \cite{basupaper}, $(2011)$ \cite{basubook}). As a result, our proposed estimators, which we refer to as the componentwise minimum DPD estimators (CMDPDEs) of multivariate location and scatter, also become highly robust and efficient along with significantly lower computational burden in all scenarios, particularly for larger datasets.

We present some important theoretical properties of our estimators in Sections \ref{sm_sec3.1} and \ref{sm_sec3.2}, such as, consistency and asymptotic normality of our estimators, asymptotic positive definiteness of our covariance matrix estimator and the influence function analysis of our functionals. Asymptotic relative efficiencies of our estimators (component means, variances and correlation estimators) and the explicit forms of the influence functions of our functionals are derived under the assumption of normality. In this scenario, we observed the superiority of CMDPDEs over the ordinary MDPDEs in terms of asymptotic efficiency, especially for the location and scale estimators; in case of the correlation also the CMDPDEs are, at the least, competitive with the ordinary MDPDEs. A simulation study is conducted to compare the componentwise method with the (non-robust) maximum likelihood method and (robust) ordinary minimum DPD method and some non-parametric methods including MCD and MVE algorithms (and others) which also suggest the possible superiority of the componentwise method in terms of bias and mean squared error. Obtaining numerical solutions (with small to moderate computing effort) for the estimates is guaranteed under the componentwise approach. Finding the ordinary MDPDE (the simultaneous minimizer) may, on the other hand, be an uphill task, particularly under growing dimensions and larger tuning parameter $\beta$; one will occasionally come across a situation, more common in small sample sizes (relative to dimensions), where practically none of the root solving techniques will be able to find numerical solutions for the simultaneous minimization problem.   

Importantly, since our newly proposed algorithm is componentwise, it can also be applied to high dimensional set-ups $(n\ll p)$ for analyzing massive datasets. However, to deal with sparsity, a crucial assumption in high dimensional literature, appropriate thresholding or regularization tools are needed to be incorporated on which we will work in future. 

In summary, the following positive points stand out among the benefits of our new proposal.

\begin{itemize}
	\item \textbf{Computational Superiority:} The principal advantage of this new componentwise method over the ordinary minimum DPD estimation procedure is the huge computational success of the componentwise philosophy. It offers fully converged estimates of the location and scale even in case of small sample sizes and large values of the tuning parameter $\beta$. The ordinary minimum DPD method, unfortunately, cannot match this performance which provides the prime motivation of this work.
	
	\item \textbf{Improved Empirical Performance:} As reflected in the simulation outputs, the new componentwise method often performs better or competitively as compared to the existing robust methodologies in terms of bias and mean squared errors, especially for the covariance matrix estimators.
	
	\item \textbf{Ease of Extension to High Dimension:} Although we develop a componentwise method for the robust and efficient estimation of multivariate location and scale under higher data dimensions in this work, our method can also be applied to standard high dimensional set-up, where number of parameters is greater than the sample size. As long as we have $4-5$ observations, our method can technically be applied to a multivariate data of any dimension; however, we recommend to apply this method with at least a sample of size around $50$ to get a reasonable accuracy of the parameter estimates.
	
\end{itemize}
Our paper is organized as follows. The method along with its statistical and probabilistic background is described in Section \ref{sm_sec2} while its theoretical properties are provided in Section \ref{sm_sec3}. The asymptotic relative efficiencies of our estimators and the detailed structures of our influence functions are presented in special cases. Simulation experiments and results are discussed in Section \ref{sm_sec4}.  In Section \ref{sm_sec5}, we apply our method on a real-life dataset on credit card transactions (with some fraudulent ones) to understand the capability of our method in maintaining desired robustness even in higher dimensions. Concluding remarks are given in Section \ref{sm_sec6}. Detailed mathematical derivations of the theoretical results, additional simulation results and other relevant algebraic calculations are provided in the Appendices.
\\
\newline 
\noindent
\textbf{Software Implementation:} The $\sf{R}$ codes of the newly proposed componentwise algorithm are available at \href{https://github.com/sc2-wbes/R-Codes-for-CMDPDE}{https://github.com/sc2-wbes/R-Codes-for-CMDPDE}.
\section{Model Set-up and the Proposed Estimation Procedure}
\label{sm_sec2}
\subsection{Elliptically Symmetric Distributions}
\label{sm_sec2.1}
Let $\boldsymbol{Z}=(Z_{1},\ldots,Z_{p})^\top$ be an absolutely continuous random vector in $\mathbb{R}^{p}$ which is spherically symmetric around the origin, that is, $\boldsymbol{Z}\overset{d}{=} \boldsymbol{P}\boldsymbol{Z}$
for any orthogonal matrix $\boldsymbol{P}$ of dimension $p \times p$. The probability density function (PDF) of $\boldsymbol{Z}$ is of the form $ k\psi(||z||)$, $z\in \mathbb{R}^p$ for a non-negative real valued function $\psi$ and normalizing constant $k>0$, where $||\cdot||$ represents the $L_{2}$-norm. The random vector $\boldsymbol{Z}$ is assumed to have finite second order moments. By choosing $\boldsymbol{P}=-\boldsymbol{I}_{p}$ it can be shown that $\text{E}(\boldsymbol{Z})=\boldsymbol{0}_{p}$. By choosing $\boldsymbol{P}$ to be different $p \times p$ permutation matrices, it can be shown that all the component variances of the random vector $\boldsymbol{Z}$ are same ($c$, say). Finally, by choosing $\boldsymbol{P}=\text{Diag}(1,\ldots,1,-1,1,\ldots,1)$ where the $i$-th diagonal element is $-1$, it can be shown that $\text{Cov}(Z_{i},Z_{j})=0$ for any $j \neq i$. Thus it gives $\text{E}(\boldsymbol{Z})=\boldsymbol{0}_{p}$ and $\text{Var}(\boldsymbol{Z})=c\boldsymbol{I}_p$. We assume $c=1$ for standardization. Let $\boldsymbol{\mu} \in \mathbb{R}^{p}$ and $\boldsymbol{\Sigma}$ be a symmetric, positive definite $p \times p$ matrix. We define the random vector $\boldsymbol{X}$ as, $\boldsymbol{X}=\boldsymbol{\mu}+\boldsymbol{\Sigma}^{\frac{1}{2}}\boldsymbol{Z}$, where $\boldsymbol{\Sigma}^{\frac{1}{2}}$ is the positive square root of $\boldsymbol{\Sigma}$. Hence, $\text{E}(\boldsymbol{X})=\boldsymbol{\mu}$ and $\text{Var}(\boldsymbol{X})=\boldsymbol{\Sigma}$. Then $\boldsymbol{X}$ is said to have an elliptically symmetric distribution, with the corresponding PDF of $\boldsymbol{X}$ being given by,
\begin{equation}
\centering
f_{\boldsymbol{\omega}}(\boldsymbol{x})=\frac{k}{|\boldsymbol{\Sigma}|^{\frac{1}{2}}}\psi\left(\sqrt{(\boldsymbol{x}-\boldsymbol{\mu})^\top\boldsymbol{\Sigma}^{-1}(\boldsymbol{x}-\boldsymbol{\mu})}\right),\;\boldsymbol{\omega}=(\boldsymbol{\mu}, \boldsymbol{\Sigma}).
\label{sm_eq3}
\end{equation}
We refer to Kelker $(1970)$ \cite{kelker} and Chmielewski $(1981)$ \cite{elliptical} for further details about spherical and elliptical distributions. The multivariate normal distribution is an example of elliptically symmetric distributions with $k=(2\pi)^{-\frac{p}{2}}$ and $\psi(x)=\text{exp}(-\frac{x^2}{2})$. Other notable examples include the multivariate $t$-distribution.
\subsection{The Minimum Density Power Divergence Estimator}
\label{sm_sec2.2}
 Let us introduce the notation $\mathcal{E}_{p}(\boldsymbol{\mu},\boldsymbol{\Sigma})$ to denote the family of elliptically symmetric distributions of dimension $p$, as described in Section \ref{sm_sec2.1}, with mean $\boldsymbol{\mu}=(\mu_1,\ldots,\mu_p)^{\top}$, covariance matrix $\boldsymbol{\Sigma}=\left[\left[\sigma_{ij}\right]\right]$ and PDF $f_{\boldsymbol{\omega}}$, as in Equation (\ref{sm_eq3}). Note that, the correlation between the $j$-th and $k$-th components is given by $\rho_{jk}=\frac{\sigma_{jk}}{\sqrt{\sigma_{jj} \sigma_{kk}}}$. Now, suppose that we have a random sample of multivariate observations  $\{\boldsymbol{X}_1,\ldots,\boldsymbol{X}_n\}$ (with $\boldsymbol{X}_{i}=(X_{i1},\ldots,X_{ip})^{\top}$ for $1\leq i\leq n$) from an unknown probability distribution with density $g$ (distribution $G$) and we model this unknown $g$ by the  $\mathcal{E}_{p}(\boldsymbol{\mu},\boldsymbol{\Sigma})$ model family. Our objective is to estimate this $\boldsymbol{\mu}$ and $\boldsymbol{\Sigma}$ and subsequently, the correlation matrix $\boldsymbol{R}=[[\rho_{jk}]]$. Thus, our parameter of interest is $\boldsymbol{\omega}=(\boldsymbol{\mu},\boldsymbol{\Sigma})$ and the parameter space is  $\boldsymbol{\Omega}=\Big\{\boldsymbol{\omega}=(\boldsymbol{\mu},\boldsymbol{\Sigma}): \boldsymbol{\mu} \in \mathbb{R}^{p},\;\boldsymbol{\Sigma} \in \mathbb{R}^{p\times p},\;\boldsymbol{\Sigma}\; \text{is symmetric and positive definite}\Big\}$. 

The MDPDE of $\boldsymbol{\omega}=(\boldsymbol{\mu},\boldsymbol{\Sigma})$ can be obtained by the proposal of Basu et al.~$(1998)$ \cite{basupaper}. More precisely, the MDPDE is defined as a minimizer of the DPD measure between the true density $g$ and the model density $f_{\boldsymbol{\omega}}$ with respect to $\boldsymbol{\omega}$, where the DPD measure is defined as 
 \begin{align}
 \centering 
 \label{sm_eq4}
 D_{\beta}(g,f_{\boldsymbol{\omega}}) =\int f_{\boldsymbol{\omega}}^{1+\beta}(\boldsymbol{x}) \;d\boldsymbol{x} - \left(1+\frac{1}{\beta}\right)\text{E}_{G}(f_{\boldsymbol{\omega}}^{\beta}(\boldsymbol{X})) + \frac{1}{\beta}\int g^{1+\beta}(\boldsymbol{x})\;d\boldsymbol{x}
 \end{align}
 with $\beta \in [0,1]$ being a tuning parameter controlling the trade-off between robustness and efficiency of the resulting estimator. Since the third term in the right hand side of Equation (\ref{sm_eq4}) is free from $\boldsymbol{\omega}$ and the second term can be estimated easily via replacing $G$ by the corresponding empirical cumulative distribution function (CDF), we can obtain the MDPDE, $\widehat{\boldsymbol{\omega}}_n$, as the minimizer of an equivalent yet simpler objective function as
\begin{equation*}
 \label{sm_eq6}
 \widehat{\boldsymbol{\omega}}_{n}=\underset{\boldsymbol{\omega} \in \boldsymbol{\Omega}}{\text{argmin}}\;\left[\int f_{\boldsymbol{\omega}}^{1+\beta}(\boldsymbol{x}) \;d\boldsymbol{x} - \left(1+\frac{1}{\beta}\right)\frac{1}{n}\sum_{i=1}^{n}f_{\boldsymbol{\omega}}^{\beta}(\boldsymbol{X}_{i})\right].
 \end{equation*}
 Note that, the estimator $\widehat{\boldsymbol{\omega}}_{n}$ is indeed a \enquote{simultaneous minimizer} of the DPD measure with respect to all components of the parameter. Further, as $\beta \rightarrow 0$, the DPD converges to the usual Kullback-Liebler divergence (Basu et al.~$(2011)$ \cite{basubook}) and, hence, the corresponding MDPDE becomes (in a limiting sense) the usual MLE. With increasing $\beta$, the MDPDE achieves greater outlier stability  with minimal loss in efficiency for most common parametric models. The MDPDE has been studied extensively for univariate and regression set-ups and has become highly popular in recent times due to its simple interpretation as a robust generalization of the MLE, high robustness against data contamination as well as high asymptotic efficiency at the assumed model family. However, the MDPDE under the multivariate set-up has not been studied in much detail; recently Martin $(2020)$ \cite{nirian} discussed the properties of the MDPDEs (robustness and efficiency) of the correlation matrices under multivariate normal model assumption while proposing a new robust version of the Rao's score test. Although, not theoretically studied, the MDPDEs of multivariate location and scale have previously been used for different applications. For example, Badsha et al.~$(2013)$ \cite{badsha} developed a robust hierarchical clustering algorithm based on these MDPDEs and applied the same for gene clustering. Recently, Chakraborty et al.~$(2023)$ \cite{arxiv_methodology}
  has also developed a robust clustering algorithm in multivariate normal mixture models by estimating the location and scatter of each multivariate normal component via minimum DPD estimation. In both of these applications, iteratively reweighted least squares (IRLS) algorithms are used to derive the multivariate MDPDEs and it has been observed that the convergences of these iterative algorithms (to the MDPDEs) are extremely dependent on the sample size $n$, tuning parameter $\beta$ and, most of all, on the data dimension $p$. Thus, the computation of MDPDE becomes problematic in higher dimensions especially with a moderate sample size and large value of $\beta$ (as needed to achieve higher robustness). The convergence rates also get slower with increase in dimension of the given multivariate data. Motivated by these observations, here we propose a new robust and efficient componentwise minimum DPD estimation method of multivariate location and scatter under elliptically symmetric models which is presented in the following subsection.
\subsection{Our Proposed Estimation Algorithm: The Componentwise MDPDE}
 \label{sm_sec2.3}
 Let us consider the problem of estimating multivariate location and scatter $(\boldsymbol{\mu},\boldsymbol{\Sigma})$ of the model distribution $\mathcal{E}_{p}(\boldsymbol{\mu},\boldsymbol{\Sigma})$ based on the random sample $\{\boldsymbol{X}_1,\ldots,\boldsymbol{X}_n\}$, as described in Section \ref{sm_sec2.2}. Now, we additionally assume that, for each $j\in\{1,\ldots,p\}$, the  marginal PDF of the $j$-th component be $f_j$ which is fully characterized by the parameters $\mu_j$ and $\sigma_{jj}$. Similarly, let the joint marginal PDF of the $j$-th and $k$-th components is $f_{jk}$ which is fully characterized by the parameters $\mu_{j},\;\sigma_{jj},\;\mu_{k},\;\sigma_{kk},\;\text{and}\;\rho_{jk}$ for all $1\leq j<k\leq p$. Such assumptions often hold for common elliptically symmetric distributions including the multivariate normal distribution; in the latter case $f_j$ and $f_{jk}$ are univariate and bivariate normals for all $1\leq j<k\leq p$, respectively. Analogously, let $g_{j}\;(G_{j})$ and $g_{jk}\;(G_{jk})$ be the true (unknown) marginal density (distribution) of the $j$-th component, $1\leq j\leq p$ and the joint density (distribution) of the $j$-th and $k$-th components, respectively.

Now, in order to propose the new componentwise algorithm for estimating $\boldsymbol{\mu}$ and $\boldsymbol{\Sigma}$, let us introduce some new notation by systematically reparametrizing $\boldsymbol{\mu}$ and $\boldsymbol{\Sigma}$. For notational simplicity, we hereonwards denote the $j$-th component variance $\sigma_{jj}$ by $\sigma^{2}_{j}$. The parameter of interest is   $\boldsymbol{\theta}=(\boldsymbol{\theta}_{1},\ldots,\boldsymbol{\theta}_{p},\boldsymbol{\rho})$ with $\boldsymbol{\theta}_{j}=(\theta_{j1},\theta_{j2})=(\mu_{j},\sigma^{2}_{j})$, $1\leq j \leq p$ and $\boldsymbol{\rho}=(\rho_{jk}:\;1\leq j < k\leq p)$ where $\rho_{jk}=\text{Cor}(X_{ij},X_{ik})=\frac{\sigma_{jk}}{\sigma_{j}\sigma_{k}}$ for $1\leq j < k\leq p$. The parameter space is given by $\boldsymbol{\Theta}=\{\boldsymbol{\theta}=(\mu_{1},\sigma^{2}_{1},\ldots,\mu_{p},\sigma^{2}_{p},\rho_{jk}:\;1\leq j < k\leq p): \mu_{j} \in \mathbb{R},\;\sigma^{2}_{j}>0,\;1\leq j \leq p,\;\rho_{jk} \in [-1,1],\;1\leq j < k\leq p\}$. Let,
\begin{align*}
H_{jn}(\boldsymbol{\theta}_{j})&=\int f_{j}^{1+\beta}(x,\boldsymbol{\theta}_{j}) \;dx - \left(1+\frac{1}{\beta}\right)\frac{1}{n}\sum_{i=1}^{n}f_{j}^{\beta}(X_{ij},\boldsymbol{\theta}_{j})=\frac{1}{n}\sum_{i=1}^{n}V_{j}(X_{ij},\boldsymbol{\theta}_{j}),\\
H_{jkn}(\boldsymbol{\theta}_{j},\boldsymbol{\theta}_{k},\rho_{jk})&=\int f_{jk}^{1+\beta}(x_1,x_2,\boldsymbol{\theta}_{j},\boldsymbol{\theta}_{k}, \rho_{jk}) \;dx_1\;dx_2 - \left(1+\frac{1}{\beta}\right)\frac{1}{n}\sum_{i=1}^{n}f_{jk}^{\beta}(X_{ij},X_{ik},\boldsymbol{\theta}_{j},\boldsymbol{\theta}_{k}, \rho_{jk})\\
&=\frac{1}{n}\sum_{i=1}^{n}V_{jk}(X_{ij},X_{ik},\boldsymbol{\theta}_{j},\boldsymbol{\theta}_{k}, \rho_{jk}),
\end{align*}
where $V_{j}(x,\boldsymbol{\theta}_{j})=\int f_{j}^{1+\beta}(t,\boldsymbol{\theta}_{j}) \;dt - \left(1+\frac{1}{\beta}\right)f_{j}^{\beta}(x,\boldsymbol{\theta}_{j}),\;1\leq j \leq p\; \text{and} \;V_{jk}(x_1,x_2,\boldsymbol{\theta}_{j},\boldsymbol{\theta}_{k},\rho_{jk})=\int f_{jk}^{1+\beta}(u,v,\boldsymbol{\theta}_{j},\boldsymbol{\theta}_{k}, \rho_{jk})\\ \;du\;dv - \left(1+\frac{1}{\beta}\right)f_{jk}^{\beta}(x_1,x_2,\boldsymbol{\theta}_{j},\boldsymbol{\theta}_{k}, \rho_{jk})$, $1\leq j < k\leq p$. The proposed componentwise minimum DPD estimators are defined as,
\begin{equation}
\centering
\label{sm_eq12}
\widehat{\boldsymbol{\theta}}_{jn}=\underset{\boldsymbol{\theta}_{j}}{\text{argmin}}\;H_{jn}(\boldsymbol{\theta}_{j}),\;\widehat{\rho}_{jkn}=\underset{\rho_{jk} \in [-1,1]}{\text{argmin}}\;H_{jkn}(\widehat{\boldsymbol{\theta}}_{jn},\widehat{\boldsymbol{\theta}}_{kn},\rho_{jk})
\end{equation}
and thus $\widehat{\boldsymbol{\theta}}_{n}=(\widehat{\boldsymbol{\theta}}_{1n},\ldots,\widehat{\boldsymbol{\theta}}_{pn},\hat{\rho}_{jkn}:1\leq j<k\leq p)$ for $n\geq 1$. We can now summarize these estimators to describe the new componentwise algorithm, as follows.
 \begin{enumerate}{}
 	\item[\null]{\textbf{Step 1:}} 
 	For each $j\in \{1,\ldots,p\}$, we  estimate the $j$-th component mean $\mu_j$ and variance $\sigma^{2}_{j}$ by minimizing the marginal DPD based objective function as 
 	\begin{equation}
 	\centering
 	\label{sm_eq8}
 	(\widehat{\mu}_{jn},\widehat{\sigma}^{2}_{jn})=(\hat{\theta}_{j1n},\hat{\theta}_{j2n})=\underset{\boldsymbol{\theta}_{j}}{\text{argmin}}\; H_{jn}(\boldsymbol{\theta}_{j}).
 	\end{equation}
 	 It gives us marginal MDPDEs of the mean and the variance corresponding to the $j$-th component. 
 	
 	\item[\null]{\textbf{Step 2:}} For any $1\leq j < k \leq p $, we  estimate the correlation coefficient of the $j$-th and $k$-th components, $\rho_{jk}$, by using the (marginal) MDPDEs of component means and variances obtained in Step $1$. In particular,we estimate the correlation coefficient $\rho_{jk}$ by minimizing the joint (bivariate) DPD based objective function corresponding to the $j$-th and $k$-th components after plugging in the (marginal) MDPDEs of component means and variances,  obtained in Step $1$, as 
 	\begin{equation}
 	\centering
 	\label{sm_eq10}
 	\widehat{\rho}_{jkn}=\underset{\rho_{jk} \in [-1,1]}{\text{argmin}}\;H_{jkn}(\boldsymbol{\hat{\theta}}_{jn},\boldsymbol{\hat{\theta}}_{kn},\rho_{jk}).
 	\end{equation}
 	\end{enumerate}
The resulting estimators of $\boldsymbol{\mu}$ and $\boldsymbol{\Sigma}$, as obtained from the above algorithm via $\widehat{\boldsymbol{\mu}}_{n}=(\widehat{\mu}_{1n},\ldots,\widehat{\mu}_{pn})'$ and $\widehat{\boldsymbol{\Sigma}}_{n} = ((\widehat{\sigma}_{jkn}))$  with $\widehat{\sigma}_{jkn}=\widehat{\sigma}_{jn}\widehat{\sigma}_{kn}\widehat{\rho}_{jkn}$ ( $\widehat{\sigma}^{2}_{jn}=\widehat{\sigma}_{jjn}$, $1\leq j \leq p$), are referred to as the \enquote{componentwise minimum DPD estimators} (CMDPDE). The corresponding CMDPDE of the correlation matrix is given by $\widehat{\boldsymbol{R}}_{n}=((\widehat{\rho}_{jkn}))$ with $\widehat{\rho}_{jjn}=1$ $\forall j\in\{1,\ldots,p\}$. Note that, we have estimated the component means and variances at first and then use these estimates to estimate the correlation coefficients further. So, the unknown correlation coefficients cannot affect the estimation of component means and variances. Following this observation, we can expect that our estimators of component means and variances will achieve greater efficiency than the usual (multivariate) MDPDE discussed in Martin $(2020)$ \cite{nirian} and Chakraborty et al.~$(2023)$ \cite{arxiv_methodology} in case of $\beta>0$; this advantage would be further illustrated in the subsequent sections, concretely under normal model distributions.
 \begin{remark}
 	 The optimization problem in Equation ($\ref{sm_eq8}$) does not have any closed form solution. Standard numerical procedures like Newton-Raphson, iteratively reweighted least squares (IRLS), etc, can be used to iteratively solve this optimization problem. In our numerical illustrations, we have used a suitable IRLS algorithm 
 	 which is a simplified version of the algorithm discussed in Chakraborty et al.~$(2023)$ \cite {arxiv_methodology} under the normality assumption; for general elliptical families the derivation of such iterative procedure is algebraically similar. The second constrained optimization problem in Equation (\ref{sm_eq10}) can be solved easily using any standard statistical software packages, e.g., the $\sf{optim}$ function in R software (R core team $(2018)$, \cite{Roptim}). In particular, we have utilised the $\sf{Brent}$ method within the $\sf{optim}$ function to solve the constrained optimization stated in Equation (\ref{sm_eq10}).
 \end{remark}
 \begin{remark}
\label{sm_highamremark} 
We have proved that our proposed CMDPDE of the covariance matrix is asymptotically positive definite (Theorem \ref{sm_pdt}). However, it is important to note that our method does not necessarily guarantee the positive definiteness of the covariance matrix estimator $\widehat{\boldsymbol{\Sigma}}_{n}$ in every finite sample case. This is a common problem of any componentwise procedures of covariance matrix estimation (see for example, Ma and Genton $(2001)$ \cite{jmv2001}). Some possible corrections were  suggested by Rousseeuw and Molenberghs $(1993)$ \cite{pd1} and Higham $(2002)$ \cite{pd2}. In particular, the second one proposed a numerical algorithm to compute the nearest positive semi-definite correlation matrix to a given symmetric matrix which is approximately a correlation matrix. Such a correction needs to be incorporated in order to make the proposed CMDPDE $\widehat{\boldsymbol{\Sigma}}_{n}$ positive definite in practical applications (i.e., real datasets with comparatively small sample sizes) if the need arises. The output of the Higham's method (if applied on the estimated correlation matrix which is not positive definite) will be positive semi-definite (with some eigenvalues as zero). A further eigenvalue truncation step can be applied to make it positive definite which replaces the zero eigenvalues by some small positive constant. This entire two-stage procedure (Higham's method followed by the eigenvalue truncation step) can be implemented in the R software through the $\sf{nearPD}$ function available in the $\sf{Matrix}$ package (Bates and Maechler $(2019)$ \cite{nearpd}) which will be utilized as necessary.
\end{remark}
In the next section, we will show that the proposed CMDPDEs are consistent even if the true distribution $(g)$ does not belong to the model family $\mathcal{E}_{p}(\boldsymbol{\mu},\boldsymbol{\Sigma})$. However, in such a case, the CMDPDE will converge to the best fitting value of the parameter $\boldsymbol{\theta^{g}}=(\boldsymbol{\theta}_{1}^{\boldsymbol{g}},\ldots,\boldsymbol{\theta}_{p}^{\boldsymbol{g}},\rho^{g}_{jk}:1\leq j<k\leq p)$, defined as, 
\begin{equation*}
\centering
\label{sm_eq11}
\boldsymbol{\theta}_{j}^{\boldsymbol{g}}=\underset{\boldsymbol{\theta}_{j}}{\text{argmin}}\;H_{j}(\boldsymbol{\theta}_{j}),\;\rho_{jk}^{g}=\underset{\rho_{jk} \in [-1,1]}{\text{argmin}}\;H(\boldsymbol{\theta}_{j}^{\boldsymbol{g}},\boldsymbol{\theta}_{k}^{\boldsymbol{g}},\rho_{jk}),
\end{equation*}
where $H_{j}(\boldsymbol{\theta}_{j})=\text{E}_{G_{j}}(V_{j}(X_{ij},\boldsymbol{\theta}_{j})),\;1\leq j \leq p,\;
H_{jk}(\boldsymbol{\theta}_{j},\boldsymbol{\theta}_{k},\rho_{jk})=\text{E}_{G_{jk}}(V_{jk}(X_{ij},X_{ik},\boldsymbol{\theta}_{j},\boldsymbol{\theta}_{k}, \rho_{jk})),\\1\leq j < k\leq p.$
\section{Theoretical Results}
\label{sm_sec3}
\subsection{Asymptotic Properties}
\label{sm_sec3.1}
\subsubsection{Technical Assumptions}
In order to establish the asymptotic properties of our estimators, such as, consistency and asymptotic normality, we need the following technical assumptions. The first three assumptions have been taken from Basu et al.~$(2011)$ \cite{basubook} (Section $9.2.1$) with little modifications.
\begin{assumption}
	\label{sm_a1}
	There is an open subset $\gamma_{0}$ of the parameter space $\boldsymbol{\Theta}$ such that for almost all $(x_{1},x_{2})\in \mathbb{R}^2$ and all $\boldsymbol{\theta} \in \gamma_{0}$, the densities $f_{j}$ $(1\leq j\leq p)$ and $f_{jk}$ $(1\leq j<k \leq p)$ are four times differentiable with respect to $\boldsymbol{\theta}$ and the fourth partial derivatives are continuous with respect to $\boldsymbol{\theta}$. 
\end{assumption}
\begin{assumption}
	\label{sm_a2}
	The integrals $\int f_{j}^{1+\beta}(x,\boldsymbol{\theta}_{j})\;dx$, $\int f_{jk}^{1+\beta}(x_1,x_2,\boldsymbol{\theta}_{j},\boldsymbol{\theta}_{k},\rho_{jk})\;dx_1dx_2$, $\int f_{j}^{\beta}(x,\boldsymbol{\theta}_{j})g_{j}(x)\;dx$ and $\int f_{jk}^{\beta}(x_1,x_2,\boldsymbol{\theta}_{j}\\,\boldsymbol{\theta}_{k}, \rho_{jk})g_{jk}(x_1,x_2)\;dx_1dx_2$ are differentiable four times with respect to $\boldsymbol{\theta}$ for $1\leq j\leq p$, $1\leq j<k \leq p$ and the derivatives can be taken under the integral signs.
\end{assumption}
\begin{assumption}
	\label{sm_a3}
	There exist $L_{1}$ bounded functions $M^{j}_{k_{1}l_{1}m_{1}}(x)$ which satisfy,
	\begin{align*}
	|\nabla_{k_{1}l_{1}m_{1}}V_{j}(x,\boldsymbol{\theta}_{j})|<M^{j}_{k_{1}l_{1}m_{1}}(x), \forall\;
    \boldsymbol{\theta}_{j},\; (1\leq j \leq p)\; \text{and}\; x\in \mathbb{R}.
	\end{align*}
\end{assumption}
 \begin{assumption}
	\label{sm_a4}
	Let, $U_{\rho_{jk}}(x_1,x_2,\rho_{jk}\;|\;\boldsymbol{\theta}_{j},\boldsymbol{\theta}_{k})=\frac{\partial \log\;f_{jk}(x_1,x_2,\boldsymbol{\theta}_{j},\boldsymbol{\theta}_{k},\rho_{jk}) }{\partial \rho_{jk}}$. Then, there exist $L_{1}$ bounded functions $M_{jk}(x_1,x_2,\rho_{jk})$ (with a finite integral) and $N_{jk}(x_1,x_2,\rho_{jk})$ which satisfy
	\begin{align*}
	\Big|\frac{\partial }{\partial \theta_{j_{1}k_{1}}}f_{jk}^{1+\beta}(x_1,x_2,\boldsymbol{\theta}_{j},\boldsymbol{\theta}_{k},\rho_{jk})U_{\rho_{jk}}(x_1,x_2,\rho_{jk}| \boldsymbol{\theta}_{j},\boldsymbol{\theta}_{k})\Big|&<M_{jk}(x_1,x_2,\rho_{jk}),\\
	\Big|\frac{\partial }{\partial \theta_{j_{1}k_{1}}}f_{jk}^{\beta}(x_1,x_2,\boldsymbol{\theta}_{j},\boldsymbol{\theta}_{k},\rho_{jk})
	U_{\rho_{jk}}(x_1,x_2,\rho_{jk}|\boldsymbol{\theta}_{j},\boldsymbol{\theta}_{k})\Big|&<N_{jk}(x_1,x_2,\rho_{jk}),
	\end{align*}
	 for all $\boldsymbol{\theta} \in \gamma_{0}$, $\rho_{jk}\in[-1,1]$ and $(x_1,x_2)\in \mathbb{R}^2$, where $(j_{1},k_{1})\in \{(j,1),(j,2),(k,1),(k,2)\}$, $1\leq j<k\leq p$.
\end{assumption}
\begin{assumption}
	\label{sm_a5}
	There exist $L_{1}$ bounded functions $v^{jk}_{1}(x_1,x_2,\boldsymbol{\theta}_{j},\boldsymbol{\theta}_{k})$ and $v^{jk}_{2}(x_1,x_2,\boldsymbol{\theta}_{j},\boldsymbol{\theta}_{k})$ which satisfy
	\begin{align*}
	\Big|\frac{\partial^{3} V_{jk}(x_1,x_2,\boldsymbol{\theta}_{j},\boldsymbol{\theta}_{k},\rho_{jk})}{\partial^{3} \rho_{jk}}\Big|<v^{jk}_{1}(x_1,x_2,\boldsymbol{\theta}_{j},\boldsymbol{\theta}_{k}),\;
	\Big|\frac{\partial^{4} V_{jk}(x_1,x_2,\boldsymbol{\theta}_{j},\boldsymbol{\theta}_{k},\rho_{jk})}{\partial^{3} \rho_{jk}\partial \theta_{j_{1}k_{1}}}\Big|<v^{jk}_{2}(x_1,x_2,\boldsymbol{\theta}_{j},\boldsymbol{\theta}_{k}),
	\end{align*}
	for all $\boldsymbol{\theta} \in \gamma_{0}$, $\rho_{jk}\in[-1,1]$ and $(x_1,x_2)\in\mathbb{R}^2$,
	where  $(j_{1},k_{1})\in \{ (j,1),(j,2),(k,1),(k,2)\}$, $1\leq j<k\leq p$.
\end{assumption}
It is important to note that, we have derived the estimator $\widehat{\boldsymbol{\theta}}_{jn}$ marginally by assuming $\{X_{1j},\ldots,\\X_{nj}\}$ as a univariate random sample ($1\leq j\leq p$). The weak consistency of $\widehat{\boldsymbol{\theta}}_{jn}$ ($1\leq j\leq p$) is already established in Basu et al.~$(2011)$ \cite{basubook} (Theorem $9.2$) and Basu et al.~$(1998)$ \cite{basupaper} (Theorem $2$) under appropriate assumptions (which include Assumptions \ref{sm_a1}-\ref{sm_a3}). Thus, with probability tending to $1$, there exists a sequence of estimators $\{\widehat{\boldsymbol{\theta}}_{jn}\}_{n=1}^{\infty}$, which satisfy,
\begin{equation}
\centering
\label{sm_eq13}
\widehat{\boldsymbol{\theta}}_{jn}\overset{p}{\rightarrow}\boldsymbol{\theta}_{j}^{\boldsymbol{g}},\;1\leq j\leq p.
\end{equation}
So, assuming the existence of these estimators, it remains to prove the consistency of $\widehat{\rho}_{jkn}$, defined as in Equation (\ref{sm_eq10}). Then, utilizing consistency of these estimators, we will establish their asymptotic normality. 
 
  Since the proposed estimating algorithm is componentwise, it is sufficient to prove the theoretical results under bivariate case ($p=2$) at first and then extend the results to general $p$-dimensional cases $(p\geq 3)$.
\subsubsection{Properties of the estimators in Bivariate Case $(p=2)$}
\label{sm_sec3.1.2}
As we have decided in the last paragraph, we now present the consistency and asymptotic normality of the CMDPDEs only for the bivariate case. So, here $p=2$ and hence the parameters are $\boldsymbol{\theta}_{1},\boldsymbol{\theta}_{2}$ and $\rho_{12}$ only. But for further notational simplicity, we will drop the suffix \enquote{$12$} from the required notations and expressions needed as the only possible pair of components is the pair of the first and second components under $p=2$. Precisely, we simply denote  $\rho=\rho_{12},\;\widehat{\rho}_{n}=\widehat{\rho}_{12n},\;\rho^{g}=\rho^{g}_{12},\; f=f_{12}$, $H_{n}=H_{n12},\;H=H_{12}$ and $V=V_{12}$. In fact, these notations will also be used in Sections \ref{sm_sec3.2} (the influence function analysis under $p=2$) and \ref{sm_sec3.3} (the normal model family). %The detailed mathematical proofs of the following theorems can be found in Appendix \ref{sm_appen0}.

Now, by the aforesaid notations, $\widehat{\rho}_{n}$ can be described as the solution of,
\begin{equation}
\label{sm_eq14}
\frac{\partial H_{n}(\widehat{\boldsymbol{\theta}}_{1n},\widehat{\boldsymbol{\theta}}_{2n},\rho)}{\partial \rho}=0.
\end{equation}
\begin{theorem}{\label{sm_consistency}}[Consistency] Under Assumptions \ref{sm_a1}-\ref{sm_a5}, Equation (\ref{sm_eq14}) has a consistent sequence of solution $\{\hat{\rho}_n\}$ with probability tending to $1$.
\end{theorem}
Our next job is to establish the asymptotic normality of $\widehat{\boldsymbol{\theta}}_{n}$ to study the asymptotic variances (thus asymptotic efficiencies) of the individual CMDPDEs. To state and prove the result, we need to introduce the following expressions. Let us define the random vector
\begin{align*}
\boldsymbol{\zeta}=\left(\frac{\partial\;V_{1}(X_{11},\boldsymbol{\theta}_{1})}{\partial\;\theta_{11}},\;\frac{\partial\;V_{1}(X_{11},\boldsymbol{\theta}_{1})}{\partial\;\theta_{12}},\;\frac{\partial\;V_{2}(X_{12},\boldsymbol{\theta}_{2})}{\partial\;\theta_{21}},\;\frac{\partial\;V_{2}(X_{12},\boldsymbol{\theta}_{2})}{\partial\;\theta_{22}},\;\frac{\partial\;V(X_{11},X_{12},\boldsymbol{\theta}_{1},\boldsymbol{\theta}_{2},\rho)}{\partial\;\rho}\right)\Bigg|_{\boldsymbol{\theta}=\boldsymbol{\theta}^{\boldsymbol{g}}}^{\top}
\end{align*}
  and its covariance matrix is given by $\boldsymbol{\Gamma_{0}}=\text{Var}_{g}(\boldsymbol{\zeta})$. Let us also define the matrix $\boldsymbol{B}$ as,
  
  \begin{align*}
  \boldsymbol{B}=\begin{bmatrix}
  b^{(1)}_{11} & b^{(1)}_{12} & 0 & 0 & 0\\
  b^{(1)}_{21} & b^{(1)}_{22} & 0 & 0 & 0\\
  0 & 0 & b^{(2)}_{11} & b^{(2)}_{12} & 0\\
  0 & 0 & b^{(2)}_{21} & b^{(2)}_{22} & 0\\ 
  e_{11} & e_{12} & e_{21} & e_{22} & a \\
  \end{bmatrix},
  \end{align*}
  where, $b^{(1)}_{jk}=\text{E}_{g}\left(\frac{\partial^{2}\;V_{1}(X_{11},\boldsymbol{\theta}_{1})}{\partial\;\theta_{1j}\;\partial\;\theta_{1k}}\right)\Big|_{\boldsymbol{\theta}_{1}=\boldsymbol{\theta}^{\boldsymbol{g}}_{1}}$, $b^{(2)}_{jk}=\text{E}_{g}\left(\frac{\partial^{2}\;V_{2}(X_{12},\boldsymbol{\theta}_{2})}{\partial\;\theta_{2j}\;\partial\;\theta_{2k}}\right)\Big|_{\boldsymbol{\theta}_{2}=\boldsymbol{\theta}^{\boldsymbol{g}}_{2}}$, \\ $e_{jk}=\text{E}_{g}\left(\frac{\partial^{2}\;V(\boldsymbol{X}_{1},\boldsymbol{\theta}_{1},\boldsymbol{\theta}_{2},\rho)}{\partial\;\theta_{jk}\;\partial\;\rho}\right)\Big|_{\boldsymbol{\theta}=\boldsymbol{\theta}^{\boldsymbol{g}}}$ for $j\in\{1,2\}$, $k\in\{1,2\}$ and $a=\text{E}_{g}\left(\frac{\partial^{2}\;V(\boldsymbol{X}_{1},\boldsymbol{\theta}_{1},\boldsymbol{\theta}_{2},\rho)}{\partial^{2} \rho}\right)\Big|_{\boldsymbol{\theta}=\boldsymbol{\theta}^{\boldsymbol{g}}}$.
  \begin{theorem}{\label{sm_normality}}[Asymptotic Normality]
  	Under Assumptions \ref{sm_a1}-\ref{sm_a5}, $\sqrt{n}(\widehat{\boldsymbol{\theta}}_{n}-\boldsymbol{\theta}^{g})$ converges in distribution to a normal distribution with zero mean and covariance matrix $\boldsymbol{\Gamma}=\boldsymbol{B}^{-1}\boldsymbol{\Gamma}_{0}{\boldsymbol{B}^{-1}}^\top$.
  \end{theorem}
\subsubsection{Properties of the Estimators under General Multivariate Set-up $(p\geq 3)$}
 Our next task is to extend the previous results in general $p$-dimensional scenarios. Under the general $p$-dimensional case, the parameter $\boldsymbol{\theta}=(\boldsymbol{\theta}_{1},\ldots,\boldsymbol{\theta}_{p},\rho_{12},\ldots,\rho_{1p},\rho_{23},\ldots,\rho_{2p},\ldots,\rho_{p-1,p})$ is $P=(p^{2}+3p)/2$ dimensional. Now we have the following result.
\begin{theorem}{\label{sm_pdim}}[Multivariate Consistency and Asymptotic Normality]
	For the general $p$-dimensional case, and under Assumptions \ref{sm_a1}-\ref{sm_a5}, the estimator $\widehat{\boldsymbol{\theta}}_{n}\overset{p}{\rightarrow}\boldsymbol{\theta^{g}}$ and $\sqrt{n}(\widehat{\boldsymbol{\theta}}_{n}-\boldsymbol{\theta^{g}})\overset{d}{\rightarrow}\;N(0,\boldsymbol{\widetilde{\Gamma}})$, where $\boldsymbol{\widetilde{\Gamma}}=\boldsymbol{\widetilde{B}}^{-1}\boldsymbol{\widetilde{\Gamma}_{0}}{\boldsymbol{\widetilde{B}}^{-1}}^\top$. 
	\end{theorem}
Here, the matrix $\boldsymbol{\widetilde{\Gamma_{0}}}$ is the covariance matrix of the random vector
\begin{align*}
&\Bigg(\frac{\partial\;V_{1}(X_{11},\boldsymbol{\theta}_{1})}{\partial\;\theta_{11}},\;\frac{\partial\;V_{1}(X_{11},\boldsymbol{\theta}_{1})}{\partial\;\theta_{12}},\ldots,\;\frac{\partial\;V_{p}(X_{1p},\boldsymbol{\theta}_{p})}{\partial\;\theta_{p1}},\;\frac{\partial\;V_{p}(X_{p2},\boldsymbol{\theta}_{p})}{\partial\;\theta_{p2}},\;\frac{\partial\;V_{12}((X_{11},X_{12}),\boldsymbol{\theta}_{1},\boldsymbol{\theta}_{2},\rho_{12})}{\partial\;\rho_{12}},\ldots,\\
&\frac{\partial\;V_{p-1,p}((X_{1,p-1},X_{1p}),\boldsymbol{\theta}_{p-1},\boldsymbol{\theta}_{p},\rho_{p-1,p})}{\partial\;\rho_{p-1,p}}\Bigg)\Bigg|_{\boldsymbol{\theta}=\boldsymbol{\theta^{g}}}^{\top}
\end{align*}
and 
\begin{align*}
\boldsymbol{\widetilde{B}}=\begin{bmatrix}
b^{(1)}_{11} & b^{(1)}_{12} & 0 & 0  & \cdots & \cdots & \cdots & \cdots & \cdots & \cdots & \cdots & \cdots & \cdots \\
b^{(1)}_{21} & b^{(1)}_{22} & 0 & 0 & \cdots & \cdots & \cdots & \cdots & \cdots & \cdots & \cdots & \cdots & \cdots \\
0 & 0 & b^{(2)}_{11} & b^{(2)}_{12} & \cdots & \cdots & \cdots & \cdots & \cdots & \cdots & \cdots & \cdots & \cdots \\
0 & 0 & b^{(2)}_{21} & b^{(2)}_{22} &  \cdots & \cdots & \cdots & \cdots & \cdots & \cdots & \cdots & \cdots & \cdots \\ 
\cdots & \cdots & \cdots&\cdots & \cdots & \cdots & \cdots & \cdots & \cdots & \cdots & \cdots & \cdots & \cdots \\
0 & 0 & \cdots & \cdots &\cdots  & b^{(p-1)}_{11} & b^{(p-1)}_{12} & \cdots & \cdots & \cdots & \cdots & \cdots & \cdots \\
0 & 0 & \cdots & \cdots & \cdots  & b^{(p-1)}_{21} & b^{(p-1)}_{22} & \cdots & \cdots & \cdots & \cdots & \cdots & \cdots\\
0 & 0 & \cdots & \cdots  & \cdots & \cdots & \cdots & b^{(p)}_{11} & b^{(p)}_{12} & \cdots & \cdots & \cdots & \cdots \\
0 & 0 & \cdots & \cdots  & \cdots & \cdots & \cdots & b^{(p)}_{21} & b^{(p)}_{22} & \cdots & \cdots & \cdots & \cdots\\
e^{(1)}_{12} & e^{(2)}_{12} & e^{(3)}_{12} & e^{(4)}_{12} & 0 & 0 & \cdots & 0 & \cdots & a_{12} & 0 & \cdots & \cdots \\
\cdots & \cdots & \cdots&\cdots & \cdots & \cdots & \cdots & \cdots & \cdots & \cdots & \cdots & \cdots & \cdots\\
0 & 0 &\cdots &	\cdots & \cdots & e^{(1)}_{p-1,p} & e^{(2)}_{p,p-1} & e^{(3)}_{p-1,p} & e^{(4)}_{p-1,p} & 0 & \cdots & \cdots & a_{p-1,p}
\end{bmatrix}.
\end{align*}
with $b^{(j)}_{k_{1}l_{1}}=\text{E}_{g_{j}}\left(\frac{\partial^{2}\;V_{j}(X_{1j},\boldsymbol{\theta}_{j})}{\partial\;\theta_{jk_{1}}\;\partial\;\theta_{jl_{1}}}\right)\Big|_{\boldsymbol{\theta}_{j}=\boldsymbol{\theta}^{\boldsymbol{g}}_{j}}$ for $k_{1},l_{1}\in \{1,2\}$,\\
$e^{(1)}_{jk}=\text{E}_{g_{jk}}\left(\frac{\partial^{2}\;V_{jk}((X_{1j},X_{1k}),\boldsymbol{\theta}_{j},\boldsymbol{\theta}_{k},\rho_{jk})}{\partial\;\theta_{j1}\;\partial\;\rho_{jk}}\right)\Big|_{(\boldsymbol{\theta}_{j},\boldsymbol{\theta}_{k})=(\boldsymbol{\theta^{g}}_{j},\boldsymbol{\theta^{g}}_{k}),\rho_{jk}=\rho_{jk}^{g}}$,\\ $e^{(2)}_{jk}=\text{E}_{g_{jk}}\left(\frac{\partial^{2}\;V_{jk}((X_{1j},X_{1k}),\boldsymbol{\theta}_{j},\boldsymbol{\theta}_{k},\rho_{jk})}{\partial\;\theta_{j2}\;\partial\;\rho_{jk}}\right)\Big|_{(\boldsymbol{\theta}_{j},\boldsymbol{\theta}_{k})=(\boldsymbol{\theta^{g}}_{j},\boldsymbol{\theta^{g}}_{k}),\rho_{jk}=\rho_{jk}^{g}}$, \\ $e^{(3)}_{jk}=\text{E}_{g_{jk}}\left(\frac{\partial^{2}\;V_{jk}((X_{1j},X_{1k}),\boldsymbol{\theta}_{j},\boldsymbol{\theta}_{k},\rho_{jk})}{\partial\;\theta_{k1}\;\partial\;\rho_{jk}}\right)\Big|_{(\boldsymbol{\theta}_{j},\boldsymbol{\theta}_{k})=(\boldsymbol{\theta^{g}}_{j},\boldsymbol{\theta^{g}}_{k}),\rho_{jk}=\rho_{jk}^{g}}$,\\ $e^{(4)}_{jk}=\text{E}_{g_{jk}}\left(\frac{\partial^{2}\;V_{jk}((X_{1j},X_{1k}),\boldsymbol{\theta}_{j},\boldsymbol{\theta}_{k},\rho_{jk})}{\partial\;\theta_{k2}\;\partial\;\rho_{jk}}\right)\Big|_{(\boldsymbol{\theta}_{j},\boldsymbol{\theta}_{k})=(\boldsymbol{\theta^{g}}_{j},\boldsymbol{\theta^{g}}_{k}),\rho_{jk}=\rho_{jk}^{g}}$, \\ $a_{jk}=\text{E}_{g_{jk}}\left(\frac{\partial^{2}\;V_{jk}((X_{1j},X_{1k}),\boldsymbol{\theta}_{j},\boldsymbol{\theta}_{k},\rho_{jk})}{\partial\;\rho_{jk}^{2}}\right)\Big|_{(\boldsymbol{\theta}_{j},\boldsymbol{\theta}_{k})=(\boldsymbol{\theta^{g}}_{j},\boldsymbol{\theta^{g}}_{k}),\rho_{jk}=\rho_{jk}^{g}}$, $1\leq j<k\leq p$.\\
Our next result is to establish the asymptotic positive definiteness of our covariance matrix estimator $\widehat{\boldsymbol{\Sigma}}_{n}=((\widehat{\sigma}_{nij}))_{i,j=1}^{p}$. 
%\begin{equation*}
%\widehat{\sigma}_{nij}= 
%\begin{cases}
% \widehat{\rho}_{nij}\widehat{\sigma}_{ni}\widehat{\sigma}_{ni},& \text{for } i \neq j\\
%\widehat{\sigma}^{2}_{ni}           & \text{for } i=j.
%\end{cases}
%\end{equation*}
The true covariance matrix is $\boldsymbol{\Sigma^{g}}=((\sigma^{g}_{ij}))_{i,j=1}^{p}$.
%\begin{equation*}
%\sigma^{g}_{ij}= 
%\begin{cases}
%\rho^{g}_{ij}\sigma^{g}_{i}\sigma^{g}_{i},& \text{for } i \neq j\\
%\widehat{\sigma^{g}}^{2}_{i}           & \text{for } i=j.
%\end{cases}
%\end{equation*}
 It will be established as a consequence of Theorem \ref{sm_consistency} under a specific assumption. We assume the following.
\begin{assumption}
	\label{sm_a6}
	The minimum eigenvalue of $\boldsymbol{\Sigma}^{\boldsymbol{g}}$ is bounded away from $0$, that is, there exists a positive constant $c$ such that $\lambda^{g}_{(1)}\geq c$ where $\lambda^{g}_{(1)}$ is the minimum eigenvalue of $\boldsymbol{\Sigma}^{\boldsymbol{g}}$.
\end{assumption}
\begin{theorem}{\label{sm_pdt}}[Asymptotic Positive Definiteness] Under Assumptions \ref{sm_a1}$-$\ref{sm_a6}, the estimator $\widehat{\boldsymbol{\Sigma}}_{n}$ is positive definite with probability tending to $1$ as $n\rightarrow \infty$.
\end{theorem}
\noindent
Mathematical proofs of the aforesaid theorems are provided in Appendix \ref{sm_appen0}.
\subsection{Influence Functions of the CMDPDEs}
\label{sm_sec3.2}
As noted previously, it is enough to study the properties of the CMDPDE in the bivariate case as our procedure is
indeed based on all possible bivariate combinations from given multivariate data; in particular, the same is true for the influence function as well. Let us use the same notations as in Section \ref{sm_sec3.1.2}. To derive the influence functions, we have to study the algebraic relationships among the statistical functionals corresponding to our CMDPDEs (i.e., the corresponding best fitting parameters, observed as functionals of the true density $g$), viz., $\theta^{g}_{11},\;\theta^{g}_{12},\;\theta^{g}_{21},\;\theta^{g}_{22},$ and $\rho^{g}$ (i.e., $\mu^{g}_{1}$, $\sigma^{g^2}_{1}$, $\mu^{g}_{2}$, $\sigma^{g^2}_{2}$ and $\rho^{g}$, respectively). By definition, $\theta^{g}_{jk}$ ($j\in\{1,2\}$ and $k\in\{1,2\}$) can be derived by solving,
\begin{align*}
\centering 
\frac{\partial}{\partial\;\theta_{jk}}\left[\int f^{1+\beta}_{j}(x,\boldsymbol{\theta}_{j})\;dx-\left(1+\frac{1}{\beta}\right)\int f^{\beta}_{j}(x,\boldsymbol{\theta}_{j})g_{j}(x)\;dx\right]=0\;\text{for}\;j\in\{1,2\},\;k\in\{1,2\}.
\end{align*}
Since the model density $f$ belongs to a location-scale family, $\int f^{1+\beta}_{j}(x,\boldsymbol{\theta}_{j})\;dx$ is independent of $\theta_{j1}$ for $j\in\{1,2\}$. So, the aforesaid equation boils down to
\begin{equation*}
\centering 
\label{sm_eq28}
\int f^{\beta}_{j}(x,\boldsymbol{\theta}_{j})U_{\theta_{j1}}(x,\boldsymbol{\theta}_{j})g_{j}(x)\;dx=0,\;j\in\{1,2\},
\end{equation*}
and 
\begin{equation*}
\centering 
\label{sm_eq29}
\left[\int f^{1+\beta}_{j}(x,\boldsymbol{\theta}_{j})U_{\theta_{j2}}(x,\boldsymbol{\theta}_{j})\;dx-\int f^{\beta}_{j}(x,\boldsymbol{\theta}_{j})U_{\theta_{j2}}(x,\boldsymbol{\theta}_{j})g_{j}(x)\;dx\right]=0,\;j\in\{1,2\},
\end{equation*}
where $U_{\theta_{jk}}(x,\boldsymbol{\theta}_{j})=\frac{\partial}{\partial\;\theta_{jk}}\log\;f_{j}(x,\boldsymbol{\theta}_{j})$ for $j\in\{1,2\},\;k\in\{1,2\}$. 
%But for $k=2$, we have, 
%Similarly, the functionals $\theta^{g}_{21}$ and $\theta^{g}_{22}$ can be derived by solving,
%\begin{equation}
%\centering 
%\label{sm_eq30}
%\int f^{\beta}_{2}(x,\boldsymbol{\theta}_{2})U_{\theta_{21}}(x,\boldsymbol{\theta}_{2})g_{2}(x)\;dx=0
%\end{equation} and
%\begin{equation}
%\centering 
%\label{sm_eq31}
%\left[\int %f^{1+\beta}_{2}(x,\boldsymbol{\theta}_{2})U_{\theta_{22}}(x,\boldsymbol{\theta}%_{2})\;dx-\int %f^{\beta}_{2}(x,\boldsymbol{\theta}_{2})U_{\theta_{22}}(x,\boldsymbol{\theta}_{%2})g_{2}(x)\;dx\right]=0.
%\end{equation}
The functional $\rho^{g}$ can be derived by solving,
\begin{equation*}
\centering 
\label{sm_eq32}
\begin{array}{l}
\int f^{1+\beta}(x_1,x_2,\boldsymbol{\theta}_{1}^{\boldsymbol{g}},\boldsymbol{\theta}_{2}^{\boldsymbol{g}},\rho)U_{\rho}(x_1,x_2,\boldsymbol{\theta}_{1}^{\boldsymbol{g}},\boldsymbol{\theta}_{2}^{\boldsymbol{g}},\rho)\;dx_1dx_2\\
-\int f^{\beta}(x_1,x_2,\boldsymbol{\theta}_{1}^{\boldsymbol{g}},\boldsymbol{\theta}_{2}^{\boldsymbol{g}},\rho)U_{\rho}(x_1,x_2,\boldsymbol{\theta}_{1}^{\boldsymbol{g}},\boldsymbol{\theta}_{2}^{\boldsymbol{g}},\rho)g(x_1,x_2)\;dx_1dx_2=0, 
\end{array}
\end{equation*}
where $U_{\rho}(x_1,x_2,\boldsymbol{\theta}_{1}^{\boldsymbol{g}},\boldsymbol{\theta}_{2}^{\boldsymbol{g}},\rho)=\frac{\partial}{\partial\;\rho}\log\;f(x_1,x_2,\boldsymbol{\theta}_{1}^{\boldsymbol{g}},\boldsymbol{\theta}_{2}^{\boldsymbol{g}},\rho)$.
% Now, we are in a position to initiate the derivation of the influence functions of the aforesaid functionals. To do that explicitly, we need to fix a model family. We derive the influence functions under bivariate normal model and also assume that the true distribution belongs to the aforesaid model family, that is, the true distribution is also bivariate normal (thus $g(\cdot)=f(\cdot,\theta^{g},\rho^{g})$).
To derive the influence functions, our first step is to contaminate the true distribution at a prefixed point in $\mathbb{R}^2$. In particular, the contaminated distribution function is given by $G_{\epsilon}=(1-\epsilon)G+\epsilon \wedge_{\boldsymbol{y}}$ where $\boldsymbol{y}=(y_{1},y_{2})$ is the point of contamination and $\wedge_{\boldsymbol{y}}$ is the probability distribution degenerate at $\boldsymbol{y}$. Note that, the marginals corresponding to $G_{\epsilon}$ would be $G_{j\epsilon}=(1-\epsilon)G_{j}+\epsilon \wedge_{y_{j}}$ for $j\in\{1,2\}$. Let, $g_{\epsilon}$, $g_{1\epsilon}$ and $g_{2\epsilon}$ are the corresponding contaminated densities, respectively. Let us also assume that $\theta_{jk\epsilon}$ for $j\in\{1,2\},\;k\in\{1,2\}$ and $\rho_{\epsilon}$ are the best fitting parameters corresponding to the contaminated distribution, i.e., $G_{\epsilon}$. Now, by definition, the influence functions of the functionals can be derived by solving the following system of equations. 
\begin{equation}
\label{sm_eq33}
\begin{array}{l}
\frac{\partial}{\partial \epsilon}\left[\int f^{\beta}_{1}(x,\boldsymbol{\theta}_{1\epsilon})U_{\theta_{11\epsilon}}(x,\boldsymbol{\theta}_{1\epsilon})g_{1\epsilon}(x)\;dx
\right]\Big|_{\epsilon=0}=0,\\
\frac{\partial}{\partial \epsilon}\left[\int f^{1+\beta}_{1}(x,\boldsymbol{\theta}_{1\epsilon})U_{\theta_{12\epsilon}}(x,\boldsymbol{\theta}_{1\epsilon})\;dx-\int f^{\beta}_{1}(x,\boldsymbol{\theta}_{1\epsilon})U_{\theta_{12\epsilon}}(x,\boldsymbol{\theta}_{1\epsilon})g_{1\epsilon}(x)\;dx
\right]\Big|_{\epsilon=0}=0,\\
\frac{\partial}{\partial \epsilon}\left[\int f^{\beta}_{2}(x,\boldsymbol{\theta}_{2\epsilon})U_{\theta_{21\epsilon}}(x,\boldsymbol{\theta}_{2\epsilon})g_{2\epsilon}(x)\;dx=0\right]\Big|_{\epsilon=0}=0,\\
\frac{\partial}{\partial \epsilon}\left[\int f^{1+\beta}_{2}(x,\boldsymbol{\theta}_{2\epsilon})U_{\theta_{22\epsilon}}(x,\boldsymbol{\theta}_{2\epsilon})\;dx-\int f^{\beta}_{2}(x,\boldsymbol{\theta}_{2\epsilon})U_{\theta_{22\epsilon}}(x,\boldsymbol{\theta}_{2})g_{2\epsilon}(x)\;dx\right]\Big|_{\epsilon=0}=0,\\
\frac{\partial}{\partial \epsilon}\Big[\int f^{1+\beta}(x_1,x_2,\boldsymbol{\theta}_{1\epsilon},\boldsymbol{\theta}_{2\epsilon},\rho_{\epsilon})U_{\rho_{\epsilon}}(x_1,x_2,\boldsymbol{\theta}_{1\epsilon},\boldsymbol{\theta}_{2\epsilon},\rho_{\epsilon})\;dx_1dx_2\\-\int f^{\beta}(x_1,x_2,\boldsymbol{\theta}_{1\epsilon},\boldsymbol{\theta}_{2\epsilon},\rho_{\epsilon})U_{\rho_{\epsilon}}(x_1,x_2,\boldsymbol{\theta}_{1\epsilon},\boldsymbol{\theta}_{2\epsilon},\rho_{\epsilon})g_{\epsilon}(x_1,x_2)\;dx_1dx_2\Big]\Big|_{\epsilon=0}=0.\\
\end{array}
\end{equation}
Further simplification of the System (\ref{sm_eq33}) is not possible in general for any arbitrary elliptically symmetric probability model but can be done in specific cases. The simplified algebraic forms of the aforesaid influence functions are derived for the common 
normal model in Section \ref{sm_sec3.3.1} where a specific example is also presented pictorially to illustrate its behaviour indicating robustness.
\subsection{Example: Normal Model Family}
\label{sm_sec3.3}
\subsubsection{Influence Functions}
\label{sm_sec3.3.1}
After simplifying and solving the system (\ref{sm_eq33}) in case of bivariate normal model family, we have the following influence functions of the respective functionals.
\begin{align*}
&\text{IF}(\theta_{j1},y_{j},g_{j})=(\theta^{g}_{j2})^{\frac{\beta}{2}}(1+\beta)^{\frac{3}{2}}(2\pi)^{\frac{\beta}{2}}(y_{j}-\theta^{g}_{j1})f^{\beta}_{j}(y_{j},\boldsymbol{\theta}^{\boldsymbol{g}}_{j})\;\text{for}\;j\in\{1,2\},\\
&\text{IF}(\theta_{j2},y_{j},g_{j})=\frac{f^{\beta}_{j}(y_{j},\boldsymbol{\theta}^{\boldsymbol{g}}_{j})(-\theta^{g}_{j2}+(y_{j}-\theta^{g}_{j1})^{2})-\int f^{1+\beta}_{j}(x,\boldsymbol{\theta}^{\boldsymbol{g}}_{j})(-\theta^{g}_{j2}+(x-\theta^{g}_{j1})^{2})\;dx }{\frac{1}{2}\int f^{1+\beta}_{j}(x,\boldsymbol{\theta}^{\boldsymbol{g}}_{j})\left(\frac{(x-\theta^{g}_{j1})^{2}}{\theta^{g}_{j2}}-1\right)^{2}\;dx}\;\text{for}\;j\in\{1,2\},\\
&\text{IF}(\rho,y_1,y_2,g)=\frac{f^{\beta}(y_1,y_2,\boldsymbol{\theta}_{1}^{\boldsymbol{g}},\boldsymbol{\theta}_{2}^{\boldsymbol{g}},\rho^{g})U_{\rho^{g}}(y_1,y_2,\boldsymbol{\theta}_{1}^{\boldsymbol{g}},\boldsymbol{\theta}_{2}^{\boldsymbol{g}},\rho^{g})}{\int B(x_1,x_2) f^{\beta}(x_1,x_2,\boldsymbol{\theta}_{1}^{\boldsymbol{g}},\boldsymbol{\theta}_{2}^{\boldsymbol{g}},\rho^{g})U_{\rho^{g}}(x_1,x_2,\boldsymbol{\theta}_{1}^{\boldsymbol{g}},\boldsymbol{\theta}_{2}^{\boldsymbol{g}},\rho^{g})\;dx_1dx_2}\\
&~~~~~~~~~~~~~~-\frac{\int (1+A(x_1,x_2,y_1,y_2)) f^{\beta}(x_1,x_2,\boldsymbol{\theta}_{1}^{\boldsymbol{g}},\boldsymbol{\theta}_{2}^{\boldsymbol{g}},\rho^{g})U_{\rho^{g}}(x_1,x_2,\boldsymbol{\theta}_{1}^{\boldsymbol{g}},\boldsymbol{\theta}_{2}^{\boldsymbol{g}},\rho^{g})\;dx_1dx_2}{\int B(x_1,x_2) f^{\beta}(x_1,x_2,\boldsymbol{\theta}_{1}^{\boldsymbol{g}},\boldsymbol{\theta}_{2}^{\boldsymbol{g}},\rho^{g})U_{\rho^{g}}(x_1,x_2,\boldsymbol{\theta}_{1}^{\boldsymbol{g}},\boldsymbol{\theta}_{2}^{\boldsymbol{g}},\rho^{g})\;dx_1dx_2}, 
\end{align*}
where, detailed algebraic expressions for $A(x_1,x_2,y_1,y_2)$, $B(x_1,x_2)$  and the integrals in the aforesaid influence functions are provided in Appendix \ref{sm_appen2}.

Let us observe that the functions $\text{IF}(\theta_{j1},y_{j},g_{j})$ and $\text{IF}(\theta_{j2},y_{j},g_{j})$ are linear in $(y_{j}-\theta^{g}_{j1})f^{\beta}_{j}(y_{j},\boldsymbol{\theta}^{\boldsymbol{g}}_{j})$, $(y_{j}-\theta^{g}_{j1})^{2}f^{\beta}_{j}(y_{j},\boldsymbol{\theta}^{\boldsymbol{g}}_{j})$ and $f^{\beta}_{j}(y_{j},\boldsymbol{\theta}^{\boldsymbol{g}}_{j})$ for $j\in\{1,2\}$. Since each of these functions are bounded in $y_{j}$, the boundedness of both $\text{IF}(\theta_{j1},y_{j},g_{j})$ and $\text{IF}(\theta_{j2},y_{j},g_{j})$ for $j\in\{1,2\}$ is trivial. By observing
\begin{align*}
U_{\rho^{g}}(y_1,y_2,\boldsymbol{\theta}_{1}^{\boldsymbol{g}},\boldsymbol{\theta}_{2}^{\boldsymbol{g}},\rho^{g})=\frac{\rho^{g}}{1-(\rho^{g})^{2}}-\frac{\rho^{g}\left[\frac{(y_{1}-\theta^{g}_{11})^{2}}{\theta^{g}_{12}}+\frac{(y_{2}-\theta^{g}_{21})^{2}}{\theta^{g}_{22}}\right]-(1+(\rho^{g})^{2})\frac{(y_{1}-\theta^{g}_{11})(y_{2}-\theta^{g}_{21})}{\sqrt{\theta^{g}_{12}\theta^{g}_{22}}}}{(1-(\rho^g)^{2})^2},
\end{align*}
\begin{figure}[h!]
	\centering
%	\small 
	\begin{tabular}{cc}
		\begin{subfigure}{0.4\textwidth}\centering\includegraphics[height=1.5 in,width=\columnwidth]{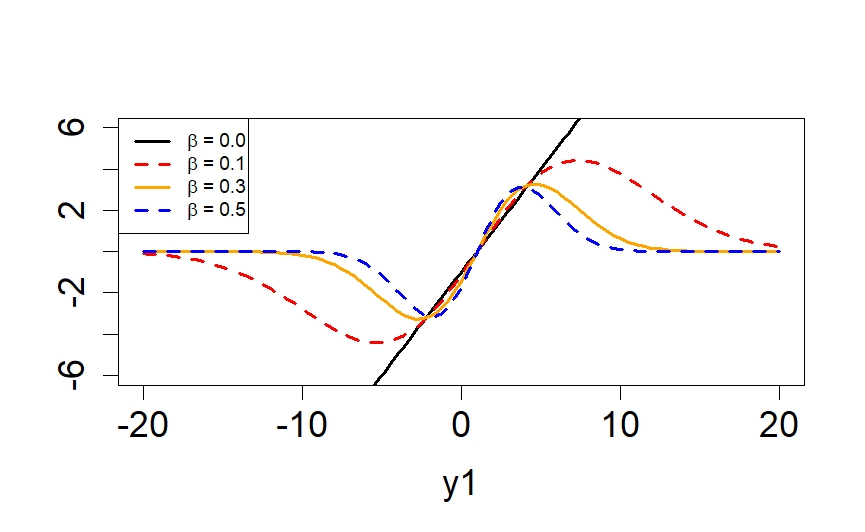}\caption{$\text{IF}(\theta_{11},y_{1},g_{1})$}\end{subfigure}&
		\begin{subfigure}{0.4\textwidth}\centering\includegraphics[height=1.5 in,width=\columnwidth]{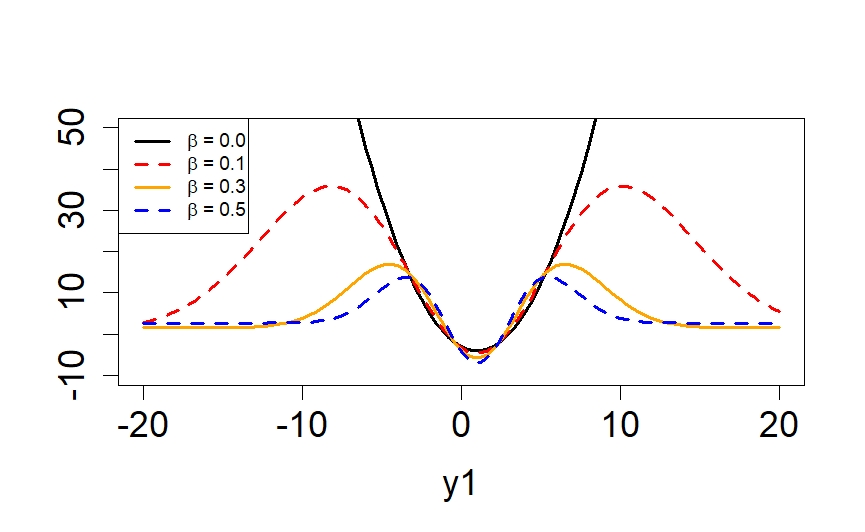}\caption{$\text{IF}(\theta_{12},y_{1},g_{1})$}\end{subfigure}\\
		\newline
		\begin{subfigure}{0.4\textwidth}\centering\includegraphics[height=1.5 in,width=\columnwidth]{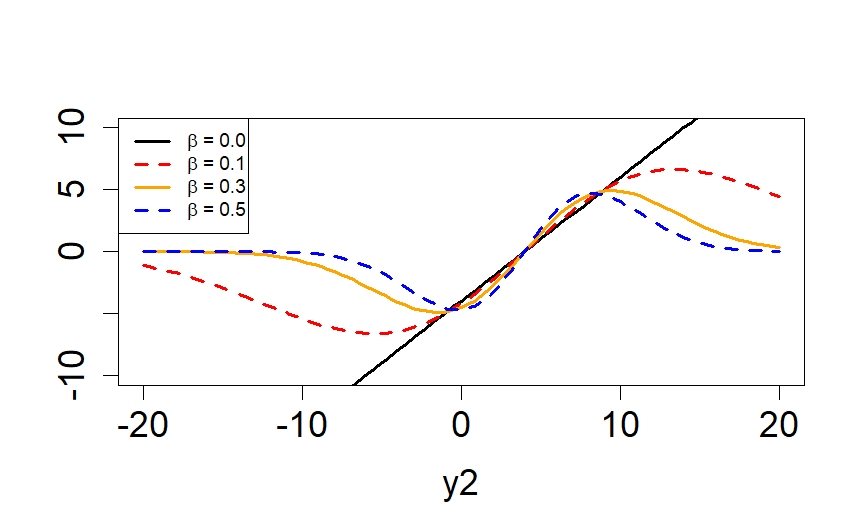}\caption{$\text{IF}(\theta_{21},y_{2},g_{2})$}\end{subfigure}&
		\begin{subfigure}{0.4\textwidth}\centering\includegraphics[height=1.5 in,width=\columnwidth]{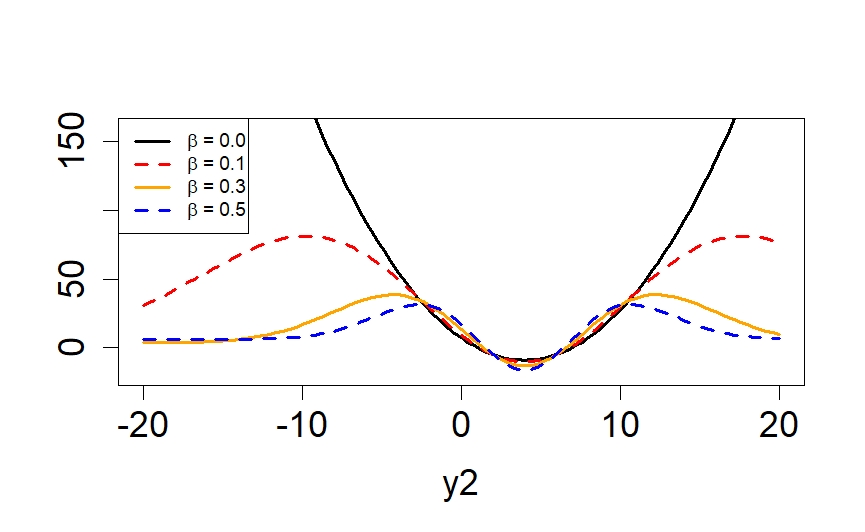}\caption{$\text{IF}(\theta_{22},y_{2},g_{2})$}\end{subfigure}\\
	\end{tabular}
	\caption{Influence functions of component means and variances.}
	\label{sm_figure1}
\end{figure}
\begin{figure}[h!]
	\centering
%	\small 
	\begin{tabular}{cc}
		\begin{subfigure}{0.4\textwidth}\centering\includegraphics[height=1.8 in,width=\columnwidth]{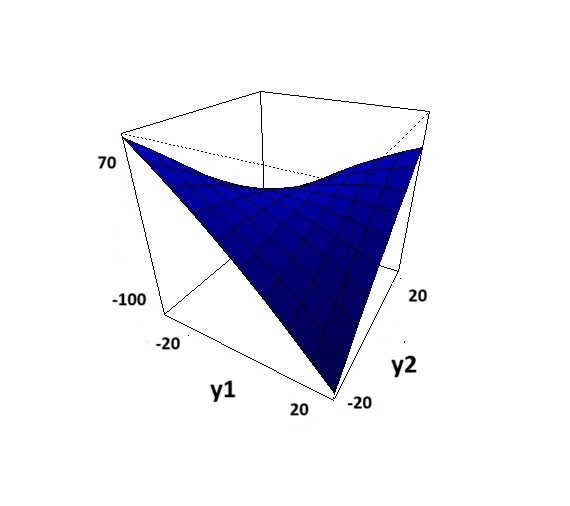}\caption{MLE}\end{subfigure}&
		\begin{subfigure}{0.4\textwidth}\centering\includegraphics[height=1.8 in,width=\columnwidth]{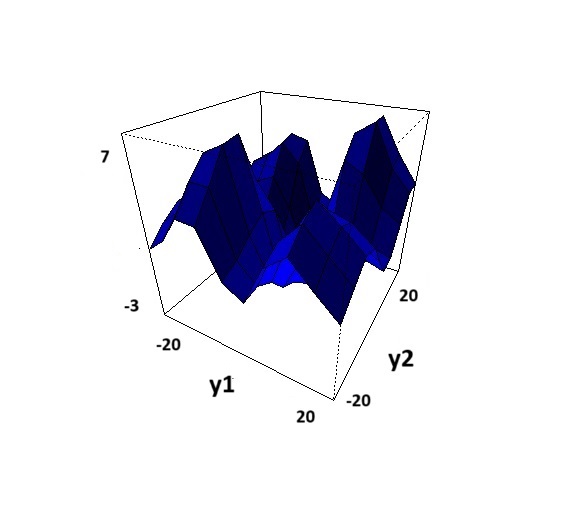}\caption{$\beta=0.1$}\end{subfigure}\\
		\newline
		\begin{subfigure}{0.4\textwidth}\centering\includegraphics[height=1.8 in,width=\columnwidth]{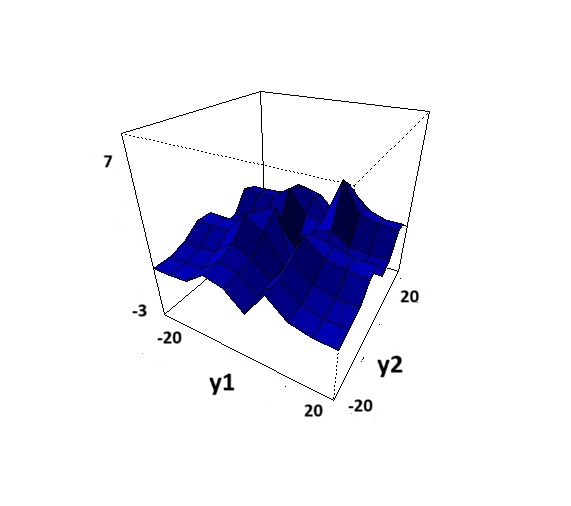}\caption{$\beta=0.3$}\end{subfigure}&
		\begin{subfigure}{0.4\textwidth}\centering\includegraphics[height=1.8 in,width=\columnwidth]{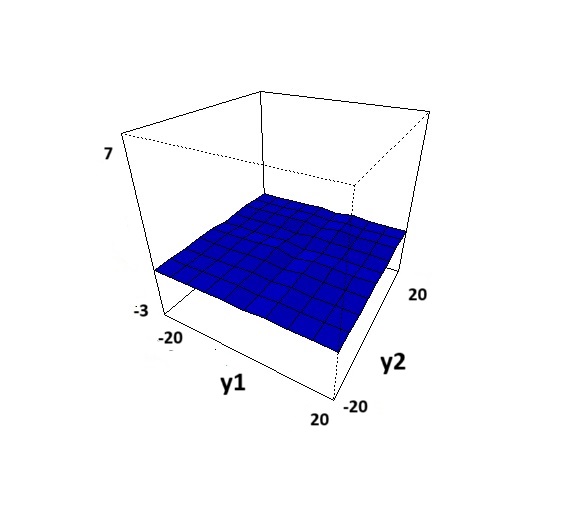}\caption{$\beta=0.5$}\end{subfigure}\\
	\end{tabular}
	\caption{Influence functions of correlation estimates.}
	\label{sm_figure2}
\end{figure}
it is easy to establish that $\text{IF}(\rho,y_1,y_2,g)$ is a linear function of $\text{IF}(\theta_{j1},y_{j},g_{j})$, $\text{IF}(\theta_{j2},y_{j},g_{j})$ for $j\in\{1,2\}$ and $(y_{j}-\theta^{g}_{j1})^{2}f^{\beta}(y_1,y_2,\boldsymbol{\theta}_{1}^{\boldsymbol{g}},\boldsymbol{\theta}_{2}^{\boldsymbol{g}},\rho^{g})$, $f^{\beta}(y_1,y_2,\boldsymbol{\theta}_{1}^{\boldsymbol{g}},\boldsymbol{\theta}_{2}^{\boldsymbol{g}},\rho^{g})$ for $j\in\{1,2\}$ and $(y_{1}-\theta^{g}_{11})(y_{2}-\theta^{g}_{21})f^{\beta}(y_1,y_2,\boldsymbol{\theta}_{1}^{\boldsymbol{g}},\boldsymbol{\theta}_{2}^{\boldsymbol{g}} ,\rho^{g})$. The boundedness of $\text{IF}(\rho,y_1,y_2,g)$ is now easily followed by the boundedness of the aforesaid components of $\text{IF}(\rho,y_1,y_2,g)$. To see the behaviour explicitly, we present a special case pictorially where the model family is bivariate normal and the true distribution is also bivariate normal with component means $\theta^{g}_{11}=1$, $\theta^{g}_{21}=4$, component variances $\theta^{g}_{12}=4$, $\theta^{g}_{22}=9$ and correlation $\rho^{g}=0.5$. The influence functions of the component means and variances are presented in Figure \ref{sm_figure1} and the influence functions (for different values of $\beta$ including the maximum likelihood case) of the correlation are presented in Figure \ref{sm_figure2}.
\subsubsection{Asymptotic Relative Efficiency}
Now we study the asymptotic relative efficiencies (with respect to the maximum likelihood estimators) of our estimators when the data are normally distributed. We assume that the true distribution is bivariate normal with component means $0$ and $0$, component variances $1$ and $1$ and correlation coefficient $\rho$. We will consider seven different values of $\rho$, namely, $-0.7,\;-0.5,\;-0.3,\;0,\;0.3,\;0.5\;\text{and}\; 0.7$ representing the true correlation coefficient. This range of the correlation coefficient is used to observe the nature of the asymptotic relative efficiency of the correlation estimator under both high and low correlation structures. 

From Basu et al.~$(2011)$ \cite{basubook}, it follows that the asymptotic variance of $\sqrt{n}\widehat{\theta}_{j1}$ is $\left(1+\frac{\beta^2}{1+2\beta}\right)^\frac{3}{2}{\theta_{j2}^{g}}$ and $\left(1+\frac{\beta^2}{1+2\beta}\right)^{2}{\theta_{j2}^{g}}$ $(j\in\{1,2\})$, respectively for the CMDPDE and MDPDE. Clearly, the asymptotic relative efficiency of the componentwise method is higher than that of the ordinary method, at least for the component mean estimators. 

Explicit calculations of the asymptotic variances of the component variance and correlation estimators in case of both CMDPDE and MDPDE are extremely cumbersome; the algebraic forms of these asymptotic variances are provided Appendix \ref{sm_appen1}. The asymptotic relative efficiencies (in percentage) of our estimators are tabulated in Tables \ref{sm_table0.1} and \ref{sm_table0.4}.
\begin{table}[h!]
	\centering 
%	\small
	\begin{tabular}{c c c c c c c  } 
		
		\hline
		Estimators & Methods &  & & $\beta$ &  &  \\  
		& &0 &0.1 & 0.3 & 0.5 &  0.7  \\
		\hline
		\multirow{ 2}{*}{Mean}& CMDPDE & 100.000 & 98.717  &  92.081 & 83.822 & 75.700  \\
		& MDPDE & 100.000 & 98.328  &  89.606 & 78.989 & 68.966
		 \\
		& & & & & & \\
		\multirow{ 2}{*}{Variance}& CMDPDE & 100.000 & 97.561 &  85.507 & 73.046 & 63.452	\\
		& MDPDE  &  100.000 & 97.135 &  83.368 & 69.300 & 58.207
		\\
		\hline
	\end{tabular}
	\caption{Asymptotic relative efficiencies (in percentage) of component mean and variance estimators. }
	\label{sm_table0.1}
\end{table}
%\vspace{-0.5in}
\begin{table}[h!]
	\centering 
%	\small
	\begin{tabular}{c c c c c c c c c } 
		
		\hline
		& & &  $\beta$  &  & & \\  
		$\rho^{g}$ & 0.0 & 0.1 & 0.3 & 0.5 & 0.7  \\
		\hline
		$-0.7$  &  100(100) &    97.378(97.378) & 84.967(84.691) &  70.845(70.270) &  59.361(57.269) \\
		$-0.5$  & 100(100) &    97.574(97.574) & 84.789(84.917) &  70.375(70.287) &  58.161(57.332)  \\
		$-0.3$  &  100(100) &    97.527(97.527) & 84.663(84.836) &  69.992(70.229) &  57.222(57.261) \\
		~~0.0  &   100(100) &    97.561(97.561) & 84.890(84.890) &  70.225(70.225) &  57.274(57.274)   \\
		~~0.3  &  100(100) &    97.527(97.527) & 84.663(84.836) &  69.992(70.229) &  57.222(57.261)    \\
		~~0.5  & 100(100) &    97.574(97.574) & 84.789(84.917) &  70.375(70.287) &  58.161(57.332)    \\
		~~0.7 &   100(100) &    97.378(97.378) & 84.967(84.691) &  70.845(70.270) &  59.361(57.269)   \\
		\hline
	\end{tabular}
	\caption{Asymptotic relative efficiencies (in percentage) of correlation estimators for the CMDPDE; the same for the corresponding MDPDEs are given in parentheses.}
	\label{sm_table0.4}
\end{table}

It can be easily observed that the CMDPDEs have higher asymptotic relative efficiencies compared to the ordinary MDPDEs both for component means and component variances at each positive $\beta$. For both correlation estimators (CMDPDE and ordinary MDPDE), the asymptotic relative efficiencies show a symmetric behaviour with respect to $\rho^{g}$, the true value of $\rho$. We may conclude aggregatively that both the CMDPDE and ordinary MDPDE are close to each other as estimates of the correlation coefficient in terms of their asymptotic relative efficiencies. 

\section{Simulation Experiments} 
\label{sm_sec4}
\subsection{Experimental Set-up and Performance Measures}
\label{sm_sim_set}
In this section, we assess the performance of our method with the other existing methods under the multivariate normal set-up. Here we assume that the true distribution is also multivariate normal, i.e., the true distribution belongs to the model family. To carry out our simulation experiments, we simulate $100$ random samples of size $n=1000$ from the multivariate normal distribution of dimension $p$ with mean vector $\boldsymbol{\mu}$ and covariance matrix $\boldsymbol{\Sigma}$; different choices of $p$, $\boldsymbol{\mu}$ and $\boldsymbol{\Sigma}$ are taken to vary the simulation set-ups. Data from $2$, $5$ and $10$ dimensions are simulated with mean vector $\boldsymbol{0}$ and two types of covariance structures are used, viz., diagonal and a special kind of non-diagonal structure. Identity matrices of appropriate dimensions are taken as diagonal covariance matrices. For the non-diagonal choice, we consider the following $p\times p$ matrix (with $p_{1}=p/2$ and $p_{2}=p-p_{1}$),
\begin{align*}
\boldsymbol{\Sigma}=\begin{bmatrix}
\boldsymbol{U}(0.7) & \boldsymbol{0}_{p_{1}\times p_{2}} \\
 \boldsymbol{0}_{p_{2}\times p_{1}} & \boldsymbol{I}_{p_2}\\
\end{bmatrix},
\end{align*}
where $\boldsymbol{U}(\rho)=[[u_{ij}]]_{i,j=1}^{p_{1}}$ with $u_{ij}=\rho^{|i-j|}$ for $1 \leq i,j \leq p_{1}$ and $-1\leq \rho \leq 1$. This structure is known as the block-banded covariance matrix. But for $p=2$ (only), we take $\boldsymbol{\Sigma}=\begin{bmatrix}
1 & 0.7 \\
0.7 & 1\\
\end{bmatrix}$. 

We are going to evaluate two measures of accuracy, viz., $L_{2}$ bias and mean squared error (MSE) for the mean and covariance matrix estimators,  separately. Let, $\widehat{\boldsymbol{\mu}}_{i}$ and $\widehat{\boldsymbol{\Sigma}}_{i}$ be the estimators of the mean vector and the covariance matrix for the $i$-th replication, $1\leq i \leq 100$. We calculate the $L_{2}$ bias and mean squared errors as follows. 
\begin{align*}
&\text{Bias of mean estimators}=\Big|\Big|\frac{1}{100}\sum_{i=1}^{100}\widehat{\boldsymbol{\mu}}_{i} - \boldsymbol{\mu}\Big|\Big|_{2},\;\text{MSE of mean estimators}=\frac{1}{100}\sum_{i=1}^{100}\Big|\Big|\widehat{\boldsymbol{\mu}}_{i} - \boldsymbol{\mu}\Big|\Big|^{2}_{2},\\
&\text{Bias of covariance matrix estimators}=\Big|\Big|\frac{1}{100}\sum_{i=1}^{100}\widehat{\boldsymbol{\Sigma}}_{i} - \boldsymbol{\Sigma}\Big|\Big|_{F},\\ 
&\text{MSE of covariance matrix estimators}=\frac{1}{100}\sum_{i=1}^{100}\Big|\Big|\widehat{\boldsymbol{\Sigma}}_{i} - \boldsymbol{\Sigma}\Big|\Big|^{2}_{F},
\end{align*}
where $||\cdot||_{2}$ is the $L_2$-norm and $||\cdot||_{F}$ is the Frobenius norm of a matrix.

To study the robustness and efficiency of our method, we consider both pure as well as contaminated datasets in each of the aforesaid set-ups (depending on data dimension and covariance structure). Pure datasets are simulated from $p$-variate normal distributions with zero mean and the aforesaid choices of the covariance matrix $\boldsymbol{\Sigma}$ for $p=2,$ $5$ and $10$. Additionally, two types of data contamination schemes are used for simulation; one with subtle outliers and the other with distant outliers. Both types of contaminated datasets are simulated from appropriate two component normal mixtures; the first component is the $p$-variate normal distribution with zero mean and the aforesaid choices of the covariance matrix $\boldsymbol{\Sigma}$ (the regular component with $90\%$ weight) and the remaining component is the $p$-variate normal distribution with mean vector $\boldsymbol{\mu}_c$ and $p$-dimensional identity covariance matrix (the contaminating component with $10\%$ weight). We consider $\boldsymbol{\mu}_c=2\lambda_{\text{min}}\boldsymbol{v}_{\text{min}}$ for the subtle outlier contamination set-up ($\lambda_{\text{min}}$ and $\boldsymbol{v}_{\text{min}}$ are the minimum eigenvalue and the corresponding unit normed eigenvector of $\boldsymbol{\Sigma}$, respectively) and $\boldsymbol{\mu}_c=(20,\ldots,20)^{\top}$ for the distant outlier contamination set-up.

%and two types contaminated datasets are simulated by generating observations from a mixture normal distribution. The main component of this mixture distribution is multivariate normal with mean $\boldsymbol{0}$ and covariance matrix $\boldsymbol{\Sigma}$ and the contaminating component is a multivariate normal with mean vector $\boldsymbol{(20,20,\cdots,\;20)}$ and identity covariance matrix. The mixing proportions are $0.9$ and $0.1$, so that, the data represent $10\%$ contamination of the pure model.

\begin{figure}[h!]
	\centering
%	\small 
	\begin{tabular}{ccc}
		\begin{subfigure}{0.4\textwidth}\centering\includegraphics[height=2 in, width=2 in]{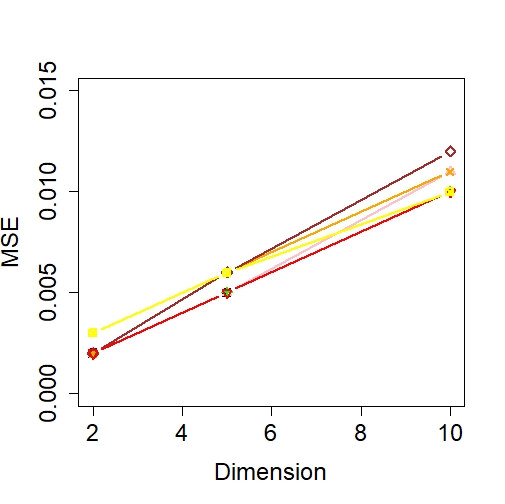}\caption{Pure data, location estimators}\end{subfigure}&
		\begin{subfigure}{0.4\textwidth}\centering\includegraphics[height=2 in, width=2 in]{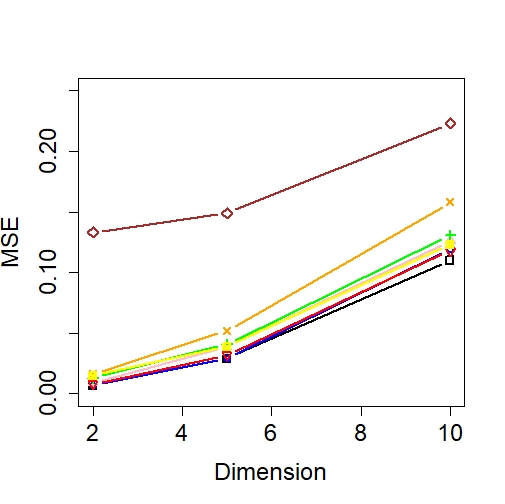}\caption{Pure data, scale estimators}\end{subfigure}\\
        \begin{subfigure}{0.4\textwidth}\centering\includegraphics[height=2 in, width=2 in]{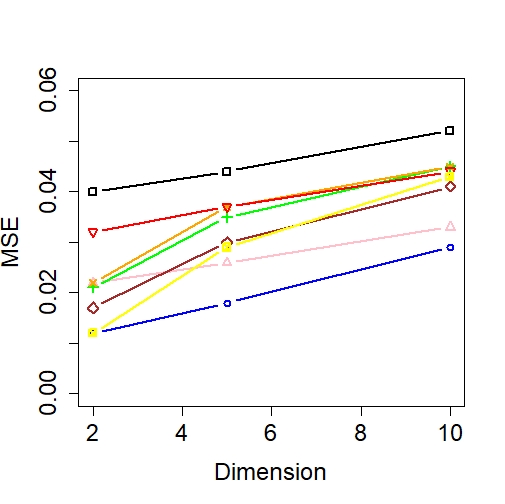}\caption{Subtle outlier contaminated data, location estimators}\end{subfigure}&
		\begin{subfigure}{0.4\textwidth}\centering\includegraphics[height=2 in, width=2 in]{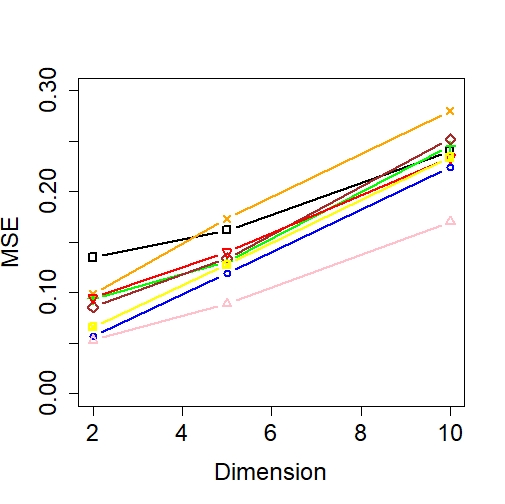}\caption{Subtle outlier contaminated data, scale estimators}\end{subfigure}& \begin{subfigure}{0.4\textwidth}\hspace{0.2 in}\includegraphics[height=2 in, width=1 in]{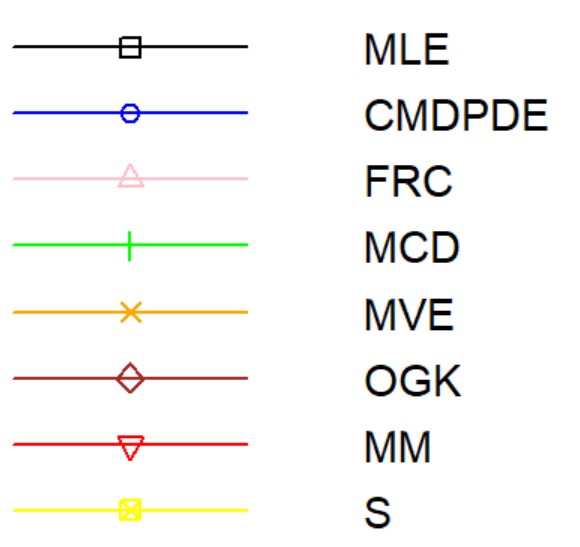}\caption*{}\end{subfigure}\\
        \begin{subfigure}{0.4\textwidth}\centering\includegraphics[height=2 in, width=2 in]{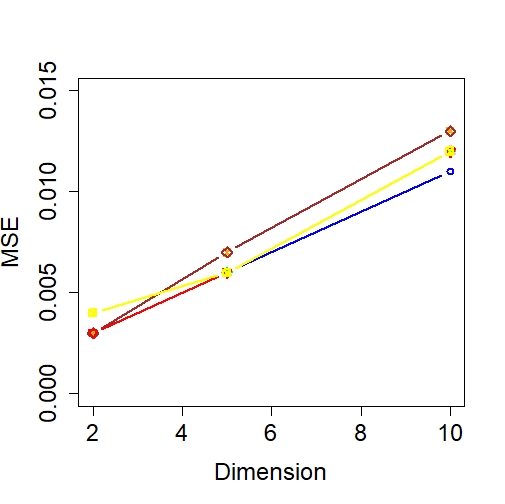}\caption{Distant outlier contaminated data, location estimators}\end{subfigure}&
		\begin{subfigure}{0.4\textwidth}\centering\includegraphics[height=2 in, width=2 in]{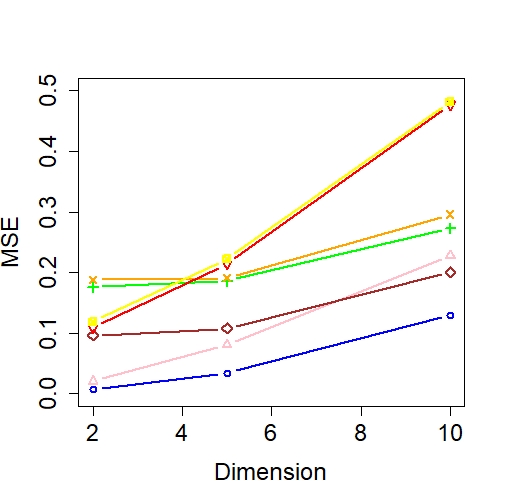}\caption{Distant outlier contaminated data, scale estimators}\end{subfigure}
		\end{tabular}
	\caption{Estimated mean squared errors of location and scale estimators in case of diagonal covariance structures.}
	\label{sm_figure_simulation_diag}
\end{figure}
\begin{figure}[h!]
	\centering
%	\small 
	\begin{tabular}{ccc}
		\begin{subfigure}{0.4\textwidth}\centering\includegraphics[height=2 in, width=2 in]{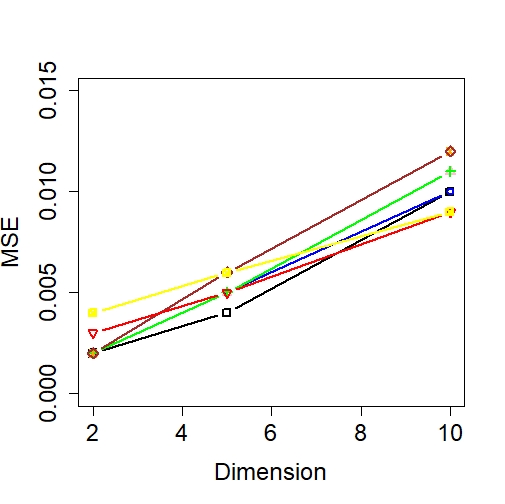}\caption{Pure data, location estimators}\end{subfigure}& 
		\begin{subfigure}{0.4\textwidth}\centering\includegraphics[height=2 in, width=2 in]{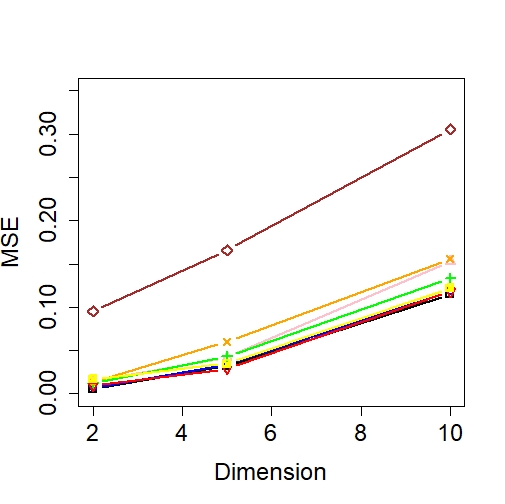}\caption{Pure data, scale estimators}\end{subfigure}\\
        \begin{subfigure}{0.4\textwidth}\centering\includegraphics[height=2 in, width=2 in]{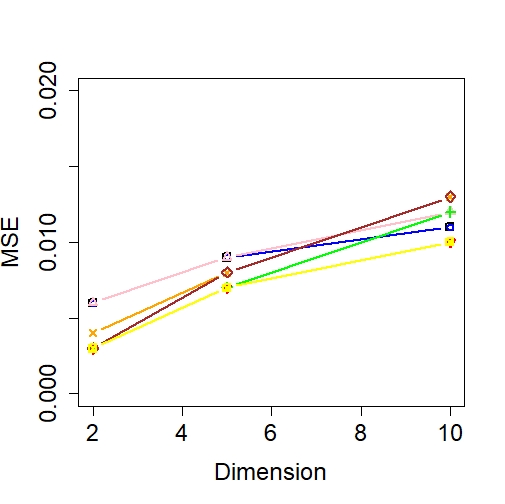}\caption{Subtle outlier contaminated data, location estimators}\end{subfigure}&  
		\begin{subfigure}{0.4\textwidth}\centering\includegraphics[height=2 in, width=2 in]{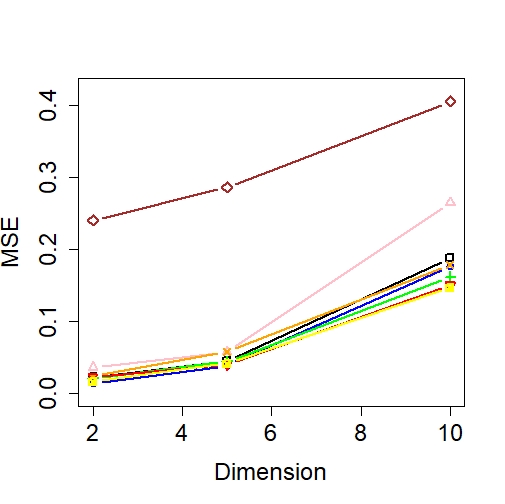}\caption{Subtle outlier contaminated data, scale estimators}\end{subfigure}&\begin{subfigure}{0.4\textwidth}\hspace{0.2 in}\includegraphics[height=2 in, width=1 in]{marker_legends.jpg}\caption*{}\end{subfigure}\\
        \begin{subfigure}{0.4\textwidth}\centering\includegraphics[height=2 in, width=2 in]{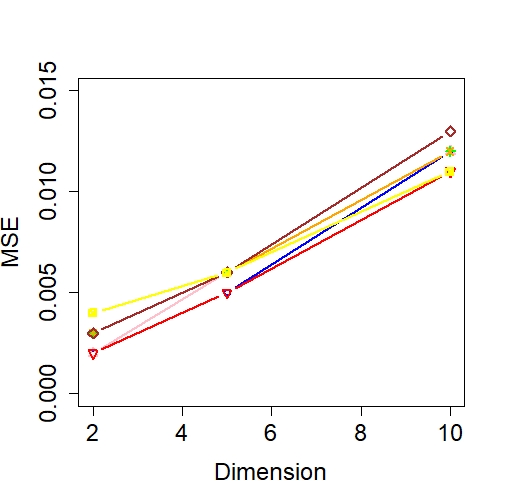}\caption{Distant outlier contaminated data, location estimators}\end{subfigure} &
		\begin{subfigure}{0.4\textwidth}\centering\includegraphics[height=2 in, width=2 in]{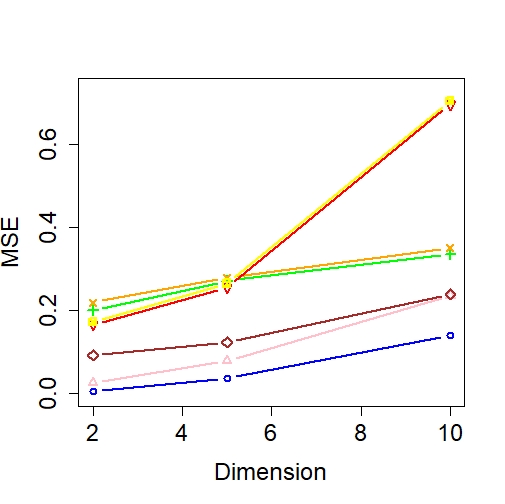}\caption{Distant outlier contaminated data, scale estimators}\end{subfigure}
		\end{tabular}
	\caption{Estimated mean squared errors of location and scale estimators in case of non-diagonal covariance structures. }
	\label{sm_figure_simulation_nondiag}
\end{figure}
In each case, we are considering various values of $\beta$ in the range $(0,1)$; higher values of $\beta$ are avoided from efficiency considerations. We are going to present the output of the simulation experiments pictorially in Figures \ref{sm_figure_simulation_diag} and \ref{sm_figure_simulation_nondiag} for diagonal and non-diagonal covariance structures, respectively, in terms of the estimated mean squared errors of the location and scale estimators. The actual values of the estimated $L_2$ bias and mean squared errors of these estimators are tabulated in Appendix \ref{sm_appen3} (Tables \ref{sm_table1.1}-\ref{sm_table2.2}). The esimated MSEs of the MLEs are much larger as compared to those of the robust methods  in case of distant outlier contaminated datasets. Thus, these MLEs cannot be presented pictorially in the same frames along with its robust alternatives in case of distant outlier contaminated datasets. 

These simulation experiments provide a comparative overview of our newly proposed estimators along with the ordinary maximum likelihood (corresponds to $\beta=0$ case), fast robust correlation (FRC) (see Raymaekers and Rousseeuw $(2019)$ \cite{frc}), MCD, MVE, Orthogonalized Gnanadesikan-Kettenring (OGK) (see Maronna and Zamar $(2002)$, \cite{GK}), MM estimators of location and scale (MM) (originally proposed by Yohai (1987) \cite{mmorg} in regression set-up and later Tatsuoka and Tyler $(2000)$ \cite{mm} developed the multivariate location-scale version) and S-estimators of location and scale (Rousseeuw and Yohai $(1984)$ \cite{sest84}, Ruppert $(1992)$ \cite{S}). Appropriate $\sf{R}$-packages \cite{cellWiseR, rrcov, robust, robustbase, GSER} are used for the computing of these estimators; the respective $\sf{R}$ functions are utilized with their default implementations. For the pictorial representation of simulation results, CMDPDEs corresponding to those $\beta$ values are considered for which the  estimated mean squared errors are minimum (for each of the aforesaid simulation set-ups).

\subsection{Discussion of Simulation Results }
The simulations that we have performed are quite extensive, and it is necessary to understand what the salient features of these numbers (Tables \ref{sm_table1.1}-\ref{sm_table2.2} in Appendix \ref{sm_appen3}) are. In the following we describe these features.
\begin{enumerate}
    \item For pure datasets, MLEs are the best compared to all its robust alternatives which is expected from the angle of asymptotic efficiency. All the robust methods are almost similar in terms of their estimated MSEs in case of location estimation, however, in case of scale estimation, CMDPDEs become the best alternative almost all the time compared to other robust estimators although both the MM and the S estimators maintain very close competition with the CMDPDEs.

    \item For subtle outlier contaminated data, our method is found to be generally competitive with the remaining robust methods. For the diagonal covariance structure, our method becomes the best one (in terms of the estimated MSEs) in case of location estimation but the FRC method becomes the best one in case of scale estimation. However, for the non-diagonal covariance structure, our method performs better than the FRC method. But in this case (i.e., non-diagonal covariance structure), although our scale estimators become the best among all the robust scale estimators in case of data dimensions $2$ and $5$, the MM and S scale estimators have less estimated MSEs as compared to the remaining robust methods in case of data dimension $10$.

    \item For distant outlier contaminated datasets, our method becomes the best one among all the robust methods, especially with lower values of $\beta$ (specifically $0.1$). The pairwise OGK method improved a lot in this case, especially for the scale estimation; the OGK and the FRC estimators become the closest competitors of the CMDPDEs in case of scale estimation.

    \item In an aggregative sense, we may conclude that our newly proposed method performs the best in case of datasets having distant outliers, and performs competitively in comparison with the aforementioned robust alternatives in case of the subtle outlier contamination set-up.

    \item None of the CMDPDEs of the covariance matrices have been found to be non-positive definite throughout our entire simulation exercise.

\end{enumerate}
A comparative simulation study on bias and mean squared errors of the mean, variance and correlation estimators of CMDPDEs and MDPDEs are presented in Appendix \ref{sm_appen4} along with the implications of these simulation outputs. 
\subsection{Cellwise Contamination}
\label{sm_cellwise}
We have only considered casewise (or rowwise) contaminations in the aforesaid simulation experiments, where some of the sample observations (rows of the data matrix) are perturbed. However, cellwise (columnwise) contamination does arise in real life datasets where some of the components (columns of the data matrix) are perturbed. 
\begin{figure}[h!]
	\centering
%	\small 
	\begin{tabular}{cc}
		\begin{subfigure}{0.4\textwidth}\centering\includegraphics[height=2 in]{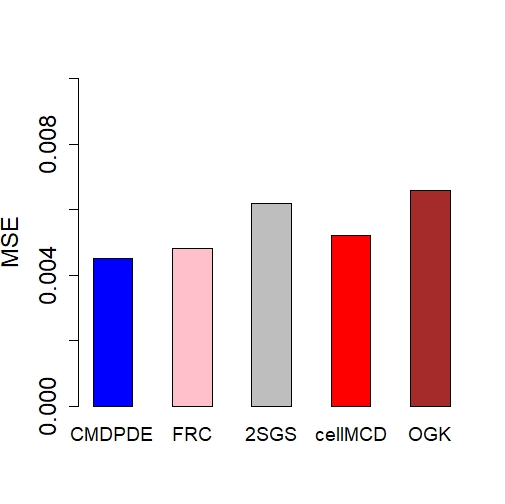}\caption{Location vector}\end{subfigure}&
		\begin{subfigure}{0.4\textwidth}\centering\includegraphics[height=2 in]{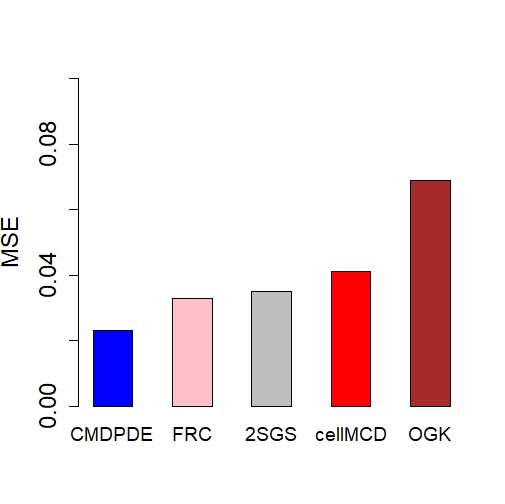}\caption{Scale matrix}\end{subfigure}\\
		\end{tabular}
	\caption{Estimated mean squared errors of the location vector and the scale matrix in case of cellwise contamination.}
	\label{sm_figure_cell}
\end{figure}
This results in the contamination of a comparatively larger proportion of the sample observations, at least partially. Discarding or downweighting all those partially contaminated observations may result in a significant loss of information which affects the efficiency of the resulting estimators. Since the proposed methodology in this work is componentwise in nature, it is intuitively expected that this procedure can tackle the cellwise contaminated datasets more efficiently as compared to the ordinary minimum DPD method. We simulate $100$ datasets (sample size $1000$) from the $4$-dimensional standard normal distribution. Then a fixed proportion ($5\%$ in this case) of observations (chosen at random) from each of the $4$ components (variables) are contaminated with a distant value ($20$ in our case).

The estimated mean squared errors of the location and scale matrix estimates using our newly proposed method along with some other componentwise robust methods including the Orthogonalized Gnanadesikan-Kettenring method, the two step generalized S-estimation (2SGS) method (Agostinelli et al.~ $(2015)$ \cite{recent}), the fast robust correlation method and the cellMCD method (Raymaekers and Rousseeuw $(2024)$ \cite{cellmcd}), which are considered suitable to tackle cellwise contamination, are presented pictorially in Figure \ref{sm_figure_cell}. The superiority of the CMDPDEs can clearly be observed from Figure \ref{sm_figure_cell} over its competitors in terms of the estimated MSEs.

\subsection{High Dimensional Data}
\label{sm_hd_sim}
The simulation set-ups considered so far are typically multivariate where data dimensions are much smaller than the respective sample sizes. We now present some simulation 
\begin{figure}[h!]
	\hspace{-2 cm}
%	\small 
	\begin{tabular}{ccc}
		\begin{subfigure}{0.4\textwidth}\centering\includegraphics[height=2 in]{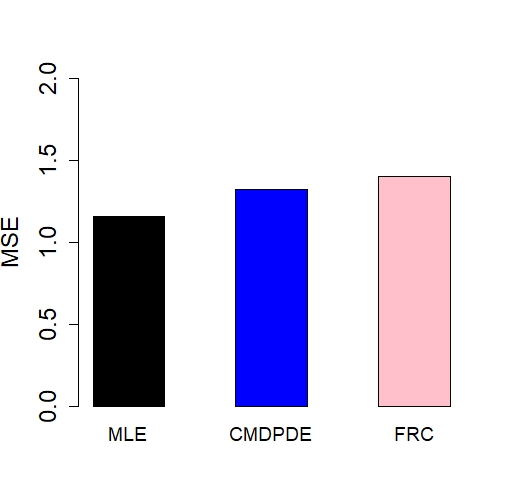}\caption{Pure data}\end{subfigure}&
		\begin{subfigure}{0.4\textwidth}\centering\includegraphics[height=2 in]{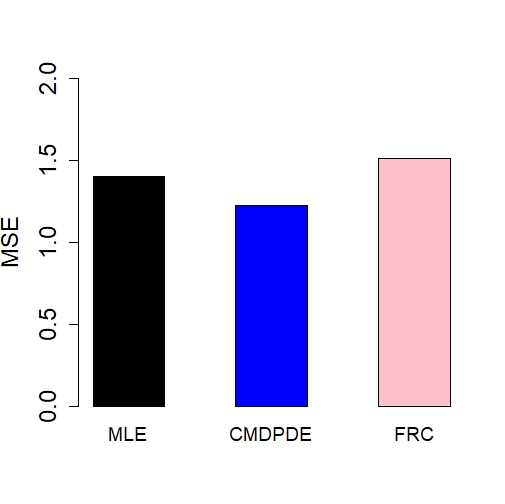}\caption{Subtle outlier contaminated}\end{subfigure}&
		\begin{subfigure}{0.4\textwidth}\centering\includegraphics[height=2 in]{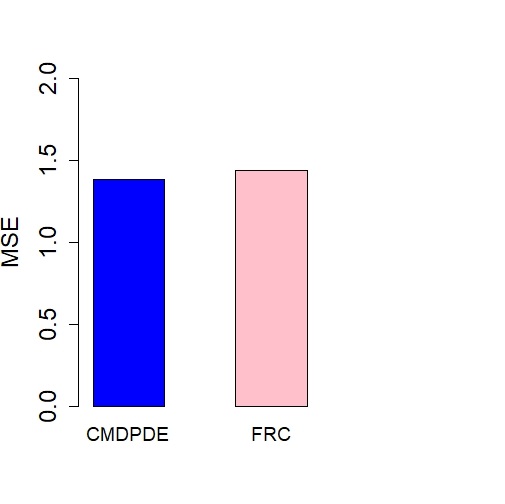}\caption{Distant outlier contaminated}\end{subfigure}\\
        \begin{subfigure}{0.4\textwidth}\centering\includegraphics[height=2 in]{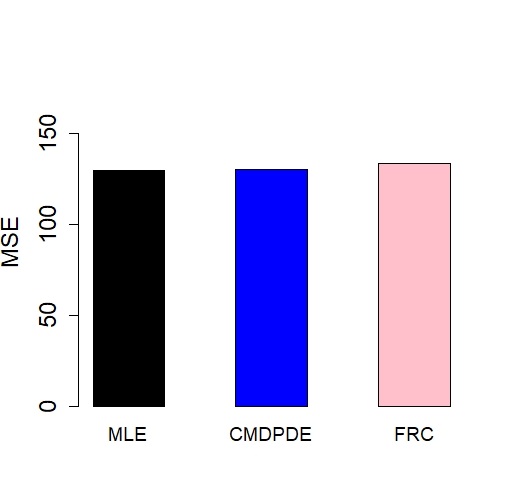}\caption{Pure data}\end{subfigure}&
		\begin{subfigure}{0.4\textwidth}\centering\includegraphics[height=2 in]{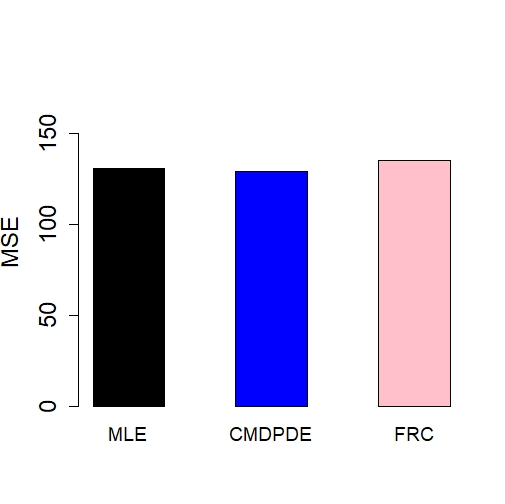}\caption{Subtle outlier contaminated}\end{subfigure}&
		\begin{subfigure}{0.4\textwidth}\centering\includegraphics[height=2 in]{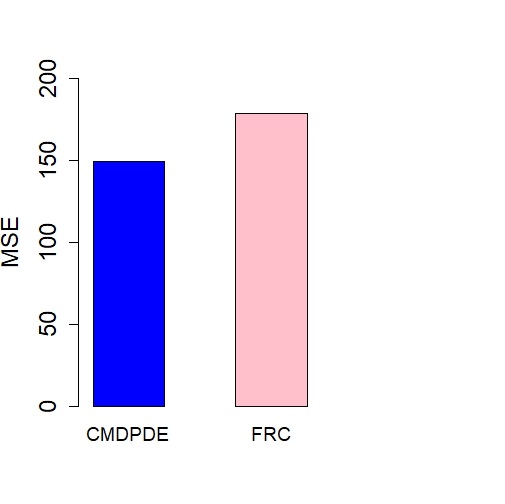}\caption{Distant outlier contaminated}\end{subfigure}\\
		\end{tabular}
	\caption{Estimated mean squared errors of location (in the first row) and scale (in the second row) estimates.}
	\label{sm_figure_hd}
\end{figure}
experiments in high dimensional set-ups where the sample sizes are less than the respective data dimensions. In particular, we consider the data dimension $p=100$ and sample size $n=80$ and simulate $100$ random samples in each of the three simulation set-ups (i.e., pure, subtle outlier contaminated and distant outlier contaminated) described in Section \ref{sm_sim_set} with the non-diagonal covariance structure. We use only the maximum likelihood method, our newly proposed method and the FRC method as the remaining robust alternatives (i.e., MCD, MVE etc.) used in the simulation experiments cannot be successful in producing the estimates of location vector and scale matrix for possible singularity as $n<p$. The estimated MSEs of the location and scale estimates are presented in Figure \ref{sm_figure_hd}. CMDPDEs corresponding to those $\beta$ values are considered (in Figure \ref{sm_figure_hd}) whose estimated MSEs are minimum (in each of the simulation set-ups). In case of the distant outlier contamination set-up, the estimated MSEs of the MLEs are so large that they cannot be presented with the MSEs of the robust methods in the same frames.

As earlier, MLEs have the least MSEs in case of pure datasets. However, the robust methods achieve lower MSEs (as compared to that of the MLE) in case of contaminated datasets. The superiority of our method is evident in case of these contaminated set-ups from Figure \ref{sm_figure_hd}.

\subsection{Empirical Breakdown Analysis}
Although we have not derived the breakdown properties of the CMDPDEs, here we empirically illustrate the possible breakdown behaviour of the proposed estimators with simulation by exploring it stability under increasing levels of contamination. 
\begin{figure}[h!]
	\centering
%	\small 
	\begin{tabular}{cc}
		\begin{subfigure}{0.4\textwidth}\centering\includegraphics[height=2.5 in, width=2 in]{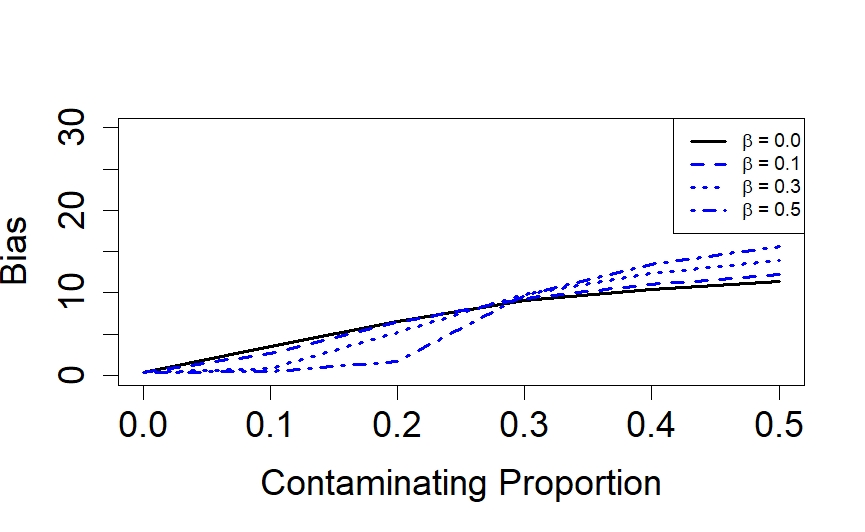}\caption{$p=2$}\end{subfigure}&
		\begin{subfigure}{0.4\textwidth}\centering\includegraphics[height=2.5 in, width=2 in]{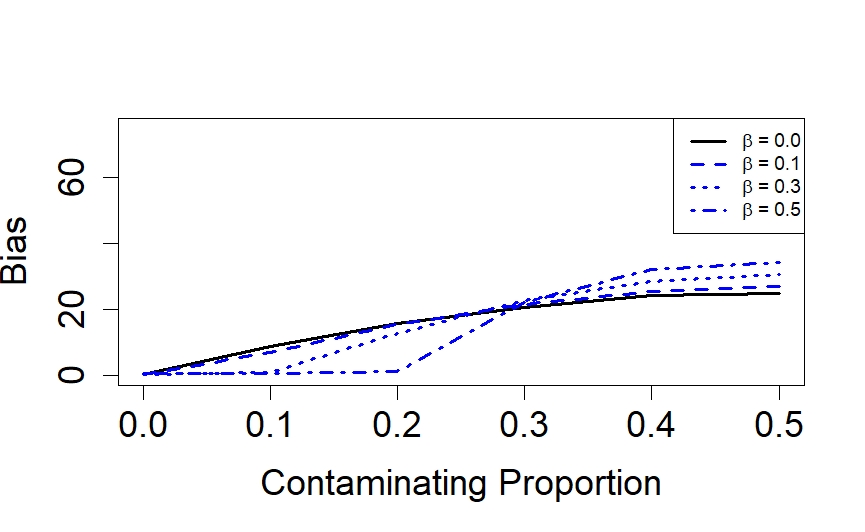}\caption{$p=5$}\end{subfigure}\\
     %   \vspace{-0.5 in}
        \begin{subfigure}{0.4\textwidth}\centering\includegraphics[height=2.5 in, width=2 in]{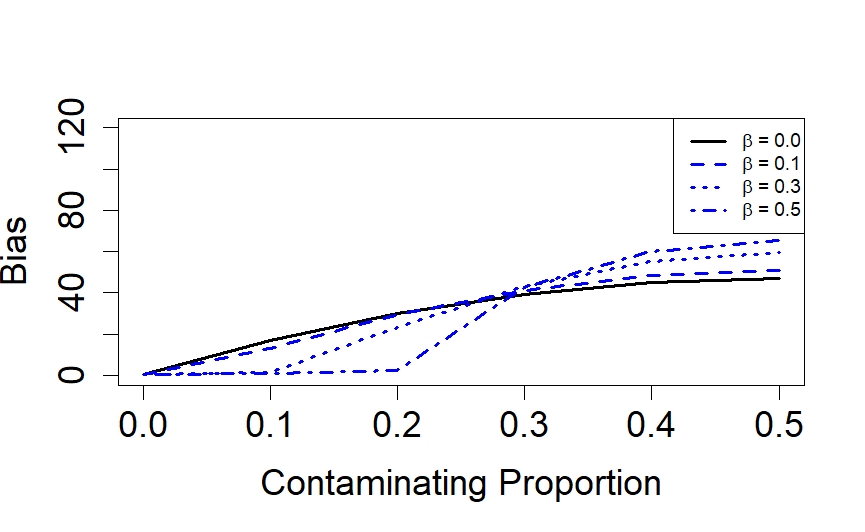}\caption{$p=10$}\end{subfigure}&
		\begin{subfigure}{0.4\textwidth}\centering\includegraphics[height=2.5 in, width=2 in]{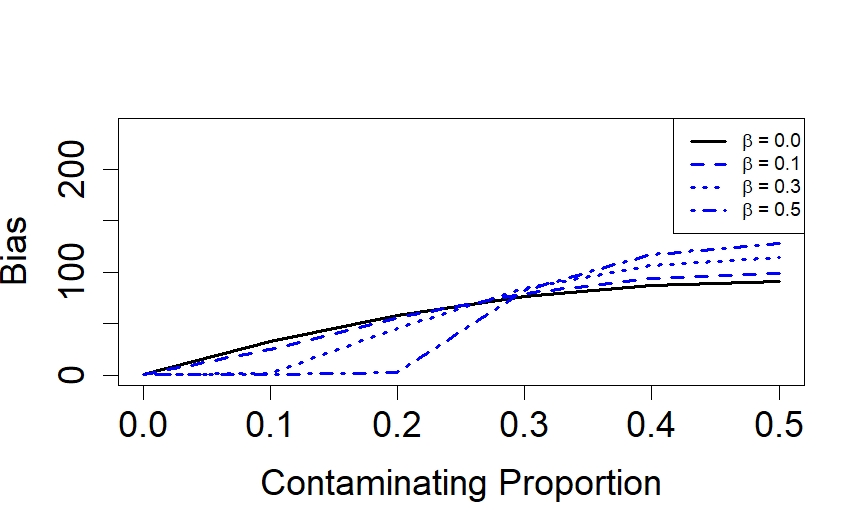}\caption{$p=20$}\end{subfigure}
		\end{tabular}
	\caption{$L_2$ bias of the CMDPDEs with varying contaminating proportions.}
	\label{sm_figure_empbp}
\end{figure}
In particular, we generate $4$ random samples (one from each dimension) from $p$-dimensional standard normal distributions for $p=2$, $5$, $10$ and $20$ with respective sample sizes $n=100$, $200$, $500$ and $1000$. These samples are then contaminated with observations from $p$-dimensional normal distributions with mean $(5,\ldots,5)$ and covariance $\boldsymbol{I}_p$ with varying contaminating proportions; $0\%$, $10\%$, $20\%$, $30\%$, $40\%$, and $50\%$ are the contaminating proportions considered. The CMDPDEs of the unknown parameters (considered as $(\boldsymbol{\mu}^{\top},\text{vech}(\boldsymbol{\Sigma}))^{\top}$) are then computed  assuming the normal model and their $L_2$ bias are plotted in Figure \ref{sm_figure_empbp}. It is to be observed that:  $(i)$ the empirical breakdown values of the CMDPDEs are greater than those of the MLEs which supports the robustness of the CMDPDEs, $(ii)$ empirical breakdown values of the CMDPDEs increase as $\beta$ increases, and $(iii)$ the empirical breakdown values of the CMDPDEs show similar patterns (at least in terms of the slopes of the empirical breakdown curves) in case of all the data dimensions which happens probably due to the componentwise nature of the newly proposed estimation procedure. 

\subsection{Scalability of the Proposed CMDPDE}
We have already realized the fact that the componentwise minimum DPD estimation is computationally far more efficient than 
\begin{table}[h!]
	\centering 
%	\small
	\begin{tabular}{c c c c c c c c c c c } 
		
		\hline
		Methods& &  & &  $p$ &  & & & \\  
		& $\beta$& 2 & 5 & 10 & 20 & 30 & 40 &  50  \\
		\hline
		Number of Parameters &  & 5 & 20 &65 & 230 & 495 & 860 & 1325  \\
		\multirow{4}{*}{}CMDPDE  & 0.1 & 100 $\%$ & 100 $\%$ & 100$\%$ & 100$\%$ & 100$\%$ & 100$\%$ & 100$\%$   \\
		& 0.3 & 100 $\%$ & 100 $\%$ &100$\%$ & 100$\%$ & 100$\%$ & 100$\%$ & 100$\%$  \\
		& 0.5 & 100 $\%$ & 100 $\%$ &100$\%$ & 100$\%$ & 100$\%$ & 100$\%$ & 100$\%$  \\
		& 0.7 &  100 $\%$ & 100 $\%$ & 100$\%$ & 100$\%$ & 100$\%$ & 100$\%$ & 100$\%$  \\
		\multirow{4}{*}{}MDPDE & 0.1 & 100 $\%$ & 100 $\%$ & 100$\%$ & 100$\%$ & 100$\%$ & 92$\%$ & 83$\%$  \\
		& 0.3 & 100 $\%$ & 100 $\%$ &100$\%$ & 94$\%$ & 81$\%$ & 32$\%$ & 5$\%$  \\
		& 0.5 & 100 $\%$ & 100 $\%$ & 78$\%$ & 46$\%$ & 0$\%$ & 0$\%$ & 0$\%$  \\
		& 0.7 & 100 $\%$ & 100 $\%$ &28$\%$ & 0$\%$ & 0$\%$ & 0$\%$ & 0$\%$   \\
%		\multirow{4}{*}{}MCD  & - &100$\%$ & 100$\%$ &100$\%$ & 100$\%$ & 100$\%$ & 100$\%$ & 100$\%$   \\
%		\multirow{4}{*}{}MVE  & -&100$\%$ & 100$\%$  &100$\%$ & 100$\%$ & 100$\%$ & 100$\%$ & 100$\%$   \\
%		\multirow{4}{*}{}GK  &-& 100$\%$ & 100$\%$  &100$\%$ & 100$\%$ & 100$\%$ & 100$\%$ & 100$\%$   \\
%		\multirow{4}{*}{}MM  & - &100$\%$ & 100$\%$ &100$\%$ & 100$\%$ & 100$\%$ & 100$\%$ & 100$\%$   \\
%		\multirow{4}{*}{}S  & - &100$\%$ & 100$\%$ &100$\%$ & 100$\%$ & 100$\%$ & 100$\%$ & 100$\%$   \\
		\hline
	\end{tabular}
	\caption{Empirical convergence rates of the indicated methods for a sample size of $n=2000$. }
	\label{sm_conrate}
\end{table}
the ordinary minimum DPD estimation in the sense that the existence of the CMDPDE is computationally guaranteed for almost all possible combinations of $n$, $p$ and $\beta$ unlike the ordinary MDPDE. Especially, in higher dimensions, the algorithm used to obtain the ordinary MDPDE may fail to converge. The empirical convergence rates of the ordinary and componentwise minimum DPD methods %(with a maximum iteration limit as $100$)
are shown in Table \ref{sm_conrate} where this fact is clearly borne out. In fact for large data dimensions and large values of $\beta$, the minimum DPD method practically never leads to convergence. Convergence rates of the other competitors are found to be perfect like the componentwise minimum DPD method.

\section{Credit Card Transactions Data}
\label{sm_sec5}
In this section, we apply the componentwise minimum DPD method on a real data containing information about credit card transactions by European cardholders on two particular days of September, $2013$; some of these transactions were fraudulent. The dataset consists of (\href{https://www.kaggle.com/mlg-ulb/creditcardfraud/version/3}{link}) $28$ features (first $28$ principal components of the original dataset which is confidential; the original dataset has been collected and analysed during a research collaboration of Worldline and the Machine Learning Group (\href{http://mlg.ulb.ac.be}{http://mlg.ulb.ac.be}) of ULB (Université Libre de Bruxelles) on big data mining and fraud detection) along with the elapsed times (from the first transaction), transaction amount and original transaction labels (i.e., genuine or fraudulent). 
\begin{figure}[h!]
	\centering
	\includegraphics[height=3 in]{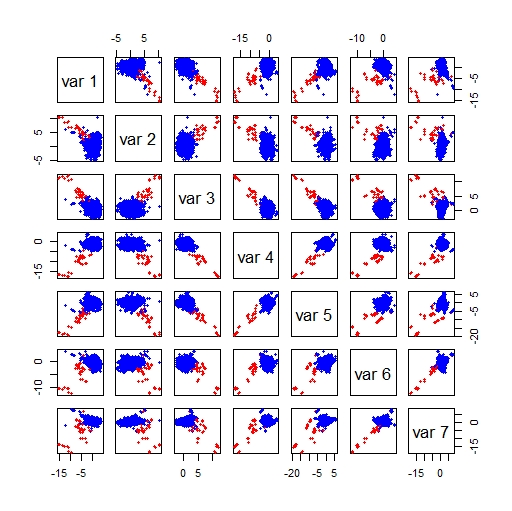}
	\caption{Pairwise scatter plots of some of the components with blue points as genuine observations and red as fraudulent ones.}
	\label{sm_ccf}
\end{figure}
This dataset has been analyzed in recent years using various machine learning tools \cite{ccf1,ccf2,ccf3}. For time and space complexity, we analyze a subset of this dataset with the first $10000$ transactions of which only $38$ are fraudulent and the remaining $9962$ are genuine. We consider these fraudulent transactions as outlying observations present in the dataset (as observed in Figure \ref{sm_ccf}). But the proportion of outliers is only $0.0038$ which implies the dataset can almost be regarded as a noise-free one and no robust method is expected to show its supremacy through drastic differences in comparison with the traditional likelihood based procedure if applied on this dataset. To make the contamination stronger, we perturb another $362$ observations which represent genuine transactions. This is done following two approaches. Let us note that, we only have to choose a sub-sample of size $362$ of genuine observations (transactions) out of a total of $9962$ genuine observations (transactions). Either we can choose this sub-sample randomly (without replacement) or we may choose those $362$ genuine observations which are mostly concentrated around their central tendency. That is, if $\boldsymbol{\tilde{\mu}}$ is the sample componentwise median of the $9962$ genuine observations, then we may choose those $362$ genuine observations whose distances from $\boldsymbol{\tilde{\mu}}$ are the least.
 \begin{table}[!h]
	\centering
%	\small
	\begin{tabular}{c c c c } 
		\hline
		Resampling  & Method &  Difference between  & Difference between   \\ [1ex] 
		Type & & estimated mean vectors & estimated covariance matrices\\
		\hline
		Most & MLE & 2.007 & 50.584\\
		Concentrated & CMDPDE $(\beta=0.1)$ & 0.268 & 1.506  \\
		& CMDPDE $(\beta=0.3)$& 0.109 & 0.378 \\
		& CMDPDE $(\beta=0.5)$ & 0.353 & 1.678  \\
		& & &\\
		Random & MLE & 2.013 & 50.857\\
		& CMDPDE $(\beta=0.1)$ & 0.740  & 10.616    \\
		& CMDPDE $(\beta=0.3)$& 0.280 & 2.213 \\
		& CMDPDE $(\beta=0.5)$ & 0.211 & 1.516   
		\\[2ex] 
		\hline
	\end{tabular}
	\caption{$L_2$ differences between estimated mean vectors (covariance matrices) based on the genuine sub-samples (size $362$) and the contaminated sub-samples (size $400$).}
	\label{sm_table1.8}
\end{table}

\noindent
We now apply our method along with the likelihood estimation procedure to fit the aforesaid sub-samples using multivariate normal distribution (observing an approximate elliptic nature of the overall dataset as observed in Figure \ref{sm_ccf}). To understand the robustness of our method, we first consider the noise-free sub-samples of the data with $362$ genuine observations and then the contaminated sub-samples of size $400$ where $38$ fraudulent transactions were added to the previous noise-free sub-samples of size $362$. We observe the differences ($L_2$ distances) between the estimated mean vectors (and the estimated covariance matrices) based on the noise-free sub-samples and the contaminated sub-samples. To model either the noise-free sub-samples or the contaminated sub-samples, we need $28$-dimensional normal distributions (comprising $434$ parameters) but we only have samples of size $362$ or $400$. Thus, none of the robust alternatives to our method (i.e., the ordinary minimum DPD method, MCD, MVE etc.) used in the simulation experiments can be successful in producing the estimates of mean vector and covariance matrix for possible singularity. The $L_2$ differences between the estimated mean vectors (and the estimated covariance matrices) based on the noise-free sub-samples and the contaminated sub-samples by the likelihood method and the componentwise minimum DPD method are presented in Table \ref{sm_table1.8}. These differences for our method are much less than those in case of the likelihood method which establish the superiority of our method in terms of robustness. Also, this application have shown the applicability of our method in case of such a higher dimensional dataset where the data dimension is $28$, so that, we need to estimate $434$ unknown parameters which is indeed greater than the sample size $400$. For this particular data example, Higham's algorithm (followed by an eigenvalue truncation step, as discussed in Remark \ref{sm_highamremark}) is utilized to find the nearest positive definite correlation matrices (of dimension $28\times 28$) to the estimated correlation matrices by our method.

    \section{Concluding Remarks}
    \label{sm_sec6}
    We have derived a robust and asymptotically efficient method to estimate the location and scatter matrices of elliptically symmetric probability distributions under multivariate set-ups. This estimation procedure is robust and computationally tractable as well as asymptotically efficient under certain regularity conditions. Computational tractability is the main theme of this method. We have established consistency and asymptotic normality of our estimators and derived the influence functions explicitly under the assumption of normality. Finally, the simulation experiments have suggested the positive features of our newly established method, componentwise minimum DPD estimation.    

    The proposed componentwise method has been shown to be theoretically novel and computationally efficient. However, there are certain issues which are associated with this proposal due to its methodological nature. For instance, the proposed componentwise estimation procedure may, in some cases involving small sample sizes, produce non-positive definite scale estimators. But, we have established the asymptotic positive definiteness of the same under certain assumptions (Theorem \ref{sm_pdt}), i.e., positive definiteness is guaranteed with high probability in large sample scenarios. In case of obtaining non-positive definite scale estimators, a possible correction was suggested by Higham $(2002)$ \cite{pd2} which proposed a numerical algorithm to compute the nearest positive semi-definite correlation matrix (followed by an eigenvalue truncation step, if required) to a given symmetric matrix which is approximately a correlation matrix. Such a correction needs to be incorporated in these pathological cases. We have applied this correction in case of the real data example. Secondly, while our method is found to clearly outperform all other methods in case of the distant outlier contamination set-up, the situation is more competitive in relation to some of the other methods in case of the subtle outlier contamination set-up. For example, the fast robust correlation method has slightly better performance compared to our method in case of scale estimators under the diagonal covariance structure for the subtle outlier contamination set-up. However, for the non-diagonal covariance structure, the MM and S estimators turn out to be marginally superior for the subtle outlier contaminated datasets. Finally, computational times may be an issue in case of higher data dimensions. 
    
    In future, it will be interesting to build new classification algorithms with applications in various machine learning and image analysis problems. Robust covariance matrix estimation plays a key role in the fields of finance and econometrics. This method can also be extended to high dimensional (although explored empirically in Section \ref{sm_hd_sim}) set-ups which is yet a quite difficult statistical problem. Apart from these applications, it will be interesting to study the behaviours of both the ordinary and componentwise minimum DPD methods under cellwise contamination. We have provided one such example in Section \ref{sm_cellwise}. We plan to study the same with more detailed analysis in a future work. \\
    %\newline
    %\noindent
    %\textbf{Software Implementation:} The $\sf{R}$ codes of the newly proposed componentwise algorithm are available at this \href{https://github.com/sc2-wbes/R-Codes-for-CMDPDE}{link}.

\begin{appendices}
\section{Proofs of the Theorems}
\label{sm_appen0}
\subsection{Proof of Theorem \ref{sm_consistency}}
\begin{proof} We follow the same approach which was taken by Basu et al.~$(2011)$ \cite{basubook}. Our plan is to show that for all sufficiently small $\epsilon$, $H_{n}(\widehat{\boldsymbol{\theta}}_{1n},\widehat{\boldsymbol{\theta}}_{2n},\rho^{g})<H_{n}(\widehat{\boldsymbol{\theta}}_{1n},\widehat{\boldsymbol{\theta}}_{2n},\rho)$ for all points $\rho \in \text{Surface}(Q_{\epsilon})$ with probability tending to $1$ where $Q_{\epsilon}$ is a sphere of radius $\epsilon>0$ with center at $\rho^{g}$. Hence, $H_{n}(\widehat{\boldsymbol{\theta}}_{1n},\widehat{\boldsymbol{\theta}}_{2n},\rho)$ has a local minima in the interior of $Q_{\epsilon}$. Now, let us observe the fact that at a local minimum, Equation (\ref{sm_eq14}) must be satisfied. Thus, for any sufficiently small $\epsilon>0$, Equation (\ref{sm_eq14}) has a solution $\widehat{\rho}_{n\epsilon}$ depending on $\epsilon$ with probability tending to $1$ as $n \rightarrow \infty$. To execute our plan, let us consider the following Taylor series expansion of $H_{n}(\widehat{\boldsymbol{\theta}}_{1n},\widehat{\boldsymbol{\theta}}_{2n},\rho)$ around $\rho=\rho^{g}$.
	\begin{equation}
	\centering
	\label{sm_eq15}
	\begin{array}{l}
	H_{n}(\widehat{\boldsymbol{\theta}}_{1n},\widehat{\boldsymbol{\theta}}_{2n},\rho)=H_{n}(\widehat{\boldsymbol{\theta}}_{1n},\widehat{\boldsymbol{\theta}}_{2n},\rho^{g})+(\rho-\rho^{g})\frac{\partial H_{n}(\widehat{\boldsymbol{\theta}}_{1n},\widehat{\boldsymbol{\theta}}_{2n},\rho)}{\partial \rho}\Big{|}_{\rho=\rho^{g}}+\frac{(\rho-\rho^{g})^{2}}{2!}\frac{\partial^{2} H_{n}(\widehat{\boldsymbol{\theta}}_{1n},\widehat{\boldsymbol{\theta}}_{2n},\rho)}{\partial^{2} \rho}\Big{|}_{\rho=\rho^{g}}+\\
	\frac{(\rho-\rho^{g})^{3}}{3!}\frac{\partial^{3} H_{n}(\widehat{\boldsymbol{\theta}}_{1n},\widehat{\boldsymbol{\theta}}_{2n},\rho)}{\partial^{3} \rho}\Big{|}_{\rho=\rho^{*}}
	= H_{n}(\widehat{\boldsymbol{\theta}}_{1n},\widehat{\boldsymbol{\theta}}_{2n},\rho^{g}) +S_{1}+S_{2}+S_{3}
	\end{array}
	\end{equation}
	
	for some $\rho^{*}$ lying between $\rho$ and $\rho^{g}$.  Now let us observe that,
	
	\begin{align*}
	\label{sm_eq16}
	\frac{1}{1+\beta}\frac{\partial H_{n}(\widehat{\boldsymbol{\theta}}_{1n},\widehat{\boldsymbol{\theta}}_{2n},\rho)}{\partial \rho}\Big{|}_{\rho=\rho^{g}}= \int f^{1+\beta}(\boldsymbol{x},\widehat{\boldsymbol{\theta}}_{1n},\widehat{\boldsymbol{\theta}}_{2n},\rho^{g})U_{\rho}(\boldsymbol{x},\rho^{g}|\widehat{\boldsymbol{\theta}}_{1n},\widehat{\boldsymbol{\theta}}_{2n})\;d\boldsymbol{x}\\
	-\frac{1}{n}\sum_{i=1}^{n}f^{\beta}(\boldsymbol{X}_{i},\widehat{\boldsymbol{\theta}}_{1n},\widehat{\boldsymbol{\theta}}_{2n},\rho^{g})U_{\rho}(\boldsymbol{X}_{i},\rho^{g}|\widehat{\boldsymbol{\theta}}_{1n},\widehat{\boldsymbol{\theta}}_{2n}),
	\end{align*}
	where $U_{\rho}(\boldsymbol{x},\rho\;|\;\widehat{\boldsymbol{\theta}}_{1n},\widehat{\boldsymbol{\theta}}_{2n})=\frac{\partial \text{log}\;f(\boldsymbol{x},\widehat{\boldsymbol{\theta}}_{1n},\widehat{\boldsymbol{\theta}}_{2n},\rho) }{\partial \rho}$. Let, $M(\boldsymbol{x},\widehat{\boldsymbol{\theta}}_{1n},\widehat{\boldsymbol{\theta}}_{2n},\rho^{g})=f^{1+\beta}(\boldsymbol{x},\widehat{\boldsymbol{\theta}}_{1n},\widehat{\boldsymbol{\theta}}_{2n},\rho^{g})U_{\rho}(\boldsymbol{x},\rho^{g}|\\\widehat{\boldsymbol{\theta}}_{1n},\widehat{\boldsymbol{\theta}}_{2n})$ and $N(\boldsymbol{X}_{i},\widehat{\boldsymbol{\theta}}_{1n},\widehat{\boldsymbol{\theta}}_{2n},\rho^{g})=f^{\beta}(\boldsymbol{X}_{i},\widehat{\boldsymbol{\theta}}_{1n},\widehat{\boldsymbol{\theta}}_{2n},\rho^{g})U_{\rho}(\boldsymbol{X}_{i},\rho^{g}|\widehat{\boldsymbol{\theta}}_{1n},\widehat{\boldsymbol{\theta}}_{2n})$. Thus, 
	\begin{equation}
	\label{sm_eq17}
	\frac{1}{1+\beta}\frac{\partial H_{n}(\widehat{\boldsymbol{\theta}}_{1n},\widehat{\boldsymbol{\theta}}_{2n},\rho)}{\partial \rho}\Big{|}_{\rho=\rho^{g}}=\int M(\boldsymbol{x},\widehat{\boldsymbol{\theta}}_{1n},\widehat{\boldsymbol{\theta}}_{2n},\rho^{g})\;d\boldsymbol{x}-\frac{1}{n}\sum_{i=1}^{n}N(\boldsymbol{X}_{i},\widehat{\boldsymbol{\theta}}_{1n},\widehat{\boldsymbol{\theta}}_{2n},\rho^{g}).
	\end{equation}
	Now, we are going to consider the first order Taylor series expansions of the functions $M(\boldsymbol{x},\boldsymbol{\theta}_{1},\boldsymbol{\theta}_{2}, \rho^{g})$ with respect to $(\boldsymbol{\theta}_{1},\boldsymbol{\theta}_{2})$ and around $(\boldsymbol{\theta}_{1},\boldsymbol{\theta}_{2})=(\boldsymbol{\theta}_{1}^{\boldsymbol{g}},\boldsymbol{\theta}_{2}^{\boldsymbol{g}})$ and  $\frac{1}{n}\sum_{i=1}^{n}N(\boldsymbol{X}_{i},\boldsymbol{\theta}_{1},\boldsymbol{\theta}_{2},\rho^{g})$ with respect to $(\boldsymbol{\theta}_{1},\boldsymbol{\theta}_{2})$ and around $(\boldsymbol{\theta}_{1},\boldsymbol{\theta}_{2})=(\boldsymbol{\theta}_{1}^{\boldsymbol{g}},\boldsymbol{\theta}_{2}^{\boldsymbol{g}})$ and evaluate the same functions at $(\boldsymbol{\theta}_{1},\boldsymbol{\theta}_{2})=(\widehat{\boldsymbol{\theta}}_{1n},\widehat{\boldsymbol{\theta}}_{2n})$. 
	\begin{align*}
	\int M(\boldsymbol{x},\widehat{\boldsymbol{\theta}}_{1n},\widehat{\boldsymbol{\theta}}_{2n},\rho^{g})\;d\boldsymbol{x}=\int M(\boldsymbol{x},\boldsymbol{\theta}_{1}^{\boldsymbol{g}},\boldsymbol{\theta}_{2}^{\boldsymbol{g}},\rho^{g})\;d\boldsymbol{x} +\sum_{j,k}(\widehat{\theta}_{njk}-\theta^{g}_{jk})\int \frac{\partial M(\boldsymbol{x},\boldsymbol{\theta}_{1},\boldsymbol{\theta}_{2},\rho^{g})}{\partial \theta_{jk}}\Big{|}_{(\boldsymbol{\theta}_{1},\boldsymbol{\theta}_{2})=(\boldsymbol{\theta}_{1}^{\boldsymbol{*}},\boldsymbol{\theta}_{2}^{\boldsymbol{*}})}, 
	\end{align*}
	where $(\boldsymbol{\theta}_{1}^{\boldsymbol{*}},\boldsymbol{\theta}_{2}^{\boldsymbol{*}})$ lies between $(\widehat{\boldsymbol{\theta}}_{1n},\widehat{\boldsymbol{\theta}}_{2n})$ and $(\boldsymbol{\theta}_{1}^{\boldsymbol{g}},\boldsymbol{\theta}_{2}^{\boldsymbol{g}})$ and 
	\begin{align*}
	\frac{1}{n}\sum_{i=1}^{n}N(\boldsymbol{X}_{i},\widehat{\boldsymbol{\theta}}_{1n},\widehat{\boldsymbol{\theta}}_{2n},\rho^{g})&=\frac{1}{n}\sum_{i=1}^{n}N(\boldsymbol{X}_{i},\boldsymbol{\theta}_{1}^{\boldsymbol{g}},\boldsymbol{\theta}_{2}^{\boldsymbol{g}},\rho^{g})\\
    &+ \sum_{j,k}(\widehat{\theta}_{njk}-\theta^{g}_{jk})\frac{1}{n}\sum_{i=1}^{n}\frac{\partial N(\boldsymbol{X}_{i},\boldsymbol{\theta}_{1},\boldsymbol{\theta}_{2},\rho^{g})}{\partial \theta_{jk}}\Big{|}_{(\boldsymbol{\theta}_{1},\boldsymbol{\theta}_{2})=(\boldsymbol{\theta}_{1}^{\boldsymbol{**}},\boldsymbol{\theta}_{2}^{\boldsymbol{**}})}, 
	\end{align*}
	where $(\boldsymbol{\theta}_{1}^{\boldsymbol{**}},\boldsymbol{\theta}_{2}^{\boldsymbol{**}})$ lies between $(\widehat{\boldsymbol{\theta}}_{1n},\widehat{\boldsymbol{\theta}}_{2n})$ and $(\boldsymbol{\theta}_{1}^{\boldsymbol{g}},\boldsymbol{\theta}_{2}^{\boldsymbol{g}})$. 
	%Now let us further assume the following.
	%	\begin{enumerate}
	%		\item[\null]{\textbf{A4}} $\Big|\frac{\partial M(x,\theta,\rho^{g})}{\partial \theta_{jk}}\Big|<M_{jk}(x,\rho_{g})$ for all $\theta \in \omega$ (recall assumption (A1)) and $x$ for some $M_{jk}(x,\rho^{g})$ which is $L_1$ bounded and $ \Big|\frac{\partial N(x,\theta,\rho^{g})}{\partial \theta_{jk}}\Big|<N_{jk}(x,\rho^{g})$ for all $\theta \in \omega$ and $x$, $j=k=1,2$ for some $N_{jk}(x,\rho^{g})$ which satisfy $E_{g}(N_{jk}(X_{1},\rho^{g}))<\infty$  for $j=k=1,2$.
	%	\end{enumerate}
	Now by weak law of large numbers (WLLN) and Assumption \ref{sm_a4},
	
	\begin{align*}
	\frac{1}{n}\sum_{i=1}^{n}N(\boldsymbol{X}_{i},\boldsymbol{\theta}_{1}^{\boldsymbol{g}},\boldsymbol{\theta}_{2}^{\boldsymbol{g}},\rho^{g}) &\overset{p}{\rightarrow} E_{g}\left(N(\boldsymbol{X}_{1},\boldsymbol{\theta}_{1}^{\boldsymbol{g}},\boldsymbol{\theta}_{2}^{\boldsymbol{g}},\rho^{g}\right),\\
	\Big|\frac{1}{n}\sum_{i=1}^{n}\frac{\partial N(\boldsymbol{X}_{i},\boldsymbol{\theta}_{1},\boldsymbol{\theta}_{2},\rho^{g})}{\partial \theta_{jk}}\Big{|}_{(\boldsymbol{\theta}_{1},\boldsymbol{\theta}_{2})=(\boldsymbol{\theta}_{1}^{\boldsymbol{**}},\boldsymbol{\theta}_{2}^{\boldsymbol{**}})}\Big|&<\frac{1}{n}\sum_{i=1}^{n}\Big|\frac{\partial N(\boldsymbol{X}_{i},\boldsymbol{\theta}_{1},\boldsymbol{\theta}_{2},\rho^{g})}{\partial \theta_{jk}}\Big{|}_{(\boldsymbol{\theta}_{1},\boldsymbol{\theta}_{2})=(\boldsymbol{\theta}_{1}^{\boldsymbol{**}},\boldsymbol{\theta}_{2}^{\boldsymbol{**}})}\\
	&< \frac{1}{n}\sum_{i=1}^{n} N_{jk}(\boldsymbol{X}_{i},\rho^{g})
	\overset{p}{\rightarrow} E_{g}(N_{jk}(\boldsymbol{X}_{1},\rho^{g}))<\infty.
	\end{align*}
	We also have $\Bigg|\int \frac{\partial M(\boldsymbol{x},\boldsymbol{\theta}_{1},\boldsymbol{\theta}_{2},\rho^{g})}{\partial \theta_{jk}}\Big{|}_{(\boldsymbol{\theta}_{1},\boldsymbol{\theta}_{2})=(\boldsymbol{\theta}_{1}^{\boldsymbol{*}},\boldsymbol{\theta}_{2}^{\boldsymbol{*}})}\;d\boldsymbol{x}\Bigg|<\int \Big|  \frac{\partial M(\boldsymbol{x},\boldsymbol{\theta}_{1},\boldsymbol{\theta}_{2},\rho^{g})}{\partial \theta_{jk}}\Big{|}_{(\boldsymbol{\theta}_{1},\boldsymbol{\theta}_{2})=(\boldsymbol{\theta}_{1}^{\boldsymbol{*}},\boldsymbol{\theta}_{2}^{\boldsymbol{*}})}\Big|\;d\boldsymbol{x}<\int M_{jk}(\boldsymbol{x},\rho_{g})\\ d\boldsymbol{x}<\infty$.
	From Equation (\ref{sm_eq13}), we have $\widehat{\theta}_{jkn}\overset{p}{\rightarrow}\theta^{g}_{jk}$ for $j\in\{1,2\}$, $k\in\{1,2\}$. Now using the aforesaid observations, we can write from Equation (\ref{sm_eq17}) that,
	\begin{align*}
	\frac{\partial H_{n}(\widehat{\boldsymbol{\theta}}_{1n},\widehat{\boldsymbol{\theta}}_{2n},\rho)}{\partial \rho}\Big{|}_{\rho=\rho^{g}} &\overset{p}{\rightarrow} \int M(\boldsymbol{x},\boldsymbol{\theta}_{1}^{\boldsymbol{g}},\boldsymbol{\theta}_{2}^{\boldsymbol{g}},\rho^{g})\;d\boldsymbol{x} - E_{g}\left(N(\boldsymbol{X}_{1},\boldsymbol{\theta}_{1}^{\boldsymbol{g}},\boldsymbol{\theta}_{2}^{\boldsymbol{g}},\rho^{g})\right) \\
	&=
	\frac{\partial H(\boldsymbol{\theta}_{1}^{\boldsymbol{g}},\boldsymbol{\theta}_{2}^{\boldsymbol{g}},\rho)}{\partial \rho}\Big|_{\rho=\rho^{g}}=0
	\end{align*}
	as $\rho^{g}=\underset{\rho}{\text{argmin}}\;H(\boldsymbol{\theta}_{1}^{\boldsymbol{g}},\boldsymbol{\theta}_{2}^{\boldsymbol{g}},\rho)$. Thus, $\frac{\partial H_{n}(\widehat{\boldsymbol{\theta}}_{1n},\widehat{\boldsymbol{\theta}}_{2n},\rho)}{\partial \rho}\Big{|}_{\rho=\rho^{g}} \overset{p}{\rightarrow} 0$ as $n\rightarrow \infty$. Since $\rho \in \text{Surface} (Q_{\epsilon})$, $|\rho-\rho^{g}|=\epsilon$ and by the fact that $\frac{\partial H_{n}(\widehat{\boldsymbol{\theta}}_{1n},\widehat{\boldsymbol{\theta}}_{2n},\rho)}{\partial \rho}\Big{|}_{\rho=\rho^{g}} \overset{p}{\rightarrow} 0$, we have 
	\begin{equation}
	\label{sm_eq18}
	|S_{1}|< \epsilon^{3}
	\end{equation}
	with probability tending to 1. Our next agenda is to handle the term $S_{2}$ in the right hand side of Equation (\ref{sm_eq15}). Let us note that,
	\begin{equation}
	\centering
	\label{sm_eq19}
	\begin{array}{l}
	\frac{1}{1+\beta}\frac{\partial^{2} H_{n}(\widehat{\boldsymbol{\theta}}_{1n},\widehat{\boldsymbol{\theta}}_{2n},\rho)}{\partial^{2} \rho}\Big{|}_{\rho=\rho^{g}}=(1+\beta)\int f^{1+\beta}(\boldsymbol{x},\widehat{\boldsymbol{\theta}}_{1n},\widehat{\boldsymbol{\theta}}_{2n},\rho^{g})U^{2}_{\rho}(\boldsymbol{x},\rho^{g}|\widehat{\boldsymbol{\theta}}_{1n},\widehat{\boldsymbol{\theta}}_{2n})\;d\boldsymbol{x}+\\
	\int f^{1+\beta}(\boldsymbol{x},\widehat{\boldsymbol{\theta}}_{1n},\widehat{\boldsymbol{\theta}}_{2n},\rho^{g})\frac{\partial\;U_{\rho}(\boldsymbol{x},\rho|\widehat{\boldsymbol{\theta}}_{1n},\widehat{\boldsymbol{\theta}}_{2n})}{\partial \rho}\Big|_{\rho=\rho^{g}}\;d\boldsymbol{x}
	-\frac{\beta}{n}\sum_{i=1}^{n}f^{\beta}(\boldsymbol{X}_{i},\widehat{\boldsymbol{\theta}}_{1n},\widehat{\boldsymbol{\theta}}_{2n},\rho^{g})U^{2}_{\rho}(\boldsymbol{X}_{i},\rho^{g}|\widehat{\boldsymbol{\theta}}_{1n},\widehat{\boldsymbol{\theta}}_{2n})\\
	-\frac{1}{n}\sum_{i=1}^{n}f^{\beta}(\boldsymbol{X}_{i},\widehat{\boldsymbol{\theta}}_{1n},\widehat{\boldsymbol{\theta}}_{2n},\rho^{g})\frac{\partial\;U_{\rho}(\boldsymbol{X}_{i},\rho|\widehat{\boldsymbol{\theta}}_{1n},\widehat{\boldsymbol{\theta}}_{2n})}{\partial \rho}\Big|_{\rho=\rho^{g}}.
	\end{array}
	\end{equation}
	Now, by using the same first order Taylor series expansion trick as in case of $S_{1}$, we can prove (after some algebra) that,
	\begin{align*}
	\frac{\partial^{2} H_{n}(\widehat{\boldsymbol{\theta}}_{1n},\widehat{\boldsymbol{\theta}}_{2n},\rho)}{\partial^{2} \rho}\Big{|}_{\rho=\rho^{g}} \overset{p}{\rightarrow} \frac{\partial^{2} H(\boldsymbol{\theta}_{1}^{\boldsymbol{g}},\boldsymbol{\theta}_{2}^{\boldsymbol{g}},\rho)}{\partial^{2} \rho}\Big{|}_{\rho=\rho^{g}}>0
	\end{align*}
	as $\rho^{g}$ is the minimizer of $H(\boldsymbol{\theta}_{1}^{\boldsymbol{g}},\boldsymbol{\theta}_{2}^{\boldsymbol{g}},\rho)$. Thus for any $\rho \in \text{Surface}(Q_{\epsilon})$, with probability tending to $1$, 
	\begin{equation}
	\label{sm_eq20}
	S_{2}>c\epsilon^{2}
	\end{equation}
	for some constant $c>0$. Now it only remains to take care of the third term $S_{3}$ in the right hand side of Equation (\ref{sm_eq15}). Let us observe that,
	\begin{align*}
	\frac{\partial^{3} H_{n}(\widehat{\boldsymbol{\theta}}_{1n},\widehat{\boldsymbol{\theta}}_{2n},\rho)}{\partial^{3} \rho}\Big{|}_{\rho=\rho^{*}}&=\frac{1}{n}\sum_{i=1}^{n}\frac{\partial^{3} V(\boldsymbol{X}_{i},\widehat{\boldsymbol{\theta}}_{1n},\widehat{\boldsymbol{\theta}}_{2n},\rho)}{\partial^{3} \rho}\Big{|}_{\rho=\rho^{*}}\\
	&=\frac{1}{n}\sum_{i=1}^{n}\frac{\partial^{3} V(\boldsymbol{X}_{i},\boldsymbol{\theta}_{1}^{\boldsymbol{g}},\boldsymbol{\theta}_{2}^{\boldsymbol{g}},\rho)}{\partial^{3} \rho}\Big{|}_{\rho=\rho^{*}}\\
	&+\sum_{j,k}(\widehat{\theta}_{njk}-\theta^{g}_{jk})\frac{1}{n}\sum_{i=1}^{n}\frac{\partial^{4} V(\boldsymbol{X}_{i},\widehat{\boldsymbol{\theta}}_{1n},\widehat{\boldsymbol{\theta}}_{2n},\rho)}{\partial^{3} \rho\partial \theta_{jk}}\Big{|}_{\rho=\rho^{*},(\boldsymbol{\theta}_{1}\boldsymbol{\theta}_{2})=(\boldsymbol{\theta}_{1}^{\boldsymbol{\dagger}},\boldsymbol{\theta}_{2}^{\boldsymbol{\dagger}})}.
	\end{align*}
	We utilize Assumption \ref{sm_a5} at this point to show that $S_{3}$ is finite in absolute sense with probability tending to $1$. 
	%    \begin{enumerate}
	%    	\item[\null]{\textbf{A5}} $\Big|\frac{\partial^{3} V(x,\theta^{g},\rho)}{\partial \rho^{3}}\Big|<v_{1}(x,\theta^{g})$ for all $\theta \in \omega$, for all $x$ for some $v_{1}(x,\theta^{g})$ satisfying $E_{g}(v_{1}(X_{1},\theta^{g}))<\infty$ and  $\Big|\frac{\partial^{4} V(x,\theta,\rho)}{\partial \rho^{3}\partial \theta_{jk}}\Big|_{\theta=\theta^{g}}\Big|<v_{2}(x,\theta^{g})$ for all $\theta \in \omega$, for all $x$ for some $v_{2}(x,\theta^{g})$ satisfying $E_{g}(v_{2}(X_{1},\theta^{g}))<\infty$ for $j=k=1,2$.
	%    \end{enumerate}
	By Assumption \ref{sm_a5} and WLLN, we have,
	\begin{align*}
	\Bigg|\frac{1}{n}\sum_{i=1}^{n}\frac{\partial^{3} V(\boldsymbol{X}_{i},\boldsymbol{\theta}_{1}^{\boldsymbol{g}},\boldsymbol{\theta}_{2}^{\boldsymbol{g}},\rho)}{\partial^{3} \rho}\Big{|}_{\rho=\rho^{*}}\Bigg|&< \frac{1}{n}\sum_{i=1}^{n}\Bigg|\frac{\partial^{3} V(\boldsymbol{X}_{i},\boldsymbol{\theta}_{1}^{\boldsymbol{g}},\boldsymbol{\theta}_{2}^{\boldsymbol{g}},\rho)}{\partial^{3} \rho}\Big{|}_{\rho=\rho^{*}}\Bigg|\\
	&<\frac{1}{n}\sum_{i=1}^{n}v_{1}(\boldsymbol{X}_{i},\boldsymbol{\theta}_{1}^{\boldsymbol{g}},\boldsymbol{\theta}_{2}^{\boldsymbol{g}})\overset{p}{\rightarrow}E_{g}v_{1}(\boldsymbol{X}_{1},\boldsymbol{\theta}_{1}^{\boldsymbol{g}},\boldsymbol{\theta}_{2}^{\boldsymbol{g}})=d \text{(Say)}<\infty,
	\end{align*}
	\begin{align*}
	\Bigg|\frac{1}{n}\sum_{i=1}^{n}\frac{\partial^{4} V(\boldsymbol{X}_{i},\widehat{\boldsymbol{\theta}}_{1n},\widehat{\boldsymbol{\theta}}_{2n},\rho)}{\partial^{3} \rho\partial \theta_{jk}}\Big{|}_{\rho=\rho^{*},(\boldsymbol{\theta}_{1}\boldsymbol{\theta}_{2})=(\boldsymbol{\theta}_{1}^{\boldsymbol{\dagger}},\boldsymbol{\theta}_{2}^{\boldsymbol{\dagger}})}\Bigg|&<\frac{1}{n}\sum_{i=1}^{n}\Bigg|\frac{\partial^{4} V(\boldsymbol{X}_{i},\widehat{\boldsymbol{\theta}}_{1n},\widehat{\boldsymbol{\theta}}_{2n},\rho)}{\partial^{3} \rho\partial \theta_{jk}}\Big{|}_{\rho=\rho^{*},(\boldsymbol{\theta}_{1}\boldsymbol{\theta}_{2})=(\boldsymbol{\theta}_{1}^{\boldsymbol{\dagger}},\boldsymbol{\theta}_{2}^{\boldsymbol{\dagger}})}\Bigg|\\
	&<\frac{1}{n}\sum_{i=1}^{n}v_{2}(\boldsymbol{X}_{i},\boldsymbol{\theta}_{1}^{\boldsymbol{g}},\boldsymbol{\theta}_{2}^{\boldsymbol{g}})\overset{p}{\rightarrow}E_{g}v_{2}(\boldsymbol{X}_{1},\boldsymbol{\theta}_{1}^{\boldsymbol{g}},\boldsymbol{\theta}_{2}^{\boldsymbol{g}})<\infty
	\end{align*}
	for $j\in\{1,2\}$, $k\in\{1,2\}$. From Equation (\ref{sm_eq13}), we have $\widehat{\theta}_{jkn}\overset{p}{\rightarrow}\theta^{g}_{jk}$ for $j\in\{1,2\}$, $k\in\{1,2\}$. Thus, for $\rho \in \text{Surface}(Q_{\epsilon})$,
	\begin{equation}
	\label{sm_eq21}
	|S_3|=\Bigg|\frac{(\rho-\rho^{g})^3}{3!}\Bigg|\Bigg|\frac{\partial^{3} H_{n}(\widehat{\boldsymbol{\theta}}_{1n},\widehat{\boldsymbol{\theta}}_{2n},\rho)}{\partial^{3} \rho}\Big{|}_{\rho=\rho^{*}}\Bigg|<d\epsilon^{3},\;\text{for some constant}\;d>0.
	\end{equation}
	Now, from Equations (\ref{sm_eq15}), (\ref{sm_eq18}), (\ref{sm_eq20}) and (\ref{sm_eq21}),
	\begin{align*}
	H_{n}(\widehat{\boldsymbol{\theta}}_{1n},\widehat{\boldsymbol{\theta}}_{2n},\rho^{g})- H_{n}(\widehat{\boldsymbol{\theta}}_{1n},\widehat{\boldsymbol{\theta}}_{2n},\rho)= -S_{1}-S_{2}-S_{3}\leq |S_{1}|-S_{2}+|S_{3}|<\epsilon^{3}-c\epsilon^{2}+d\epsilon^{3}, 
	\end{align*}
	which is less than $0$ if $\epsilon<\frac{c}{1+d}$. Hence, for any sufficiently small $\epsilon$, we have $H_{n}(\widehat{\boldsymbol{\theta}}_{1n},\widehat{\boldsymbol{\theta}}_{2n},\rho^{g})< H_{n}(\widehat{\boldsymbol{\theta}}_{1n},\widehat{\boldsymbol{\theta}}_{2n},\rho)$ for all $\rho \in \text{Surface}(Q_{\epsilon})$ with probability tending to $1$. This implies, with probability tending to $1$, there exists a sequence $\{\widehat{\rho}_{n\epsilon}\} \in \text{Interior}(Q_{\epsilon})$ which minimizes $H_{n}(\widehat{\boldsymbol{\theta}}_{1n},\widehat{\boldsymbol{\theta}}_{2n},\rho)$ for all sufficiently small $\epsilon$. 
	
	Let us define $\widehat{\rho}^{*}_{n}$ to be the minimizer which is closest to $\rho^{g}$ among all $\widehat{\rho}_{n\epsilon}$ for any sufficiently small $\epsilon$. Thus $\widehat{\rho}^{*}_{n} \in \text{Interior}(Q_{\epsilon})$ for all small $\epsilon$ with probability tending to $1$ as $n\rightarrow \infty$. Then,
	\begin{align*}
	\text{Pr}(|\widehat{\rho}^{*}_{n}-\rho^{g}|<\epsilon)=\text{Pr}(\widehat{\rho}^{*}_{n} \in \text{Interior}(Q_{\epsilon}))\rightarrow 1
	\end{align*} 
	as $n\rightarrow \infty$ for all sufficiently small $\epsilon>0$. This completes the proof of the existence of a consistent sequence of minimizer of the function $H_{n}(\widehat{\boldsymbol{\theta}}_{1n},\widehat{\boldsymbol{\theta}}_{2n},\rho)$.
\end{proof}
\subsection{Proof of Theorem \ref{sm_normality}}
\begin{proof}
	Here also we are going to use the same Taylor series expansion trick as we did in Equation (\ref{sm_eq15}).
	\begin{align*}
	\frac{\partial H_{n}(\widehat{\boldsymbol{\theta}}_{1n},\widehat{\boldsymbol{\theta}}_{2n},\rho)}{\partial \rho}\Big{|}_{\rho=\widehat{\rho}_{n}}=\frac{\partial H_{n}(\widehat{\boldsymbol{\theta}}_{1n},\widehat{\boldsymbol{\theta}}_{2n},\rho)}{\partial \rho}\Big{|}_{\rho=\rho^{g}} + (\widehat{\rho}_{n}-\rho^{g})\frac{\partial^{2} H_{n}(\widehat{\boldsymbol{\theta}}_{1n},\widehat{\boldsymbol{\theta}}_{2n},\rho)}{\partial^{2} \rho}\Big{|}_{\rho=\rho^{g}}\\+
	\frac{(\widehat{\rho}_{n}-\rho^{g})^{2}}{2!}\frac{\partial^{3} H_{n}(\widehat{\boldsymbol{\theta}}_{1n},\widehat{\boldsymbol{\theta}}_{2n},\rho)}{\partial^{3} \rho}\Big{|}_{\rho=\rho^{**}}
	\end{align*}
	for some $\rho^{**}$ lies between $\widehat{\rho}_{n}$ and $\rho^{g}$. Hence, from Equation (\ref{sm_eq14}), we can rewrite the above as,
	\begin{align}
	\label{sm_eq22}
	-\sqrt{n}\frac{\partial H_{n}(\widehat{\boldsymbol{\theta}}_{1n},\widehat{\boldsymbol{\theta}}_{2n},\rho)}{\partial \rho}\Big{|}_{\rho=\rho^{g}}=\sqrt{n} (\widehat{\rho}_{n}-\rho^{g})\left[\frac{\partial^{2} H_{n}(\widehat{\boldsymbol{\theta}}_{1n},\widehat{\boldsymbol{\theta}}_{2n},\rho)}{\partial^{2} \rho}\Big{|}_{\rho=\rho^{g}}+
	\frac{(\widehat{\rho}_{n}-\rho^{g})}{2!}\frac{\partial^{3} H_{n}(\widehat{\boldsymbol{\theta}}_{1n},\widehat{\boldsymbol{\theta}}_{2n},\rho)}{\partial^{3} \rho}\Big{|}_{\rho=\rho^{**}}\right].
	\end{align}
	Now, let us consider the term in the left hand side of Equation (\ref{sm_eq22}). We can expand it as,
	\begin{align*}
	&\sqrt{n}\frac{\partial H_{n}(\widehat{\boldsymbol{\theta}}_{1n},\widehat{\boldsymbol{\theta}}_{2n},\rho)}{\partial \rho}\Big{|}_{\rho=\rho^{g}}=	\sqrt{n}\frac{\partial H_{n}(\boldsymbol{\theta}_{1}^{\boldsymbol{g}},\boldsymbol{\theta}_{2}^{\boldsymbol{g}},\rho)}{\partial \rho}\Big{|}_{\rho=\rho^{g}}+\sum_{j,k}\sqrt{n}(\widehat{\theta}_{njk}-\theta^{g}_{jk})\frac{\partial^{2} H_{n}(\boldsymbol{\theta}_{1},\boldsymbol{\theta}_{2},\rho)}{\partial \rho\;\partial \theta_{jk}}\Big{|}_{(\boldsymbol{\theta}_{1},\boldsymbol{\theta}_{2})=(\boldsymbol{\theta}_{1}^{\boldsymbol{g}},\boldsymbol{\theta}_{2}^{\boldsymbol{g}}),\rho=\rho^{g}}\\
	&+\sum_{j,k}\sum_{j^{'},k^{'}}\sqrt{n}(\widehat{\theta}_{njk}-\theta^{g}_{jk})(\widehat{\theta}_{nj^{'}k^{'}}-\theta^{g}_{j^{'}k^{'}})\frac{\partial^{3} H_{n}(\boldsymbol{\theta}_{1},\boldsymbol{\theta}_{2},\rho)}{\partial \rho\;\partial \theta_{jk}\;\partial \theta_{j^{'}k^{'}}}\Big{|}_{(\boldsymbol{\theta}_{1},\boldsymbol{\theta}_{2})=(\boldsymbol{\theta}_{1}^{\boldsymbol{a*}},\boldsymbol{\theta}_{2}^{\boldsymbol{a*}}),\rho=\rho^{g}}\\
	&=	\sqrt{n}\frac{\partial H_{n}(\boldsymbol{\theta}_{1}^{\boldsymbol{g}},\boldsymbol{\theta}_{2}^{\boldsymbol{g}},\rho)}{\partial \rho}\Big{|}_{\rho=\rho^{g}} + \sum_{j,k}\sqrt{n}(\widehat{\theta}_{njk}-\theta^{g}_{jk})\Bigg[\frac{\partial^{2} H_{n}(\boldsymbol{\theta}_{1},\boldsymbol{\theta}_{2},\rho)}{\partial \rho\;\partial \theta_{jk}}\Big{|}_{(\boldsymbol{\theta}_{1},\boldsymbol{\theta}_{2})=(\boldsymbol{\theta}_{1}^{\boldsymbol{g}},\boldsymbol{\theta}_{2}^{\boldsymbol{g}}),\rho=\rho^{g}}\\
	&+\sum_{j^{'},k^{'}}(\widehat{\theta}_{nj^{'}k^{'}}-\theta^{g}_{j^{'}k^{'}})\frac{\partial^{3} H_{n}(\boldsymbol{\theta}_{1},\boldsymbol{\theta}_{2},\rho)}{\partial \rho\;\partial \theta_{jk}\;\partial \theta_{j^{'}k^{'}}}\Big{|}_{(\boldsymbol{\theta}_{1},\boldsymbol{\theta}_{2})=(\boldsymbol{\theta}_{1}^{\boldsymbol{a*}},\boldsymbol{\theta}_{2}^{\boldsymbol{a*}}),\rho=\rho^{g}}\Bigg].
	\end{align*}
	Hence, Equation (\ref{sm_eq22}) can be rewritten as,
	\begin{equation}
	\centering
	\label{sm_eq23}
	\begin{array}{l}
	-\sqrt{n}\frac{\partial H_{n}(\boldsymbol{\theta}_{1}^{\boldsymbol{g}},\boldsymbol{\theta}_{2}^{\boldsymbol{g}},\rho)}{\partial \rho}\Big{|}_{\rho=\rho^{g}}=\\
	\sum_{j,k}\sqrt{n}(\widehat{\theta}_{njk}-\theta^{g}_{jk})\Bigg[\frac{\partial^{2} H_{n}(\boldsymbol{\theta}_{1},\boldsymbol{\theta}_{2},\rho)}{\partial \rho\;\partial \theta_{jk}}\Big{|}_{(\boldsymbol{\theta}_{1},\boldsymbol{\theta}_{2})=(\boldsymbol{\theta}_{1}^{\boldsymbol{g}},\boldsymbol{\theta}_{2}^{\boldsymbol{g}}),\rho=\rho^{g}}
	+\sum_{j^{'},k^{'}}(\widehat{\theta}_{nj^{'}k^{'}}-\theta^{g}_{j^{'}k^{'}})\frac{\partial^{3} H_{n}(\boldsymbol{\theta}_{1},\boldsymbol{\theta}_{2},\rho)}{\partial \rho\;\partial \theta_{jk}\;\partial \theta_{j^{'}k^{'}}}\Big{|}_{(\boldsymbol{\theta}_{1},\boldsymbol{\theta}_{2})=(\boldsymbol{\theta}_{1}^{\boldsymbol{a*}},\boldsymbol{\theta}_{2}^{\boldsymbol{a*}}),\rho=\rho^{g}}\Bigg]+\\
	\sqrt{n} (\widehat{\rho}_{n}-\rho^{g})\left[\frac{\partial^{2} H_{n}(\widehat{\boldsymbol{\theta}}_{1n},\widehat{\boldsymbol{\theta}}_{2n},\rho)}{\partial^{2} \rho}\Big{|}_{\rho=\rho^{g}}+\frac{(\widehat{\rho}_{n}-\rho^{g})}{2!}\frac{\partial^{3} H_{n}(\widehat{\boldsymbol{\theta}}_{1n},\widehat{\boldsymbol{\theta}}_{2n},\rho)}{\partial^{3} \rho}\Big{|}_{\rho=\rho^{**}}\right].
	\end{array}
	\end{equation}
	Next, let us consider the estimators $\widehat{\boldsymbol{\theta}}_{1n}$ and $\widehat{\boldsymbol{\theta}}_{2n}$ which were derived separately before finding $\widehat{\rho}_{n}$. To do that, let us first introduce the following notations. Let $H^{j}_{1n}(\boldsymbol{\theta}_{1})=\frac{\partial\;H_{1n}(\boldsymbol{\theta}_{1})}{\theta_{1j}},
	H^{j}_{2n}(\boldsymbol{\theta}_{2})=\frac{\partial\;H_{2n}(\boldsymbol{\theta}_{2})}{\theta_{2j}}\;\text{for}\;j\in\{1,2\}.$ 
	Similarly, $H^{jk}_{1n}(\boldsymbol{\theta}_{1})$, $H^{jk}_{2n}(\boldsymbol{\theta}_{2})$ and $H^{jkl}_{1n}(\boldsymbol{\theta}_{1})$, $H^{jkl}_{2n}(\boldsymbol{\theta}_{2})$ denote the second and third order partial derivatives of $H_{1n}(\boldsymbol{\theta}_{1})$ and $H_{2n}(\boldsymbol{\theta}_{2})$, respectively. Let us consider the following Taylor series expansion,
	\begin{align*}
	H^{j}_{1n}(\widehat{\boldsymbol{\theta}}_{1n})=H^{j}_{1n}(\boldsymbol{\theta}^{\boldsymbol{g}}_{1})+\sum_{k}(\widehat{\theta}_{1kn}-\theta^{g}_{1k})H^{jk}_{1n}(\boldsymbol{\theta}^{\boldsymbol{g}}_{1})+\frac{1}{2!}\sum_{k,l}(\widehat{\theta}_{1kn}-\theta^{g}_{1k})(\widehat{\theta}_{1ln}-\theta^{g}_{1l})H^{jkl}_{1n}(\boldsymbol{\theta}^{\boldsymbol{*}}_{1})
	\end{align*}
	for some $\boldsymbol{\theta}^{\boldsymbol{*}}_{1}$ lying between $\widehat{\boldsymbol{\theta}}_{1n}$ and $\boldsymbol{\theta}^{\boldsymbol{g}}_{1}$. But from Equation (\ref{sm_eq12}), we have $H^{j}_{1n}(\widehat{\boldsymbol{\theta}}_{1n})=0$. Thus the aforesaid equation can be rewritten as,
	\begin{equation}
	\label{sm_eq24}
	-\sqrt{n}H^{j}_{1n}(\boldsymbol{\theta}^{\boldsymbol{g}}_{1})=\sum_{k}\sqrt{n}(\widehat{\theta}_{1kn}-\theta^{g}_{1k})\left[H^{jk}_{1n}(\boldsymbol{\theta}^{\boldsymbol{g}}_{1})+\frac{1}{2!}\sum_{l}(\widehat{\theta}_{1ln}-\theta^{g}_{1l})H^{jkl}_{1n}(\boldsymbol{\theta}^{\boldsymbol{*}}_{1})\right].
	\end{equation}
	Similarly, we have  
	\begin{equation}
	\label{sm_eq25}
	-\sqrt{n}H^{j}_{2n}(\boldsymbol{\theta}^{\boldsymbol{g}}_{2})=\sum_{k}\sqrt{n}(\widehat{\theta}_{2kn}-\theta^{g}_{2k})\left[H^{jk}_{2n}(\boldsymbol{\theta}^{\boldsymbol{g}}_{2})+\frac{1}{2!}\sum_{l}(\widehat{\theta}_{2ln}-\theta^{g}_{2l})H^{jkl}_{2n}(\boldsymbol{\theta}^{\boldsymbol{*}}_{2})\right]
	\end{equation}
	for some $\boldsymbol{\theta}^{\boldsymbol{*}}_{2}$ lying between $\widehat{\boldsymbol{\theta}}_{2n}$ and $\boldsymbol{\theta}^{\boldsymbol{g}}_{2}$. Now let us consider the following system of linear equations which is derived by assembling Equations (\ref{sm_eq23}), (\ref{sm_eq24}) and (\ref{sm_eq25}).
	\begin{equation}
	\centering
	\label{sm_eq26}
	\begin{array}{l}
	\sum_{k}\sqrt{n}(\widehat{\theta}_{1kn}-\theta^{g}_{1k})b^{(1)}_{jkn}=-\sqrt{n}H^{j}_{1n}(\boldsymbol{\theta}^{g}_{\boldsymbol{1}}),\;j\in\{1,2\},\\
	\\
	\sum_{k}\sqrt{n}(\widehat{\theta}_{2kn}-\theta^{g}_{2k})b^{(2)}_{jkn}=-\sqrt{n}H^{j}_{2n}(\boldsymbol{\theta}^{\boldsymbol{g}}_{2}),\;j\in\{1,2\},\\
	\\
	\sum_{j,k}\sqrt{n}(\widehat{\theta}_{njk}-\theta^{g}_{jk})e_{jkn}+\sqrt{n} (\widehat{\rho}_{n}-\rho^{g})a_{n}=	-\sqrt{n}\frac{\partial H_{n}(\boldsymbol{\theta}_{1}^{\boldsymbol{g}},\boldsymbol{\theta}_{2}^{\boldsymbol{g}}
		,\rho)}{\partial \rho}\Big{|}_{\rho=\rho^{g}}, 
	\end{array}
	\end{equation}
	where $b^{(1)}_{jkn}=\left[H^{jk}_{1n}(\boldsymbol{\theta}^{\boldsymbol{g}}_{1})+\frac{1}{2!}\sum_{l}(\widehat{\theta}_{1ln}-\theta^{g}_{1l})H^{jkl}_{1n}(\boldsymbol{\theta}^{\boldsymbol{*}}_{1})\right]$, $b^{(2)}_{jkn}=\left[H^{jk}_{2n}(\boldsymbol{\theta}^{\boldsymbol{g}}_{2})+\frac{1}{2!}\sum_{l}(\widehat{\theta}_{2ln}-\theta^{g}_{2l})H^{jkl}_{2n}(\boldsymbol{\theta}^{\boldsymbol{*}}_{2})\right]$, $e_{jkn}=\Bigg[\frac{\partial^{2} H_{n}(\boldsymbol{\theta}_{1},\boldsymbol{\theta}_{2},\rho)}{\partial \rho\;\partial \theta_{jk}}\Big{|}_{(\boldsymbol{\theta}_{1},\boldsymbol{\theta}_{2})=(\boldsymbol{\theta}_{1}^{\boldsymbol{g}},\boldsymbol{\theta}_{2}^{\boldsymbol{g}})
		,\rho=\rho^{g}}
	+\sum_{j^{'},k^{'}}(\widehat{\theta}_{nj^{'}k^{'}}-\theta^{g}_{j^{'}k^{'}})\frac{\partial^{3} H_{n}(\boldsymbol{\theta}_{1},\boldsymbol{\theta}_{2},\rho)}{\partial \rho\;\partial \theta_{jk}\;\partial \theta_{j^{'}k^{'}}}\Big{|}_{(\boldsymbol{\theta}_{1},\boldsymbol{\theta}_{2})=(\boldsymbol{\theta}_{1}^{\boldsymbol{a*}},\boldsymbol{\theta}_{2}^{\boldsymbol{a*}}),\rho=\rho^{g}}\Bigg]$ for $j\in\{1,2\}$, $k\in\{1,2\}$ and $a_{n}=\Bigg[\frac{\partial^{2} H_{n}(\widehat{\boldsymbol{\theta}}_{1n},\widehat{\boldsymbol{\theta}}_{2n},\rho)}{\partial^{2} \rho}\Big{|}_{\rho=\rho^{g}}+\frac{(\widehat{\rho}_{n}-\rho^{g})}{2!}\frac{\partial^{3} H_{n}(\widehat{\boldsymbol{\theta}}_{1n},\widehat{\boldsymbol{\theta}}_{2n},\rho)}{\partial^{3} \rho}\Big{|}_{\rho=\rho^{**}}\Bigg]$. Now, following the consistency of the estimators $\widehat{\boldsymbol{\theta}}_{1n}$, $\widehat{\boldsymbol{\theta}}_{2n}$ and $\widehat{\rho}_{n}$ and similar kind of arguments and assumptions we have made to prove Theorem \ref{sm_consistency}, we can show that, $b^{(l)}_{jkn}\overset{p}{\rightarrow}b^{(l)}_{jk}$  for $j\in\{1,2\}$, $k\in\{1,2\}$ and $l\in\{1,2\}$ (recall the elements of the matrix $\boldsymbol{B}$, $e_{jkn}\overset{p}{\rightarrow}e_{jk}$, $j\in\{1,2\}$, $k\in\{1,2\}$ and $a_{n}\overset{p}{\rightarrow}a$ as $n\rightarrow \infty$.
	
	By the central limit theorem (CLT), we have
	\begin{align*}
	\sqrt{n}\left(-\sqrt{n}H^{1}_{1n}(\boldsymbol{\theta}^{\boldsymbol{g}}_{1}),-\sqrt{n}H^{2}_{1n}(\boldsymbol{\theta}^{\boldsymbol{g}}_{1}),-\sqrt{n}H^{1}_{2n}(\boldsymbol{\theta}^{\boldsymbol{g}}_{2}),-\sqrt{n}H^{2}_{2n}(\boldsymbol{\theta}^{\boldsymbol{g}}_{2}),-\sqrt{n}\frac{\partial H_{n}(\boldsymbol{\theta}_{1}^{\boldsymbol{g}},\boldsymbol{\theta}_{2}^{\boldsymbol{g}}
		,\rho)}{\partial \rho}\Big{|}_{\rho=\rho^{g}}\right)^\top\overset{d}{\rightarrow}\;N(\boldsymbol{0},\boldsymbol{\Gamma}_{0}).
	\end{align*}
	Now, using the aforesaid observations and Lemma $4.1$ of Lehmann $(1983)$ \cite{lehmann}, we have 
	\begin{align*}
	\sqrt{n}(\widehat{\boldsymbol{\theta}}_{n}-\boldsymbol{\theta^{g}})\overset{d}{\rightarrow}\; N(\boldsymbol{0},\boldsymbol{B}^{-1}\boldsymbol{\Gamma}_{0}{\boldsymbol{B}^{-1}}^\top),
	\end{align*}
	where $\widehat{\boldsymbol{\theta}}_{n}=(\widehat{\theta}_{11n},\widehat{\theta}_{12n},\widehat{\theta}_{21n},\widehat{\theta}_{22n},\widehat{\rho}_{n})$ and $\boldsymbol{\theta^{g}}=(\theta^{g}_{11},\theta^{g}_{12},\theta^{g}_{21},\theta^{g}_{22},\rho^{g})$.
\end{proof}
\subsection{Proof of Theorem \ref{sm_pdim}}
\begin{proof}
	According to our algorithm, we first derive $\widehat{\mu}_{jn}$ and $\widehat{\sigma}^{2}_{jn}$ marginally from each of the $p$ components. From Equation (\ref{sm_eq13}), we have, $\hat{\mu}_{jn}\overset{p}{\rightarrow} \mu^{g}_{j}$ and  $\widehat{\sigma}^{2}_{jn}\overset{p}{\rightarrow} {\sigma^{g}}^{2}_{jn}$ as $n\rightarrow\infty$ for $1\leq j \leq p$. In the next step, our algorithm finds the estimators $\widehat{\rho}_{jkn}$ for each pair of components separately. By Theorem \ref{sm_consistency}, we have, $\widehat{\rho}_{jkn}\overset{p}{\rightarrow}\rho^{g}_{jk}$. For a fixed $\epsilon>0$,
	\begin{align*}
	P[||\widehat{\boldsymbol{\theta}}_{n}-\boldsymbol{\theta^{g}}||_{2}>\epsilon]= 	P[||\widehat{\boldsymbol{\theta}}_{n}-\boldsymbol{\theta^{g}}||^{2}_{2}>\epsilon^{2}]&=P\left[\sum_{l=1}^{P}(\widehat{\boldsymbol{\theta}}_{nl}-\boldsymbol{\theta^{g}}_{l})^{2}>\epsilon^{2}\right]\\
	&\leq \sum_{l=1}^{P}P\left[(\widehat{\boldsymbol{\theta}}_{nl}-\boldsymbol{\theta^{g}}_{l})^{2}>\frac{\epsilon^{2}}{P}\right]\;\text{by Boole's inequality}\\
	&\rightarrow 0\;\text{as}\;n\rightarrow \infty
	\end{align*}
	and this is true for all $\epsilon>0$. Thus, $\widehat{\boldsymbol{\theta}}_{n}\overset{p}{\rightarrow}\boldsymbol{\theta^{g}}$. Now the asymptotic normality of $\widehat{\boldsymbol{\theta}}_{n}$ can be similarly proved but the form of the asymptotic covariance matrix of $\sqrt{n}\widehat{\boldsymbol{\theta}}_{n}$ is quite large which is provided just after stating Theorem \ref{sm_pdim}.
	The proof is exactly the same as that of Theorem \ref{sm_normality}.
\end{proof}

\subsection{Proof of Theorem \ref{sm_pdt}} 
\begin{proof}
	Let, $S_{n}=\{\boldsymbol{x}=(x_{1},\ldots,x_{p})\in\mathbb{R}^{p}:\boldsymbol{x}^{\top}\widehat{\boldsymbol{\Sigma}}_{n}\boldsymbol{x}>0\;\text{and}\;\boldsymbol{x}^{\top}\boldsymbol{x}=1\}$. It is enough to show that $\text{Pr}(S_{n})\rightarrow1$ as $n\rightarrow \infty$.
	
	Since  $\boldsymbol{\Sigma}^{\boldsymbol{g}}$ is positive definite, $\boldsymbol{x}^{\top}\boldsymbol{\Sigma}^{\boldsymbol{g}} \boldsymbol{x}>0\;\forall\;\boldsymbol{x} \neq 0$.
	By considering the spectral decomposition of the symmetric matrix $\boldsymbol{\Sigma}^{\boldsymbol{g}}$, we have (from Assumption \ref{sm_a6})
	\begin{align}
	\label{sm_eq27}
	\boldsymbol{x}^{\top}\boldsymbol{\Sigma}^{\boldsymbol{g}} \boldsymbol{x}\geq \lambda^{g}_{(1)}\geq c.
	\end{align}
	From Theorem \ref{sm_consistency}, we have, $\widehat{\sigma}_{ijn} \overset{p}{\rightarrow}\sigma^{g}_{ij}\;\text{as}\;n\rightarrow \infty,\;1\leq i,j\leq p.$
	Let us fix an $\epsilon \in (0,\frac{c}{p})$. So, 
	\begin{align*}
	|\widehat{\sigma}_{ijn} -\sigma^{g}_{ij}|\leq\epsilon &\implies\;|x_{i}x_{j}\widehat{\sigma}_{ijn} -x_{i}x_{j}\sigma^{g}_{ij}|=|x_{i}||x_{j}||\widehat{\sigma}_{ijn} -\sigma^{g}_{ij}|\leq|x_{i}||x_{j}|\epsilon\\
	&\implies\; x_{i}x_{j}\widehat{\sigma}_{ijn}\geq x_{i}x_{j}\sigma^{g}_{ij}-|x_{i}||x_{j}|\epsilon
	\end{align*}
	with probability tending to $1$ for all $1\leq i,j \leq p$ (here $|x_{i}|$ is the $i$-th component of the unit vector $\boldsymbol{x}$). Now,
	\begin{align*}
	\boldsymbol{x}^{\top}\widehat{\boldsymbol{\Sigma}}_{n}\boldsymbol{x}&=\sum_{i,j}x_{i}x_{j}\widehat{\sigma}_{ijn}\geq\sum_{i,j}x_{i}x_{j}\sigma^{g}_{ij}-\epsilon\sum_{i,j}|x_{i}||x_{j}|\\
	&=\boldsymbol{x}^{\top}\boldsymbol{\Sigma}^{\boldsymbol{g}} \boldsymbol{x}-\epsilon\left(\sum_{i=1}^{p}|x_i|\right)^{2}\geq c-p\epsilon\;\text{,by Equation (\ref{sm_eq27}) and Cauchy-Schwarz inequality}\\
	&\;\;\;\;\;\;\;\;\;\;\;\;\;\;\;\;\;\;\;\;\;\;\;\;\;\;\;\;\;\;\;\;\;\;\;\;\;\;\;\;>0
	\end{align*}
	with probability tending to $1$ for all unit vector $\boldsymbol{x}(\neq 0)\in \mathbb{R}^{p}$. Thus $\text{Pr}(S_n)\rightarrow 1$ as $n\rightarrow \infty$.
\end{proof}
\section{Asymptotic Variances of the Correlation estimators}
\label{sm_appen1}
\subsection{CMDPDE}
\label{sm_appen1.1}
Let us first introduce a set of algebraic expressions, namely, $d_{22}=\frac{\beta^{2}+2}{4(2\pi)^{\beta/2}(1+\beta)^{\frac{3}{2}}\sigma_{1}^{\beta+4}}$,
\\ $d_{44}=\frac{\beta^{2}+2}{4(2\pi)^{\beta/2}(1+\beta)^{\frac{3}{2}}\sigma_{2}^{\beta+4}}$, $E_{12}=\rho(1+\beta^{2})+\rho\beta\Big(1-\beta-\frac{2\rho^4-4\rho^2+2}{(1+\beta)(1-\rho^2)^2}\Big)$, $e_{1}=-\frac{E_{12}}{2\sigma_{1}^{2}(\sigma_{1}\sigma_{2})^{\beta}(2\pi)^{\beta}(1-\rho^2)^{1+\frac{\beta}{2}}(1+\beta)}$, $e_{2}=-\frac{E_{12}}{2\sigma_{2}^{2}(\sigma_{1}\sigma_{2})^{\beta}(2\pi)^{\beta}(1-\rho^2)^{1+\frac{\beta}{2}}(1+\beta)}$, $F=1+\rho^2(1+\beta)-\frac{\beta(2\rho^6-3\rho^4+1)}{(1+\beta)(1-\rho^2)^2}$ and  $a=\frac{F}{(\sigma_{1}\sigma_{2})^{\beta}(2\pi)^{\beta}(1-\rho^2)^{2+\frac{\beta}{2}}(1+\beta)}$. Let us consider the matrix 
\begin{align*}
\boldsymbol{B}=\begin{bmatrix}
d_{22} & 0 & 0\\
0 & d_{44} & 0 \\
e_{1} & e_{2} &  a \\
\end{bmatrix}.
\end{align*}
Let us also define, $\gamma_{22}=\frac{1}{\sqrt{1+2\beta}}\Big[\frac{1}{2}+\frac{3}{2(1+2\beta)^2}-\frac{1}{1+2\beta}\Big]-\frac{\beta^2}{2(1+\beta)^3}$, $\Gamma_{22}=\frac{(1+\beta)^2}{2(2\pi)^\beta\sigma_{1}^{2\beta+4}}\gamma_{22}$, $\Gamma_{44}=\frac{(1+\beta)^2}{2(2\pi)^\beta\sigma_{2}^{2\beta+4}}\gamma_{22}$, $\gamma_{24}=\frac{1}{\sqrt{(\beta+1)^2-(\beta\rho)^2}}\Big[\Big(1-\frac{1+\beta(1-\rho^2)}{(\beta+1)^2-(\beta\rho)^2}\Big)^2+\frac{2\rho^2}{((\beta+1)^2-(\beta\rho)^2)^2}\Big]-\frac{\beta^2}{(1+\beta)^3}$, $\Gamma_{24}=\frac{(1+\beta)^2 \gamma_{24}}{4(2\pi)^{\beta}(\sigma_{1}\sigma_{2})^{\beta+2}}$, $\Gamma_{25}=\frac{\beta^2 \rho}{2(2\pi)^{\frac{3\beta}{2}}\sigma_{1}^{2\beta+2}\sigma_{2}^{\beta}(1-\rho^2)^{1+\frac{\beta}{2}}(1+\beta)^{\frac{3}{2}}}\Big[1-\frac{2(1+\beta)^2}{(1+2\beta)^{\frac{3}{2}}}\Big]$, $\Gamma_{45}=\frac{\beta^2 \rho}{2(2\pi)^{\frac{3\beta}{2}}\sigma_{2}^{2\beta+2}\sigma_{1}^{\beta}(1-\rho^2)^{1+\frac{\beta}{2}}(1+\beta)^{\frac{3}{2}}}\Big[1-\frac{2(1+\beta)^2}{(1+2\beta)^{\frac{3}{2}}}\Big]$, $\gamma_{55}=\frac{\rho^2}{1+2\beta}+\frac{1-3\rho^4+2\rho^6}{(1-\rho^2)^{2}(1+2\beta)^{3}}-\frac{2\rho^2}{(1+2\beta)^2}-\frac{\beta^2 \rho^2}{(1+\beta)^4}$ and $\Gamma_{55}=\frac{(1+\beta)^2 \gamma_{55}}{(2\pi)^{2\beta}(\sigma_{1}\sigma_{2})^{2\beta}(1-\rho^2)^{\beta+2}}$. Let us consider the matrix
\begin{align*}
\boldsymbol{\Gamma}_{0}=\begin{bmatrix}
\Gamma_{22} & \Gamma_{24} & \Gamma_{25}\\
\Gamma_{24} &  \Gamma_{44} &  \Gamma_{45} \\
\Gamma_{25} &  \Gamma_{45} &   \Gamma_{55} \\
\end{bmatrix}.
\end{align*}
For the componentwise minimum DPD estimation, the asymptotic variance of $\sqrt{n}\widehat{\rho}_{n}$ is the third diagonal element of the matrix $\boldsymbol{B}^{-1}\boldsymbol{\Gamma}_{0} {\boldsymbol{B}^{-1}}^{\top}$.
\subsection{MDPDE}
\label{sm_appen1.2}
Similarly, for the asymptotic variance of the ordinary MDPDE correlation estimates, let us define, $J_{22}=\frac{1}{4\sigma_{1}^4(2\pi)^{\beta}(\sigma_{1}\sigma_{2})^{\beta}(1-\rho^2)^{\frac{\beta}{2}}(1+\beta)^2}\Big[\beta^2 + \frac{2-\rho^2}{1-\rho^2}\Big]$, $J_{44}=\frac{1}{4\sigma_{2}^4(2\pi)^{\beta}(\sigma_{1}\sigma_{2})^{\beta}(1-\rho^2)^{\frac{\beta}{2}}(1+\beta)^2}\Big[\beta^2 + \frac{2-\rho^2}{1-\rho^2}\Big]$,\\ $J_{55}=\frac{1}{(\sigma_{1}\sigma_{2})^{\beta}(2\pi)^{\beta}(1-\rho^2)^{2+\frac{\beta}{2}}(1+\beta)}\Big[\rho^2(1+\beta)+\frac{2\rho^6-3\rho^4+1}{(1+\beta)(1-\rho^2)^2}-2\rho^2\Big]$, $J_{24}=\frac{1}{4(\sigma_{1}\sigma_{2})^{\beta+2}(2\pi)^{\beta}(1-\rho^2)^{\frac{\beta}{2}}}\Big[1-\frac{2}{1+\beta}+\frac{1-2\rho^2}{(1-\rho^2)(1+\beta)^2}\Big]$, $J_{25}=\frac{\rho}{2\sigma_{1}^{2}(\sigma_{1}\sigma_{2})^{\beta}(2\pi)^{\beta}(1-\rho^2)^{1+\frac{\beta}{2}}(1+\beta)}\Big[1-\beta-\frac{2}{1+\beta}\Big]$ and $J_{45}=\frac{\rho}{2\sigma_{2}^{2}(\sigma_{1}\sigma_{2})^{\beta}(2\pi)^{\beta}(1-\rho^2)^{1+\frac{\beta}{2}}(1+\beta)}\Big[1-\beta-\frac{2}{1+\beta}\Big]$. Let us consider the matrix
\begin{align*}
\boldsymbol{J}=\begin{bmatrix}
J_{22} & J_{24} & J_{25}\\
J_{24} &  J_{44} &  J_{45} \\
J_{25} &  J_{45} &   J_{55} \\
\end{bmatrix}.
\end{align*}
Let us also define, $K_{22}=\frac{1}{4\sigma_{1}^4(2\pi)^{2\beta}(\sigma_{1}\sigma_{2})^{2\beta}(1-\rho^2)^{\beta}}\Big[\frac{(1+\beta)^2}{(1+2\beta)^3}\Big(4\beta^2 + \frac{2-\rho^2}{1-\rho^2}\Big)-\Big(\frac{\beta}{1+\beta}\Big)^{2}\Big]$,
\\
$K_{44}=\frac{1}{4\sigma_{2}^4(2\pi)^{2\beta}(\sigma_{1}\sigma_{2})^{2\beta}(1-\rho^2)^{\beta}}\Big[\frac{(1+\beta)^2}{(1+2\beta)^3}\Big(4\beta^2 + \frac{2-\rho^2}{1-\rho^2}\Big)-\Big(\frac{\beta}{1+\beta}\Big)^{2}\Big]$, $K_{55}=\frac{\gamma_{55}}{(2\pi)^{2\beta}(\sigma_{1}\sigma_{2})^{2\beta}(1-\rho^2)^{2+\beta}}$, $K_{24}=\frac{1}{4\sigma_{1}^2(2\pi)^{2\beta}(\sigma_{1}\sigma_{2})^{2\beta}(1-\rho^2)^{\beta}}\Big[\frac{(1+\beta)^2}{(1+2\beta)^3}\Big(4\beta^2 - \frac{\rho^2}{1-\rho^2}\Big)-\Big(\frac{\beta}{1+\beta}\Big)^{2}\Big]$, $K_{25}=\frac{\rho}{2\sigma_{1}^2(2\pi)^{2\beta}(\sigma_{1}\sigma_{2})^{2\beta}(1-\rho^2)^{1+\beta}}\Big[\frac{(1+\beta)^2}{(1+2\beta)^2}\Big(1-2\beta - \frac{2}{1+2\beta}\Big)+\Big(\frac{\beta}{1+\beta}\Big)^{2}\Big]$ and $K_{45}=\frac{\rho}{2\sigma_{2}^2(2\pi)^{2\beta}(\sigma_{1}\sigma_{2})^{2\beta}(1-\rho^2)^{1+\beta}}\Big[\frac{(1+\beta)^2}{(1+2\beta)^2}\Big(1-2\beta - \frac{2}{1+2\beta}\Big)+\Big(\frac{\beta}{1+\beta}\Big)^{2}\Big]$. Let us consider the matrix
\begin{align*}
\boldsymbol{K}=\begin{bmatrix}
K_{22} & K_{24} & K_{25}\\
K_{24} &  K_{44} &  K_{45} \\
K_{25} &  K_{45} &   K_{55} \\
\end{bmatrix}.
\end{align*}
For the ordinary MDPDE method, the asymptotic variances of $\sqrt{n}\widehat{\sigma}^{2}_{1n}$, $\sqrt{n}\widehat{\sigma}^{2}_{2n}$ and $\sqrt{n}\widehat{\rho}_{n}$ are the first, second and third diagonal elements of the matrix $\boldsymbol{J}^{-1}\boldsymbol{KJ}^{-1}$, respectively.

\section{Algebraic Details of the Influence Functions}
\label{sm_appen2}
The expressions for $A(x_1,x_2,y_1,y_2)$ and $B(x_1,x_2)$ are given by,
\begin{align*}
&A(x_1,x_2,y_1,y_2)=-\frac{1}{2}\left[\frac{IF(\theta_{12},y_1,g_1)}{\theta^{g}_{12}}+\frac{IF(\theta_{22},y_2,g_2)}{\theta^{g}_{22}}\right]-\frac{1}{2(1-(\rho^{g})^2)}\frac{\partial z_{\epsilon}}{\partial \epsilon}\Big|_{\epsilon=0}\;\text{and}\\
&B(x_1,x_2)=\frac{\rho^{g}}{(1-(\rho^{g})^2)}-\frac{z^{g}\rho^{g}}{(1-(\rho^{g})^2)^2}+\frac{1}{(1-(\rho^{g})^2)}\frac{(x_1-\theta^{g}_{11})(x_2-\theta^{g}_{21})}{\sqrt{\theta^{g}_{12}\theta^{g}_{22}}},
\end{align*}
where, $z_{\epsilon}=\frac{(x_1-\theta_{11\epsilon})^2}{\theta_{12\epsilon}}+\frac{(x_2-\theta_{21\epsilon})^2}{\theta_{22\epsilon}}-\frac{2\rho_{\epsilon}(x_1-\theta_{11\epsilon})(x_2-\theta_{21\epsilon})}{\sqrt{\theta_{12\epsilon}\theta_{12\epsilon}}}$ and $z^{g}=z_{\epsilon}|_{\epsilon=0}$.
After tedious algebraic manipulations, it can be found that,
\begin{align*}
&  \int f^{1+\beta}_{j}(x,\boldsymbol{\theta}_{j})(-\theta_{j2}+(x-\theta_{j1})^{2})\;dx=\frac{-\beta}{(2\pi)^{\frac{\beta}{2}} (\theta_{j2})^{\frac{\beta}{2}-1}(1+\beta)^\frac{3}{2}},            \\
& \int f^{1+\beta}_{j}(x,\boldsymbol{\theta}_{j})\left(\frac{(x-\theta_{j1})^{2}}{\theta_{j2}}-1\right)^{2}\;dx= \frac{\beta^2 +2}{(2\pi)^{\frac{\beta}{2}} (\theta_{j2})^{\frac{\beta}{2}}(1+\beta)^\frac{5}{2}},    \\
&\int (1+A(x_1,x_2,y_1,y_2)) f^{\beta}(x_1,x_2,\boldsymbol{\theta},\rho)U_{\rho}(x_1,x_2,\boldsymbol{\theta},\rho)\;dx_1dx_2\\
&=\int f^{\beta}(x_1,x_2,\boldsymbol{\theta}_{1},\boldsymbol{\theta}_{2},\rho)U_{\rho}(x_1,x_2,\boldsymbol{\theta}_{1},\boldsymbol{\theta}_{2},\rho)\;dx_1dx_2\\
&+\int A(x_1,x_2,y_1,y_2) f^{\beta}(x_1,x_2,\boldsymbol{\theta}_{1},\boldsymbol{\theta}_{2},\rho)U_{\rho}(x_1,x_2,\boldsymbol{\theta}_{1},\boldsymbol{\theta}_{2},\rho)\;dx_1dx_2,\;\text{where}
\end{align*}
\begin{align*}
&\int f^{\beta}(x,\boldsymbol{\theta}_{1},\boldsymbol{\theta}_{2},\rho)U_{\rho}(x_1,x_2,\boldsymbol{\theta}_{1},\boldsymbol{\theta}_{2},\rho)\;dx_1dx_2=\frac{\rho\beta}{(2\pi)^\beta (\theta_{12}\theta_{22})^{\frac{\beta}{2}}(1-\rho^2)^{1+\frac{\beta}{2}}(1+\beta)^2},\\
&\int A(x_1,x_2,y_1,y_2) f^{\beta}(x_1,x_2,\boldsymbol{\theta}_{1},\boldsymbol{\theta}_{2},\rho)U_{\rho}(x_1,x_2,\boldsymbol{\theta}_{1},\boldsymbol{\theta}_{2},\rho)\;dx_1dx_2\\
&=\frac{-\rho(1+\beta^2)}{2(2\pi)^\beta (\theta_{12}\theta_{22})^{\frac{\beta}{2}}(1-\rho^2)^{1+\frac{\beta}{2}}(1+\beta)^3}\Big[\frac{IF(\theta_{12},y_{1},g_{1})}{\theta_{12}}+\frac{IF(\theta_{22},y_{2},g_{2})}{\theta_{22}}\Big],\\
&\int B(x_1,x_2) f^{\beta}(x_1,x_2,\boldsymbol{\theta}_{1},\boldsymbol{\theta}_{2},\rho)U_{\rho}(x_1,x_2,\boldsymbol{\theta}_{1},\boldsymbol{\theta}_{2},\rho)\;dx_1dx_2\\
&=\frac{1}{(2\pi)^\beta (\theta_{12}\theta_{22})^{\frac{\beta}{2}}(1-\rho^2)^{2+\frac{\beta}{2}}(1+\beta)}\Big[\rho^2 +\frac{1-3\rho^4+2\rho^6}{(1-\rho^2)^2(1+\beta)^2} -\frac{2\rho^2}{1+\beta}\Big].
\end{align*}

\section{Simulation Outputs}
\label{sm_appen3}
The estimated $L_2$ bias and mean squared errors of location and scale estimators in case of the pure, subtle outlier contaminated and distant outlier contaminated datasets are presented in the following Tables.

\begin{table}[!h]
	\centering 
%	\small
	\begin{tabular}{c c c c c c c c c } 
		
		\hline
		 Dimension & Different & & \hspace{-5.5em} Location Vector &  & \hspace{-5.5em}Scatter Matrix\\  
		 $(p)$ &Methods &Bias & MSE & Bias & MSE  \\
		\hline
		\multirow{ 4}{*}{} 2 &MLE &    0.005  &  0.002      &          0.006   &  0.007 \\
		 &CMDPDE $(\beta=0.1)$ & 0.004 &  0.002 & 0.007 & 0.007\\
		 & CMDPDE $(\beta=0.3)$ & 0.004 & 0.002 & 0.007 & 0.008\\
		 & CMDPDE $(\beta=0.5)$ & 0.004 & 0.002 & 0.006 & 0.009 \\
         & FRC &  0.003 & 0.002 & 0.005 & 0.009  \\
		 & MCD &  0.003 & 0.002 & 0.019 &  0.013 \\
		 & MVE & 0.004 & 0.002 &  0.022  & 0.016\\
		 & OGK & 0.003 & 0.002 & 0.355 &  0.133
		\\
		 & MM & 0.006 & 0.002 & 0.006 & 0.008   \\
		 & S & 0.004 & 0.003 & 0.016 & 0.015 \\
		 & & & & &  \\
		\multirow{ 4}{*}{}  5 & MLE &  0.009    & 0.005   &            0.013 &  0.029 \\
		 & CMDPDE $(\beta=0.1)$ & 0.009 & 0.005 & 0.014 & 0.029\\
		 & CMDPDE $(\beta=0.3)$ & 0.009 & 0.005 & 0.017 & 0.034 \\
		 & CMDPDE $(\beta=0.5)$ &  0.008 & 0.006 & 0.020 & 0.041\\
         & FRC  &  0.011 & 0.005 & 0.019 & 0.039   \\
		 & MCD  &  0.005 & 0.005 & 0.022 & 0.041\\
		 &  MVE &   0.007 & 0.006 & 0.040  & 0.052\\
		 & OGK &    0.005 & 0.006 & 0.341 & 0.149\\
		 & MM &  0.005 & 0.005 & 0.012 & 0.032 \\
		 & S &  0.005 & 0.006 & 0.013 & 0.039\\
		 & & & & & & & \\
		\multirow{ 4}{*}{}10 & MLE &  0.013   & 0.010   &            0.028  &  0.110\\
		& CMDPDE $(\beta=0.1)$ & 0.013 & 0.010 & 0.028 & 0.119 \\
		& CMDPDE $(\beta=0.3)$ & 0.013 & 0.011 & 0.029 &  0.130 \\
		& CMDPDE $(\beta=0.5)$ & 0.013 & 0.012 & 0.031 & 0.156 \\
        & FRC &  0.010 & 0.011 & 0.031 & 0.126 \\
		& MCD &   0.008 & 0.010 & 0.038 & 0.131  \\
		& MVE  &  0.007 & 0.011 & 0.049 & 0.158  \\
		& OGK  &   0.009 & 0.012 & 0.323 & 0.223\\
		& MM &  0.011 & 0.010 & 0.032 & 0.118\\
		& S & 0.011 & 0.010 & 0.033 & 0.123 \\
		%& & & & & \\
		%\multirow{ 4}{*}{}20 & MLE &    0.011    &  0.020      &           0.062 &   0.425\\
		%& CMDPDE $(\beta=0.1)$ & 0.010 & 0.020 & 0.063 &  0.435\\
		%& CMDPDE $(\beta=0.3)$ & 0.010 &  0.022  & 0.067 &  0.497 \\
		%&  CMDPDE $(\beta=0.5)$ & 0.010 & 0.022 & 0.073 & 0.598  \\
		%& 	MCD  & 0.013 & 0.020 &  0.082 & 0.489  \\
		%& MVE &   0.015 & 0.022 & 0.085 &  0.520  \\
		%& GK  &  0.015 &  0.023 & 0.317 &  0.555\\
		%& MM &  0.014 & 0.020 & 0.066 & 0.441  \\
		%& S &  0.014 &  0.019 & 0.066 & 0.434 \\
		
		%&  ~~~~~~~~~ & ~~~~~   ~~~~~~~  &                      ~~~~~   ~~~~~~~  & ~~~~~   ~~~~~~~   &                      ~~~~~   ~~~~~~~ \\
		%\multirow{ 4}{*}{}30 & MLE &  0.027 &  0.030 &  0.177 &  0.932\\
		%& CMDPDE $(\beta=0.1)$ & 0.027 &  0.030 &  0.183 & 0.955 \\
		%& CMDPDE $(\beta=0.3)$ &  0.031 &  0.036 &  0.204 & 1.074 \\
		%&  CMDPDE $(\beta=0.5)$ &  0.031 &  0.035 &  0.225 &  1.328\\
		%& 	MCD  & 0.039 &  0.031 &  0.217 &  1.094\\
		%& MVE &   0.040 &  0.033 &  0.222 &  1.141 \\
		%& GK  & 0.041 &  0.034 &  0.382 &  1.146 \\
		%& MM &   0.035 &  0.032 &  0.203 &  0.991  \\
		%& S &   0.035 &  0.032 &  0.200 &  0.958\\
		
		\hline 
	\end{tabular}
	\caption{Estimated bias and mean squared errors in case of diagonal covariance structures under pure data. }
	\label{sm_table1.1}
\end{table}

\begin{table}[!h]
	\centering 
%	\small
	\begin{tabular}{c c c c c c c c c } 
		
		\hline
		 Dimension & Different & & \hspace{-5.5em} Location Vector &  & \hspace{-5.5em}Scatter Matrix\\  
		 $(p)$ &Methods &Bias & MSE & Bias & MSE  \\
		\hline
		\multirow{ 4}{*}{} 2 & MLE &  0.193 & 0.040 & 0.354 & 0.135   \\
		 &CMDPDE $(\beta=0.7)$ & 0.103 & 0.013 & 0.223 & 0.061  \\
		 & CMDPDE $(\beta=0.9)$ &  0.091 & 0.012 & 0.210 & 0.057   \\
		 & FRC &  0.139 & 0.022 & 0.196 & 0.053   \\
		 & MCD & 0.132 & 0.021 & 0.275 & 0.094 \\
		 & MVE & 0.136 & 0.022 & 0.282 & 0.099 \\
		 & OGK &  0.115 & 0.017 & 0.280 & 0.086 \\
		 & MM &  0.171 & 0.032 & 0.294 & 0.095  \\
		 & S &  0.084 & 0.012 & 0.227 & 0.066  \\
		 & & & & &  \\
		\multirow{ 4}{*}{}  5 & MLE & 0.196 & 0.044 &  0.355 & 0.162  \\
		 & CMDPDE $(\beta=0.7)$ & 0.104 & 0.018 & 0.241 & 0.119  \\
		 & CMDPDE $(\beta=0.9)$ & 0.092 & 0.016 & 0.229 & 0.125 \\
	     & FRC  &  0.142 & 0.026 & 0.208 & 0.089  \\
		 & MCD  &  0.169 & 0.035 & 0.292 & 0.131 \\
		 &  MVE & 0.174 & 0.037 & 0.324 & 0.173  \\
		 & OGK &  0.156 & 0.030 & 0.310 & 0.134   \\
		 & MM & 0.176 & 0.037 & 0.320  &  0.140   \\
		 & S &   0.150 & 0.029 & 0.289 & 0.128  \\
		 & & & & & & & \\
		\multirow{ 4}{*}{}10 & MLE &  0.204 & 0.052 & 0.355 & 0.240   \\
		& CMDPDE $(\beta=0.5)$ &  0.130 & 0.029 & 0.259 & 0.224  \\
		& CMDPDE $(\beta=0.7)$ &  0.113 & 0.026 & 0.239 & 0.248  \\
		& FRC & 0.150 & 0.033 & 0.205 & 0.170 \\
		& MCD &  0.181 & 0.045 & 0.332 & 0.245  \\
		& MVE  & 0.179 & 0.045 & 0.348 & 0.280  \\
		& OGK  &  0.169 & 0.041 & 0.358 & 0.252  \\
		& MM &  0.185 & 0.044 & 0.332 & 0.234  \\
		& S &  0.179 & 0.043 & 0.325 & 0.234    \\
		
		\hline 
	\end{tabular}
	\caption{Estimated bias and mean squared errors in case of diagonal covariance structures under subtle outlier contaminated data. }
	\label{sm_table1.15}
\end{table}

\begin{table}[!h]
	\centering 
%	\small
	\begin{tabular}{c c c c c c c c c } 
			\hline
		Dimension & Different & & \hspace{-5.5em} Location Vector &  & \hspace{-5.5em}Scatter Matrix\\  
		$(p)$ &Methods &Bias & MSE & Bias & MSE  \\
		\hline
		\multirow{ 4}{*}{}2 & MLE &  2.806   &   7.939  &              71.435    &     5136.492  \\
		& CMDPDE $(\beta=0.1)$ & 0.001 & 0.003 & 0.012 & 0.007 \\
		& CMDPDE $(\beta=0.3)$ & 0.002 & 0.003 & 0.054 & 0.011 \\
		&  CMDPDE $(\beta=0.5)$ & 0.002 & 0.003 & 0.102 & 0.021 \\
        & FRC  &  0.009 &  0.003 & 0.102 & 0.021   \\
		& MCD    &  0.006 & 0.003 & 0.408 &   0.176 \\
		& MVE  &   0.006 & 0.003 & 0.408  &  0.188  \\
		& OGK  &     0.014 & 0.003 & 0.297 &  0.096
		\\
		& MM &  0.004 & 0.003 & 0.310 & 0.109 \\
		& S & 0.004 & 0.004 & 0.313 & 0.119 \\
		& & & & &  \\
		\multirow{ 4}{*}{}5 & MLE &  4.497  &    20.410     &         180.956  &      32986.971  \\
		& CMDPDE $(\beta=0.1)$ & 0.010 & 0.006 & 0.031 & 0.034 \\
		& CMDPDE $(\beta=0.3)$ & 0.010 & 0.006 & 0.098 & 0.049 \\
		& CMDPDE $(\beta=0.5)$ & 0.010 & 0.007 & 0.175 & 0.081 \\
        & FRC  &  0.008 & 0.006 & 0.185 & 0.081  \\
		&MCD  &    0.008 & 0.006 & 0.345 &  0.186 \\
		& MVE  &    0.007  & 0.007 & 0.349 &   0.191  \\
		& OGK  &     0.010 & 0.007 & 0.262 &  0.108 \\
		& MM & 0.005 & 0.006 & 0.406 & 0.214
		\\
		& S & 0.007 & 0.006 & 0.408 & 0.222 \\
		& & & & & & & \\
		\multirow{ 4}{*}{}10 & MLE & 6.281   &   39.757    &         357.579     &    128617.184 \\
		&  CMDPDE $(\beta=0.1)$ &     0.007 & 0.011 & 0.057 & 0.130\\
		& CMDPDE $(\beta=0.3)$ &  0.007 & 0.012 & 0.144 & 0.175 \\
		& CMDPDE $(\beta=0.5)$ &  0.006 & 0.013 & 0.250 & 0.259  \\
        & FRC & 0.003 & 0.012 & 0.258 & 0.228  \\
		& MCD   &   0.009 & 0.012 &  0.319   & 0.273  \\
		& MVE   &   0.008 & 0.013 &  0.324  & 0.296 \\
		& OGK   &    0.014 & 0.013 & 0.260  & 0.201  \\
		& MM & 0.008 & 0.012 & 0.549 & 0.477
		\\
		& S & 0.009 & 0.012 & 0.550 & 0.482 \\
		%& & & & & \\
		%\multirow{ 4}{*}{}20 & MLE &   8.971   &   81.329       &        721.061   &      524251.729\\
		%& CMDPDE $(\beta=0.1)$ &    0.015 & 0.024 & 0.090 & 0.500
		%\\
		%&  CMDPDE $(\beta=0.3)$ &      0.016 & 0.026 & 0.210 & 0.640
		%\\
		%& CMDPDE $(\beta=0.5)$ &   0.017 & 0.028 & 0.362 & 0.889
		%\\
		%& MCD    &           0.0131 & 0.026 & 0.330 &  0.709
		%\\
		%& MVE  &            0.014 & 0.026 & 0.321  & 0.721
		%\\
		%& GK    &      0.017 & 0.028 & 0.248  & 0.570  \\
		%& MM & 0.015 & 0.023 & 0.732 & 1.198 \\
		%& S &  0.015 & 0.023 & 0.731 & 1.189\\
		%&  ~~~~~~~~~ & ~~~~~   ~~~~~~~  &                      ~~~~~   ~~~~~~~  & ~~~~~   ~~~~~~~   &                      ~~~~~   ~~~~~~~ \\
		%\multirow{ 4}{*}{}30 & MLE &  11.077  &  123.377  & 1089.409  &  1191872.749\\
		%& CMDPDE $(\beta=0.1)$ & 0.040 & 0.034 & 0.225 & %1.101\\
		%& CMDPDE $(\beta=0.3)$ & 0.043 &  0.035 & 0.328 & 1.395 \\
		%&  CMDPDE $(\beta=0.5)$ & 0.046 &  0.038 & 0.494 &  1.896 \\
		%& 	MCD  & 0.033 & 0.037 & 0.379  & 1.392 \\
		%& MVE &   0.030 & 0.037 & 0.377 &  1.402 \\
		%& GK  &  0.034 & 0.039 &  0.327 &  1.196\\
		%& MM &   0.033 &  0.036 &  0.922 &  2.234 \\
		%& S &   0.033 & 0.036 &  0.919 &  2.202\\
		\hline 
	\end{tabular}
	\caption{Estimated bias and mean squared errors in case of diagonal covariance structures under distant outlier contaminated data. }
	\label{sm_table1.2}
	
\end{table}
\begin{table}[!h]
	\centering 
%	\small
	\begin{tabular}{c c c c c c c c c } 
		\hline
		Dimension & Different & & \hspace{-5.5em} Location Vector &  & \hspace{-5.5em}Scatter Matrix\\  
		$(p)$ &Methods &Bias & MSE & Bias & MSE  \\
		\hline
		2	&	MLE	&	0.004	&	0.002	&	0.006	&	0.006	\\
		&	CMDPDE $(\beta=0.1)$	&	0.004	&	0.002	&	0.006	&	0.007	\\
		&	CMDPDE $(\beta=0.3)$	&	0.004	&	0.002	&	0.005	&	0.008	\\
		&	CMDPDE $(\beta=0.5)$	&	0.004	&	0.003	&	0.004	&	0.009	\\
        & FRC &  0.007 & 0.002 &  0.038 & 0.015  \\
		&	MCD	&	0.004	&	0.002	&	0.020	&	0.012	\\
		&	MVE	&	0.001	&	0.002	&	0.008	&	0.013	\\
		&	OGK	&	0.005	&	0.002	&	0.434	&	0.195	\\
		&	MM	&  0.004 & 0.003 & 0.006 & 0.009 \\
		&	S	& 0.006 &  0.004 & 0.008 & 0.017 \\
		&	&	&	&	\\
		5	&	MLE	&	0.004	&	0.004	&	0.029	&	0.032	\\
		&	CMDPDE $(\beta=0.1)$	&	0.004	&	0.005	&	0.029	&	0.034	\\
		&	CMDPDE $(\beta=0.3)$	&	0.004	&	0.005	&	0.030	&	0.039	\\
		&	CMDPDE $(\beta=0.5)$	&	0.004	&	0.006	&	0.030	&	0.046	\\
        & FRC &  0.012 & 0.005 & 0.039 & 0.042 \\
		&	MCD	&	0.013	&	0.005	&	0.019	&	0.043	\\
		&	MVE	&	0.013	&	0.006	&	0.042	&	0.060	\\
		&	OGK	&	0.014	&	0.006	&	0.364	&	0.166	\\
		&	MM	& 0.007 & 0.005 & 0.018 & 0.029
		\\
		&	S	&  0.007 & 0.006 & 0.021 & 0.035 \\
		&	&	&	&	&	&	&	\\
		10	&	MLE	&	0.007	&	0.010	&	0.036	&	0.115	\\
		&	CMDPDE $(\beta=0.1)$	&	0.006	&	0.010	&	0.037	&	0.118	\\
		&	CMDPDE $(\beta=0.3)$	&	0.006	&	0.010	&	0.040	&	0.136	\\
		&	CMDPDE $(\beta=0.5)$	&	0.006	&	0.011	&	0.044	&	0.163	\\
        &  FRC &  0.009 & 0.011 & 0.120 &  0.153  \\
		&	MCD	&	0.006	&	0.011	&	0.037	&	0.133	\\
		&	MVE	&	0.006	&	0.012	&	0.050	&	0.156	\\
		&	OGK	&	0.006	&	0.012	&	0.428	&	0.305	\\
		&	MM	&  0.011 & 0.009 & 0.027 & 0.118  \\
		&	S	& 0.011 & 0.009 & 0.026 & 0.122 \\
		%&	&	&	&	&		\\
		%20	&	MLE	&	0.014	&	0.02	&	0.066	&	0.443	\\
		%&	CMDPDE $(\beta=0.1)$	&	0.014	&	0.02	&	0.064	&	0.455	\\
		%&	CMDPDE $(\beta=0.3)$	&	0.015	&	0.021	&	0.066	&	0.525	\\
		%&	CMDPDE $(\beta=0.5)$	&	0.016	&	0.023	&	0.072	&	0.631	\\
		%&	MCD	&	0.018	&	0.021	&	0.073	&	0.498	\\
		%&	MVE	&	0.019	&	0.022	&	0.081	&	0.527	\\
		%&	GK	&	0.018	&	0.022	&	0.45	&	0.672	\\
		%&	MM	& 0.019 & 0.021 & 0.070 & 0.455 \\
		%&	S	& 0.019 &  0.020 & 0.070 & 0.447  \\
		
		%&  ~~~~~~~~~ & ~~~~~   ~~~~~~~  &                      ~~~~~   ~~~~~~~  & ~~~~~   ~~~~~~~   &                      ~~~~~   ~~~~~~~ \\
		%\multirow{ 4}{*}{}30 & MLE & 0.032 & 0.032 & 0.191 & 0.973 \\
		%& CMDPDE $(\beta=0.1)$ & 0.032 & 0.032 & 0.195 & 0.999 \\
		%& CMDPDE $(\beta=0.3)$ & 0.034 & 0.033 & 0.209 & 1.142 \\
		%&  CMDPDE $(\beta=0.5)$ & 0.036 & 0.037 & 0.227 & 1.365 \\
		%& 	MCD  & 0.036 & 0.032 & 0.203 & 1.114\\
		%& MVE &   0.037 & 0.034 & 0.211 & 1.136
		% \\
		%& GK  &  0.037 & 0.035 & 0.494 & 1.269\\
		%& MM &   0.039 & 0.032 & 0.213 & 1.071 \\
		%& S &   0.038 & 0.032 & 0.209 & 1.037\\
		\hline
	\end{tabular}
	\caption{Estimated bias and mean squared errors in case of non-diagonal covariance structures under pure data. }
	\label{sm_table2.1}
\end{table}

\begin{table}[!h]
	\centering 
%	\small
	\begin{tabular}{c c c c c c c c c } 
		
		\hline
		 Dimension & Different & & \hspace{-5.5em} Location Vector &  & \hspace{-5.5em}Scatter Matrix\\  
		 $(p)$ &Methods &Bias & MSE & Bias & MSE  \\
		\hline
		\multirow{ 4}{*}{} 2 &MLE & 0.063 & 0.006 & 0.129 & 0.023   \\
		 &CMDPDE $(\beta=0.5)$ &   0.062 & 0.006 & 0.056 & 0.013     \\
		 & CMDPDE $(\beta=0.7)$ &   0.061 & 0.006 & 0.050 & 0.014  \\
		 & FRC &   0.063 & 0.006 & 0.152 & 0.037
     \\
		 & MCD & 0.024 & 0.003 & 0.065 & 0.021 \\
		 & MVE & 0.024 & 0.004 & 0.067 & 0.024\\
		 & OGK &   0.024 & 0.003 & 0.483 & 0.240 \\
		 & MM &  0.040 & 0.003 & 0.123 & 0.023  \\
		 & S &  0.016 & 0.003 & 0.049 & 0.017  \\
		 & & & & &  \\
		\multirow{ 4}{*}{}  5 & MLE & 0.060 & 0.009 & 0.119 & 0.045  \\
		 & CMDPDE $(\beta=0.1)$ & 0.059 & 0.009 & 0.088 & 0.039  \\
		 & CMDPDE $(\beta=0.3)$ & 0.057 & 0.009 & 0.059 & 0.039 \\
		 & FRC  &  0.058 & 0.009 & 0.136 & 0.058 \\
		 & MCD  &  0.036 & 0.007 & 0.073 & 0.045 \\
		 &  MVE & 0.036 & 0.008 & 0.081 & 0.058  \\
		 & OGK &  0.032 & 0.008 & 0.493 & 0.286   \\
		 & MM &  0.045 & 0.007 & 0.095 & 0.040  \\
		 & S &  0.034 & 0.007 & 0.071 & 0.042   \\
		 & & & & & & & \\
		\multirow{ 4}{*}{}10 & MLE &  0.037 & 0.011 & 0.264 & 0.188\\
		& CMDPDE $(\beta=0.1)$ &   0.037 & 0.011 & 0.230 & 0.174   \\
		& CMDPDE $(\beta=0.3)$ &  0.036 & 0.011 & 0.195 & 0.176  \\
	& FRC &  0.036 & 0.012 & 0.353 & 0.265 \\
		& MCD &  0.019 & 0.012 & 0.119 & 0.162  \\
		& MVE  &  0.020 & 0.013 & 0.120 & 0.179 \\
		& OGK  &  0.021 & 0.013 & 0.528 & 0.405  \\
		& MM &  0.025 & 0.010 & 0.194 & 0.151   \\
		& S &  0.022 & 0.010 & 0.166 & 0.147   \\
		
		\hline 
	\end{tabular}
	\caption{Estimated bias and mean squared errors in case of non-diagonal covariance structures under subtle outlier contaminated data. }
	\label{sm_table2.15}
\end{table}

\begin{table}[h!]
	\centering 
%	\small
	\begin{tabular}{c c c c c c c c c } 
		\hline
		Dimension & Different & & \hspace{-5.5em} Location Vector &  & \hspace{-5.5em}Scatter Matrix\\  
		$(p)$ &Methods &Bias & MSE & Bias & MSE  \\
		\hline
		2	&	MLE	&	2.829	&	8.071	&	71.912	&	5205.541	&	\\
		&	CMDPDE $(\beta=0.1)$	&	0.005	&	0.002	&	0.009	&	0.007	&	\\
		&	CMDPDE $(\beta=0.3)$	&	0.005	&	0.002	&	0.063	&	0.012	&	\\
		&	CMDPDE $(\beta=0.5)$	&	0.006	&	0.003	&	0.125	&	0.025	&	\\
         &  FRC & 0.005 & 0.002 & 0.125 & 0.027  \\
		&	MCD	&	0.002	&	0.003	&	0.505	&	0.201	&	\\
       &	MVE	&	0.003	&	0.003	&	0.507	&	0.220	&	\\
		&	OGK	&	0.011	&	0.003	&	0.290	&	0.092	&	\\
		&	MM	&  0.006 & 0.002 & 0.392 & 0.167 \\
		&	S	& 0.009 & 0.004 & 0.392 & 0.173  \\
		&	&	&	&	&	&	&	&	\\
		5	&	MLE	&	4.434	&	19.877	&	178.341	&	32072.472	&	\\
		&	CMDPDE $(\beta=0.1)$	&	0.009	&	0.005	&	0.034	&	0.037	&	\\
		&	CMDPDE $(\beta=0.3)$	&	0.010	&	0.006	&	0.107	&	0.055	&	\\
		&	CMDPDE $(\beta=0.5)$	&	0.011	&	0.006	&	0.191	&	0.091	&	\\
        & FRC &  0.008 & 0.006 & 0.191 & 0.080 \\
		&	MCD	&	0.002	&	0.006	&	0.379	&	0.272	&	\\
		&	MVE	&	0.003	&	0.006	&	0.388	&	0.279	&	\\
		&	OGK	&	0.011	&	0.006	&	0.285	&	0.124	&	\\
		&	MM	&  0.008 & 0.005 & 0.452 & 0.256 \\
		&	S	&  0.008 &  0.006 & 0.454 & 0.266 \\
		&	&	&	&	&	&	&	&		\\
		10	&	MLE	&	6.265	&	39.664	&	357.366	&	128759.552	&	\\
		&	CMDPDE $(\beta=0.1)$	&	0.025	&	0.012	&	0.043	&	0.140	&	\\
		&	CMDPDE $(\beta=0.3)$	&	0.025	&	0.013	&	0.157	&	0.191	&	\\
		&	CMDPDE $(\beta=0.5)$	&	0.024	&	0.014	&	0.298	&	0.300	&	\\
        & FRC &  0.010 & 0.012 & 0.260 & 0.236  \\
		&	MCD	&	0.012	&	0.012	&	0.405	&	0.337	&	\\
		&	MVE	&	0.010	&	0.012	&	0.411	&	0.352	&	\\
		&	OGK	&	0.009	&	0.013	&	0.311	&	0.240	&	\\
		&	MM	&   0.009 &  0.011 & 0.706 & 0.697  \\
		&	S	&  0.009 & 0.011 & 0.708 & 0.705  \\
		&	&	&	&	&	&	&	&	\\
		%20	&	MLE	&	8.858	&	79.225	&	712.878	&	512080.801	&	\\
		%&	CMDPDE $(\beta=0.1)$	&	0.012	&	0.022	&	0.08	&	0.529	&	\\
		%&	CMDPDE $(\beta=0.3)$	&	0.013	&	0.024	&	0.236	&	0.683	&	\\
		%&	CMDPDE $(\beta=0.5)$	&	0.014	&	0.026	&	0.442	&	0.986	&	\\
		%&	MCD	&	0.013	&	0.024	&	0.395	&	0.76	&	\\
		%&	MVE	&	0.014	&	0.024	&	0.39	&	0.775	&	\\
		%&	GK	&	0.013	&	0.025	&	0.349	&	0.641	&	\\
		%&	MM	&  0.016 & 0.024 & 0.952 & 1.579 \\
		%&	S	&   0.016 & 0.024 & 0.951 & 1.571 \\
		
		%&  ~~~~~~~~~ & ~~~~~   ~~~~~~~  &                      ~~~~~   ~~~~~~~  & ~~~~~   ~~~~~~~   &                      ~~~~~   ~~~~~~~ \\
		%\multirow{ 4}{*}{}30 & MLE & 11.061  &  123.419 & 1088.910 &   1193694.140 \\
		%& CMDPDE $(\beta=0.1)$ & 0.031 &   0.035 & 0.231 & 1.116  \\
		%& CMDPDE $(\beta=0.3)$ & 0.033 &  0.037 & 0.393 & 1.450 \\
		%&  CMDPDE $(\beta=0.5)$ & 0.034  &  0.039 & 0.645 & 2.073 \\
		%& 	MCD  & 0.042 &   0.037 & 0.441 &  1.487\\
		%& MVE &    0.045 &  0.037 & 0.436 & 1.487\\
		%& GK  &  0.045 &  0.040 & 0.419 & 1.331\\
		%& MM &   0.034  &   0.035 & 1.207 & 2.952 \\
		%& S &    0.034 &   0.035 & 1.202 & 2.907\\
		\hline 
	\end{tabular}
	\caption{Estimated bias and mean squared errors in case of non-diagonal covariance structures under distant outlier contaminated data. }
	\label{sm_table2.2}
\end{table}
\newpage
\textcolor{white}{,}
\newpage
\textcolor{white}{,}
\newpage
\textcolor{white}{,}
\newpage
\textcolor{white}{,}
\newpage
\textcolor{white}{,}
\section{Comparison between CMDPDE and MDPDE}
\label{sm_appen4}
We consider only distant outlier contamination set-up (along with pure data) for comparing CMDPDEs with MDPDEs.
%\newpage 

\begin{table}[!t]
	\centering 
	\small
	\begin{tabular}{c c c c c c c c  } 
		\hline
		Data & Covariance & Dimension &  & & \hspace{-4em} CMDPDE &  & \hspace{-4em}MDPDE\\  
		Type & Structure & $(p)$ & $\beta$ &Bias & MSE & Bias & \hspace{1em}MSE  \\
		\hline
		Pure	&	Diagonal	&	2	&	0.1	&	0.004	&	0.002	&	0.005	&	0.002	\\
		&		&		&	0.3	&	0.004	&	0.002	&	0.004	&	0.002	\\
		&		&		&	0.5	&	0.004	&	0.002	&	0.004	&	0.002	\\
		&		&		&		&	&	&	&		\\
		&		&	5	&	0.1	&	0.009	&	0.005	&	0.009	&	0.005	\\
		&		&		&	0.3	&	0.009	&	0.005	&	0.008	&	0.006	\\
		&		&		&	0.5	&	0.008	&	0.006	&	0.007	&	0.008	\\
		&		&		&		&	&	&	&		\\
		&		&	10	&	0.1	&	0.013	&	0.010	&	0.012	&	0.010	\\
		&		&		&	0.3	&	0.013	&	0.011	&	0.011	&	0.014	\\
		&		&		&	0.5	&	0.013	&	0.012	&	0.011	&	0.020	\\
		&		&		&		&	&	&	&		\\
		&	Non-Diagonal	&	2	&	0.1	&	0.004	&	0.002	&	0.004	&	0.002	\\
		&		&		&	0.3	&	0.004	&	0.002	&	0.004	&	0.002	\\
		&		&		&	0.5	&	0.004	&	0.003	&	0.004	&	0.003	\\
		&		&		&		&	&	&	&		\\
		&		&	5	&	0.1	&	0.004	&	0.005	&	0.005	&	0.005	\\
		&		&		&	0.3	&	0.004	&	0.005	&	0.006	&	0.006	\\
		&		&		&	0.5	&	0.004	&	0.006	&	0.007	&	0.008	\\
		&		&		&		&	&	&	&		\\
		&		&	10	&	0.1	&	0.006	&	0.010	&	0.007	&	0.010	\\
		&		&		&	0.3	&	0.006	&	0.010	&	0.009	&	0.013	\\
		&		&		&	0.5	&	0.006	&	0.011	&	0.011	&	0.020	\\
		&		&		&		&	&	&	&	\\
		&		&		&		&	&	&	&		\\
		Contaminated	&	Diagonal	&	2	&	0.1	&	0.001	&	0.003	&	0.001	&	0.003	\\
		&		&		&	0.3	&	0.002	&	0.003	&	0.001	&	0.003	\\
		&		&		&	0.5	&	0.002	&	0.003	&	0.001	&	0.003	\\
		&		&		&		&	&	&	&		\\
		&		&	5	&	0.1	&	0.010	&	0.006	&	0.009	&	0.006	\\
		&		&		&	0.3	&	0.010	&	0.006	&	0.010	&	0.007	\\
		&		&		&	0.5	&	0.010	&	0.007	&	0.010	&	0.008	\\
		&		&		&		&	&	&	&		\\
		&		&	10	&	0.1	&	0.007	&	0.011	&	0.008	&	0.012	\\
		&		&		&	0.3	&	0.007	&	0.012	&	0.009	&	0.014	\\
		&		&		&	0.5	&	0.006	&	0.013	&	0.012	&	0.020	\\
		&		&		&		&	&	&	&		\\
		&	Non-Diagonal	&	2	&	0.1	&	0.005	&	0.002	&	0.005	&	0.002	\\
		&		&		&	0.3	&	0.005	&	0.002	&	0.006	&	0.002	\\
		&		&		&	0.5	&	0.006	&	0.003	&	0.006	&	0.003	\\
		&		&		&		&	&	&	&		\\
		&		&	5	&	0.1	&	0.009	&	0.005	&	0.011	&	0.005	\\
		&		&		&	0.3	&	0.010	&	0.006	&	0.013	&	0.007	\\
		&		&		&	0.5	&	0.011	&	0.006	&	0.015	&	0.008	\\
		&		&		&		&	&	&	&		\\
		&		&	10	&	0.1	&	0.025	&	0.012	&	0.026	&	0.013	\\
		&		&		&	0.3	&	0.025	&	0.013	&	0.026	&	0.017	\\
		&		&		&	0.5	&	0.024	&	0.014	&	0.027	&	0.024	\\
		\hline 
	\end{tabular}
	\caption{Estimated bias and mean squared errors of mean estimators for the componentwise and ordinary minimum DPD methods. }
	\label{sm_table5}
	
\end{table}
\begin{table}[!t]
	\centering 
	\small 
	\begin{tabular}{c c c c c c c c  } 
		\hline
		Data & Covariance & Dimension &  & & \hspace{-4em} CMDPDE &  & \hspace{-4em}MDPDE\\  
		Type & Structure & $(p)$ & $\beta$ &Bias & MSE & Bias & \hspace{1em}MSE  \\
		\hline
		Pure	&	Diagonal	&	2	&	0.1	&	0.006	&	0.004	&	0.006	&	0.004	\\
		&		&		&	0.3	&	0.007	&	0.005	&	0.006	&	0.005	\\
		&		&		&	0.5	&	0.006	&	0.006	&	0.005	&	0.006	\\
		&		&		&		&	&	&	&		\\
		&		&	5	&	0.1	&	0.006	&	0.010	&	0.007	&	0.010	\\
		&		&		&	0.3	&	0.007	&	0.011	&	0.007	&	0.013	\\
		&		&		&	0.5	&	0.008	&	0.013	&	0.009	&	0.017	\\
		&		&		&		&	&	&	&		\\
		&		&	10	&	0.1	&	0.008	&	0.021	&	0.009	&	0.022	\\
		&		&		&	0.3	&	0.009	&	0.024	&	0.010	&	0.030	\\
		&		&		&	0.5	&	0.010	&	0.028	&	0.013	&	0.046	\\
		&		&		&		&	&	&	&		\\
		&	Non-Diagonal	&	2	&	0.1	&	0.005	&	0.004	&	0.005	&	0.004	\\
		&		&		&	0.3	&	0.005	&	0.005	&	0.005	&	0.005	\\
		&		&		&	0.5	&	0.004	&	0.006	&	0.005	&	0.006	\\
		&		&		&		&	&	&	&		\\
		&		&	5	&	0.1	&	0.014	&	0.011	&	0.015	&	0.011	\\
		&		&		&	0.3	&	0.016	&	0.012	&	0.018	&	0.014	\\
		&		&		&	0.5	&	0.017	&	0.015	&	0.022	&	0.018	\\
		&		&		&		&	&	&	&		\\
		&		&	10	&	0.1	&	0.019	&	0.021	&	0.018	&	0.021	\\
		&		&		&	0.3	&	0.020	&	0.024	&	0.013	&	0.028	\\
		&		&		&	0.5	&	0.020	&	0.028	&	0.014	&	0.044	\\
		&		&		&		&	&	&	&		\\
		&		&		&		&	&	&	&		\\
		Contaminated	&	Diagonal	&	2	&	0.1	&	0.012	&	0.005	&	0.012	&	0.005	\\
		&		&		&	0.3	&	0.054	&	0.008	&	0.051	&	0.008	\\
		&		&		&	0.5	&	0.102	&	0.017	&	0.090	&	0.015	\\
		&		&		&		&	&	&	&		\\
		&		&	5	&	0.1	&	0.028	&	0.011	&	0.027	&	0.011	\\
		&		&		&	0.3	&	0.097	&	0.022	&	0.077	&	0.019	\\
		&		&		&	0.5	&	0.174	&	0.046	&	0.11	&	0.030	\\
		&		&		&		&	&	&	&		\\
		&		&	10	&	0.1	&	0.046	&	0.024	&	0.045	&	0.025	\\
		&		&		&	0.3	&	0.139	&	0.046	&	0.093	&	0.043	\\
		&		&		&	0.5	&	0.247	&	0.094	&	0.107	&	0.067	\\
		&		&		&		&	&	&	&	\\
		&	Non-Diagonal	&	2	&	0.1	&	0.008	&	0.004	&	0.008	&	0.004	\\
		&		&		&	0.3	&	0.051	&	0.008	&	0.048	&	0.007	\\
		&		&		&	0.5	&	0.100	&	0.016	&	0.088	&	0.014	\\
		&		&		&		&	&	&	&		\\
		&		&	5	&	0.1	&	0.031	&	0.013	&	0.030	&	0.013	\\
		&		&		&	0.3	&	0.098	&	0.024	&	0.080	&	0.023	\\
		&		&		&	0.5	&	0.174	&	0.048	&	0.113	&	0.035	\\
		&		&		&		&	&	&	&		\\
		&		&	10	&	0.1	&	0.030	&	0.026	&	0.029	&	0.026	\\
		&		&		&	0.3	&	0.126	&	0.046	&	0.081	&	0.041	\\
		&		&		&	0.5	&	0.234	&	0.092	&	0.097	&	0.061	\\
		
		\hline 
	\end{tabular}
	\caption{Estimated bias and mean squared errors of variance estimators for the componentwise and ordinary minimum DPD methods. }
	\label{sm_table6}
	
\end{table}
\begin{table}[!t]
	\centering 
	\small 
	\begin{tabular}{c c c c c c c c  } 
		\hline
		Data & Covariance & Dimension &  & & \hspace{-4em} CMDPDE &  & \hspace{-4em}MDPDE\\  
		Type & Structure & $(p)$ & $\beta$ &Bias & MSE & Bias & \hspace{1em}MSE  \\
		\hline
		Pure	&	Diagonal	&	2	&	0.1	&	0.002	&	0.001	&	0.002	&	0.001	\\
		&		&		&	0.3	&	0.001	&	0.001	&	0.001	&	0.001	\\
		&		&		&	0.5	&	0.001	&	0.002	&	0.001	&	0.002	\\
		&		&		&		&	&	&	&		\\
		&		&	5	&	0.1	&	0.009	&	0.010	&	0.009	&	0.010	\\
		&		&		&	0.3	&	0.011	&	0.011	&	0.013	&	0.012	\\
		&		&		&	0.5	&	0.013	&	0.014	&	0.017	&	0.017	\\
		&		&		&		&	&	&	&		\\
		&		&	10	&	0.1	&	0.019	&	0.046	&	0.019	&	0.047	\\
		&		&		&	0.3	&	0.020	&	0.053	&	0.022	&	0.066	\\
		&		&		&	0.5	&	0.021	&	0.064	&	0.026	&	0.105	\\
		&		&		&		&	&	&	&	\\
		&	Non-Diagonal	&	2	&	0.1	&	0.001	&	0.001	&	0.001	&	0.001	\\
		&		&		&	0.3	&	0.001	&	0.001	&	0.001	&	0.001	\\
		&		&		&	0.5	&	0.001	&	0.001	&	0.001	&	0.001	\\
		&		&		&		&	&	&	&	\\
		&		&	5	&	0.1	&	0.016	&	0.009	&	0.016	&	0.009	\\
		&		&		&	0.3	&	0.015	&	0.011	&	0.015	&	0.012	\\
		&		&		&	0.5	&	0.015	&	0.013	&	0.016	&	0.016	\\
		&		&		&		&	&	&	&	\\
		&		&	10	&	0.1	&	0.019	&	0.041	&	0.018	&	0.042	\\
		&		&		&	0.3	&	0.021	&	0.048	&	0.023	&	0.059	\\
		&		&		&	0.5	&	0.025	&	0.058	&	0.034	&	0.096	\\
		&		&		&		&	&	&	&		\\
		&		&		&		&	&	&	&		\\
		Contaminated	&	Diagonal	&	2	&	0.1	&	0.003	&	0.001	&	0.003	&	0.001	\\
		&		&		&	0.3	&	0.004	&	0.002	&	0.004	&	0.002	\\
		&		&		&	0.5	&	0.005	&	0.002	&	0.005	&	0.002	\\
		&		&		&		&	&	&	&		\\
		&		&	5	&	0.1	&	0.009	&	0.011	&	0.009	&	0.011	\\
		&		&		&	0.3	&	0.010	&	0.013	&	0.011	&	0.014	\\
		&		&		&	0.5	&	0.011	&	0.015	&	0.014	&	0.019	\\
		&		&		&		&	&	&	&	\\
		&		&	10	&	0.1	&	0.023	&	0.052	&	0.024	&	0.053	\\
		&		&		&	0.3	&	0.025	&	0.059	&	0.029	&	0.073	\\
		&		&		&	0.5	&	0.028	&	0.071	&	0.035	&	0.115	\\
		&		&		&		&	&	&	&		\\
		&	Non-Diagonal	&	2	&	0.1	&	0.001	&	0.001	&	0.001	&	0.001	\\
		&		&		&	0.3	&	0.001	&	0.001	&	0.001	&	0.001	\\
		&		&		&	0.5	&	0.001	&	0.001	&	0.001	&	0.001	\\
		&		&		&		&	&	&	&		\\
		&		&	5	&	0.1	&	0.006	&	0.010	&	0.005	&	0.010	\\
		&		&		&	0.3	&	0.006	&	0.012	&	0.004	&	0.013	\\
		&		&		&	0.5	&	0.006	&	0.014	&	0.004	&	0.017	\\
		&		&		&		&	&	&	&		\\
		&		&	10	&	0.1	&	0.019	&	0.045	&	0.020	&	0.047	\\
		&		&		&	0.3	&	0.022	&	0.052	&	0.026	&	0.066	\\
		&		&		&	0.5	&	0.026	&	0.063	&	0.033	&	0.103	\\
		
		\hline 
	\end{tabular}
	\caption{Estimated bias and mean squared errors of correlation estimators for the componentwise and ordinary minimum DPD methods. }
	\label{sm_table7}
	
\end{table}

\begin{table}[!t]
	\centering 
	\small 
	\begin{tabular}{c c c c c c c c  } 
		\hline
		Data & Covariance & Dimension &  & & \hspace{-4em} CMDPDE &  & \hspace{-4em}MDPDE\\  
		Type & Structure & $(p)$ & $\beta$ &Bias & MSE & Bias & \hspace{1em}MSE  \\
		\hline
		Pure	&	Diagonal	&	2	&	0.1	&	0.007	&	0.007	&	0.007	&	0.007	\\
		&		&		&	0.3	&	0.007	&	0.008	&	0.006	&	0.008	\\
		&		&		&	0.5	&	0.006	&	0.009	&	0.005	&	0.010	\\
		&		&		&		&	&	&	&		\\
		&		&	5	&	0.1	&	0.014	&	0.029	&	0.015	&	0.030	\\
		&		&		&	0.3	&	0.017	&	0.034	&	0.020	&	0.037	\\
		&		&		&	0.5	&	0.020	&	0.041	&	0.025	&	0.050	\\
		&		&		&		&	&	&	&		\\
		&		&	10	&	0.1	&	0.028	&	0.113	&	0.029	&	0.117	\\
		&		&		&	0.3	&	0.029	&	0.130	&	0.033	&	0.161	\\
		&		&		&	0.5	&	0.031	&	0.156	&	0.039	&	0.257	\\
		&		&		&		&	&	&	&		\\
		&	Non-Diagonal	&	2	&	0.1	&	0.006	&	0.007	&	0.006	&	0.007	\\
		&		&		&	0.3	&	0.005	&	0.008	&	0.006	&	0.008	\\
		&		&		&	0.5	&	0.004	&	0.009	&	0.005	&	0.010	\\
		&		&		&		&	&	&	&		\\
		&		&	5	&	0.1	&	0.029	&	0.034	&	0.030	&	0.034	\\
		&		&		&	0.3	&	0.030	&	0.039	&	0.032	&	0.042	\\
		&		&		&	0.5	&	0.030	&	0.046	&	0.035	&	0.054	\\
		&		&		&		&	&	&	&	\\
		&		&	10	&	0.1	&	0.037	&	0.118	&	0.035	&	0.121	\\
		&		&		&	0.3	&	0.040	&	0.136	&	0.037	&	0.168	\\
		&		&		&	0.5	&	0.044	&	0.163	&	0.050	&	0.270	\\
		&		&		&		&	&	&	&		\\
		&		&		&		&	&	&	&		\\
		Contaminated	&	Diagonal	&	2	&	0.1	&	0.012	&	0.007	&	0.012	&	0.007	\\
		&		&		&	0.3	&	0.054	&	0.011	&	0.052	&	0.011	\\
		&		&		&	0.5	&	0.102	&	0.021	&	0.090	&	0.019	\\
		&		&		&		&	&	&	&		\\
		&		&	5	&	0.1	&	0.031	&	0.034	&	0.030	&	0.034	\\
		&		&		&	0.3	&	0.098	&	0.049	&	0.079	&	0.049	\\
		&		&		&	0.5	&	0.175	&	0.081	&	0.111	&	0.071	\\
		&		&		&		&	&	&	&		\\
		&		&	10	&	0.1	&	0.057	&	0.130	&	0.057	&	0.134	\\
		&		&		&	0.3	&	0.144	&	0.175	&	0.102	&	0.197	\\
		&		&		&	0.5	&	0.250	&	0.259	&	0.119	&	0.313	\\
		&		&		&		&	&	&	&	\\
		&	Non-Diagonal	&	2	&	0.1	&	0.009	&	0.007	&	0.009	&	0.007	\\
		&		&		&	0.3	&	0.063	&	0.012	&	0.059	&	0.011	\\
		&		&		&	0.5	&	0.125	&	0.025	&	0.108	&	0.022	\\
		&		&		&		&	&	&	&		\\
		&		&	5	&	0.1	&	0.034	&	0.037	&	0.033	&	0.038	\\
		&		&		&	0.3	&	0.107	&	0.055	&	0.087	&	0.056	\\
		&		&		&	0.5	&	0.191	&	0.091	&	0.124	&	0.080	\\
		&		&		&		&	&	&	&		\\
		&		&	10	&	0.1	&	0.043	&	0.140	&	0.043	&	0.143	\\
		&		&		&	0.3	&	0.157	&	0.191	&	0.105	&	0.207	\\
		&		&		&	0.5	&	0.298	&	0.300	&	0.127	&	0.320	\\
		
		\hline 
	\end{tabular}
	\caption{Estimated bias and mean squared errors of covariance matrix estimators for the componentwise and ordinary minimum DPD methods. }
	\label{sm_table7.1}
	
\end{table}
%The ordinary MDPDE can perform better than the CMDPDE in lower dimensions. 
A comparative study on bias and mean squared errors of the mean, variance and correlation estimators are presented separately in Tables \ref{sm_table5}, \ref{sm_table6} and \ref{sm_table7}, respectively. Let us discuss the implications of these comparative studies in the following points.
\begin{enumerate}
	\item For the mean estimators (Table \ref{sm_table5}), it can be observed that both MDPDEs and CMDPDEs have almost similar bias and MSEs in lower dimension $(p=2)$. But as $p$ increases to $5$ or $10$, the CMDPDEs tend to have less bias and MSE, especially under higher values of $\beta$ ($0.5$, in particular) for pure as well as contaminated datasets. 
	
	\item In case of the variance estimators (Table \ref{sm_table6}), the bias and MSEs of both CMDPDEs and MDPDEs are similar in lower dimension but as dimension increases the CMDPDEs tend to have more bias for pure datasets and more bias and MSEs for contaminated datasets with higher $\beta$ values ($0.3$ and $0.5$, in particular).
	
	\item In case of correlation estimators (Table \ref{sm_table7}), both the CMDPDEs and MDPDEs are comparable in terms of bias and MSEs under lower dimension but as the dimension grows up, the CMDPDEs become more accurate (in terms of bias and MSEs) under higher values of $\beta$ under both pure and contaminated datasets. 
	
	\item Now, for the entire covariance matrix estimators (Table \ref{sm_table7.1}), both CMDPDEs and MDPDEs perform quite similarly for smaller values of $\beta$. As $\beta$ increases to $0.3$ or $0.5$, the MDPDEs become slightly less biased, although their MSEs still remain comparable. 
	%This observation is quite ambiguous and it should be analysed from two different angles. Firstly, it should be observed that the MDPDEs are more or less better in terms of bias and equivalent in terms of MSEs with the CMDPDEs under contaminated datasets. This possibly indicates the superiority of the MDPDEs over the CMDPDEs in terms of robustness. On the other hand, it should also be noted that, the aforesaid observation clearly indicates that the variances of these estimated covariance matrix elements are less in case of CMDPDEs which in turn describes the superiority of CMDPDEs over MDPDEs in terms of efficiency that was already established theoretically. So, the latest modification of ordinary minimum DPD method (i.e., SMPDPD method) achieves greater efficiency with a little loss in robustness.   
\end{enumerate}

\end{appendices}

\end{document}